%% file: Salary_History_Ban_Strategic_Disclosure_and_Pay_Disparity.tex
\documentclass[11pt]{article}
\usepackage{authblk}
\usepackage{enumitem}
\usepackage[latin1]{inputenc}
\usepackage{graphics,tikz}
\usepackage{framed,color}
\usepackage{hyperref}
\usepackage{amsmath,enumerate}
\usepackage{amsbsy, amssymb, amsthm,bm}
\usepackage{mathrsfs}
\usepackage[left=0.8in,top=1in,right=0.8in,bottom=0.7in]{geometry}
\usepackage{framed}
\usepackage{array, multirow}
\usepackage{tikz}
\usepackage{float}
\usetikzlibrary{shapes,arrows,matrix,positioning,chains}
\usepackage{fancyhdr}
\usepackage{graphicx}
\usepackage{caption}
\usepackage{subfigure}
\usepackage{booktabs,dcolumn}
\usepackage{tabularx}
\usepackage{rotating}
\graphicspath{{D:/Work/Research/Box Sync/Salary History Disclosure/Visualization}}
\DeclareGraphicsExtensions{.pdf,.jpeg,.png}
\usepackage{epstopdf}
\usepackage{setspace}
\usepackage[authormarkup=superscript,final]{changes}
\definechangesauthor[name={Sourav},color=red]{Sinha}
\usepackage{xspace}
\usepackage{mwe}
\let\OldS\S
\renewcommand{\S}{\OldS\xspace}
\usepackage{tabularx,ragged2e,booktabs,caption}
\newcolumntype{C}[1]{>{\Centering}m{#1}}

\usepackage{comment}
\usetikzlibrary{matrix,shapes,arrows,fit,tikzmark}
\newcolumntype{.}{D{.}{.}{-1}}
\tikzset{   
        every picture/.style={remember picture,baseline},
        every node/.style={anchor=base,align=center,outer sep=1.5pt},
        every path/.style={thick},
        }
\newtheorem{assumption}{Assumption}

\newcommand\marktopleft[1]{%
    \tikz[overlay,remember picture] 
        \node (marker-#1-a) at (.1em,.3em) {};%
}
\newcommand\markbottomright[1]{%
    \tikz[overlay,remember picture] 
        \node (marker-#1-b) at (.1em,.3em) {};%
    \tikz[overlay,remember picture,inner sep=3pt]
        \node[draw=red,rounded corners,fit=(marker-#1-a.north west) (marker-#1-b.south east)] {};%
}



\newtheorem*{proposition*}{\propositionnumber}
\providecommand{\propositionnumber}{}
\makeatletter
\newenvironment{proposition}[2]
 {%
  \renewcommand{\propositionnumber}{Proposition #1 $\mathcal{#2}$}%
  \begin{proposition*}%
  \protected@edef\@currentlabel{#1 $\mathcal{#2}$}%
 }
 {%
  \end{proposition*}
 }
\makeatother

\usepackage[
citestyle=authoryear-comp,
bibstyle=authoryear,
maxbibnames=5,		
minbibnames=1, 		
maxcitenames=2,		
mincitenames=1,		
datezeros=false,	
date=long,
natbib=true,			
backend=bibtex		
]{biblatex}
\begin{document}

\input{Title_Page_and_Author_Details}

\input{Abstract}

\newpage 

\section{Introduction \label{Introduction}}
\renewcommand{\thefootnote}{\arabic{footnote}}

In a world where productivity is not directly observable, employers have every incentive to draw inferences about worker productivity from auxiliary signals like employment history, tenure, unemployment duration, test scores, and past wages (\citet{AltonjiPierret2001}; \citet{Lange2007}; \citet{Zhang2007}; \citet{Oyeretal2011}; \citet{DevaroWaldman2012}; \citet{Farberetal2015}; \citet{HoffmanKahnLi2017}; \citet{JaroschPilossoph2018}). 
Employers seek information on past wages to gauge not only productivity, but the applicant's outside option, and also to determine whether the worker can fit into the pay structures of the new firm. 
About 50\% of workers in the US report that their current employer had learned about pay in their earlier jobs, before making an offer (\citet{HallKrueger2012}). 
In another nationally representative survey, \citet{BarachHorton2020} find that 29.4\% of respondents were asked about their pay history during interviews and among them, 82.6\% report that the firm had asked about compensation before extending a job offer. 

While information about past wages has screening value for employers, access to such information raises potential concerns about history dependence in wage growth. 
More specifically, if employers rely on past wages in deciding compensation, workers who earn low wages early in their career, either because of entry-level mismatches or career interruptions, would see their wage growth stymied. 
These concerns are more acute for women who face major career discontinuities over their lifetime, primarily because of child-birth (\citet{BertrandGoldinKatz2010}), and several studies have shown that the gender pay gap widens in the first 15 to 20 years after entering the workforce (\citet{Goldin2014}; \citet{BlauKahn2017}; \citet{Goldinetal2017}). 
How have policymakers responded to these growing concerns about workplace pay inequality? 
On one hand, a number of countries in the EU have enacted policies that increase pay transparency by giving workers access to their employers' private information about workplace earnings distributions. 
These policies have had varying levels of success across countries (\citet{Bennedsenetal2019}, \citet{Gulyasetal2020}, \citet{Blundell2020}, \citet{Duchinietal2020})\footnote{Countries like the UK, Germany, Austria, Finland and Denmark have enacted different variants of gender pay reporting laws, which require firms to publish gender differences in pay and employment at different aggregate levels. In 2016, the then US president announced executive action that would have required firms with government contracts to disclose the average wages of employees by gender, a move that was subsequently rolled back by President Trump  (\href{https://obamawhitehouse.archives.gov/the-press-office/2016/01/29/fact-sheet-new-steps-advance-equal-pay-seventh-anniversary-lilly}{\underline{Obama-Executive\_Order-EEOC\_Action\_on\_Pay\_Data\_Collection}}). President Obama also signed Executive Order 13665 in 2014 which prohibits federal contractors from retaliating against employees who choose to discuss their compensation (\href{https://obamawhitehouse.archives.gov/the-press-office/2014/04/08/executive-order-non-retaliation-disclosure-compensation-information}{\underline{Obama\_Executive\_Order-Non-Retaliation\_For\_Disclosure\_of\_Compensation\_Information}}).}. 
On the other hand, several states and local jurisdictions in the US, starting in early 2017, have passed salary history bans (SHBs) which prohibit employers from inquiring about job applicants' current or previous salary, before making an offer including compensation. 
By restricting prospective employers access to what is now workers' private information, these bans seek to prevent low wages from following women throughout their career and improve the pay gap in the process\footnote{``Salary history based on unequal treatment then becomes the bases for the next salary'', Mr. de Blasio, a Democrat, said at a news conference. ``We have to break that cycle.'' - \href{https://www.nytimes.com/2016/11/05/nyregion/bill-de-blasio-salary-history-executive-order.html?action=click&module=RelatedCoverage&pgtype=Article&region=Footer}{\underline{New York Times (Nov 4, 2016)}}; ``(c) The problematic practice of seeking salary history from job applicants and relying on their current or past salaries to set employees' pay rates contribute to the gender pay gap by perpetrating wage inequalities across the occupational spectrum'' - \href{https://sfgov.org/olse/consideration-salary-history}{\underline{Parity in Pay Ordinance 142-17, City and County of San Francisco}}.}. 
However, one important caveat of these bans is that they do not restrict job applicants from voluntarily disclosing pay history and this has the potential to unravel any intended effects\footnote{For other explanations for why this policy might fail, check out this \href{https://hbr.org/2017/09/why-banning-questions-about-salary-history-may-not-improve-pay-equity}{\underline{2017 article}} from PayScale.}. 
Therefore, ex-ante it is unclear whether SHBs would have any effects on the gender pay gap and in this paper I provide a systematic study of this policy's effects and explain how SHBs have helped in closing this gap. 

As of December, 2019, 18 states and many other local jurisdictions at city and county levels have some version of salary history bans. 
These laws restrict employers from asking applicants about their pay history or seeking such information from public records, background checks, and current and former employers. 
Ten of these 18 states have bans for all employers, while the rest apply only to state employers and agencies. 
In this paper I exploit the two sources of variation induced by the staggered roll-out of SHBs across states in the US, to study its effects using a difference-in-difference methodology. 
I use earnings data from the Current Population Survey's (CPS) Outgoing Rotation Group (also known as the earner study) for the period January, 2010 to December, 2019 along with the CPS Basic Monthly Files. 
Since different states apply these bans to different employers,
in my main estimation sample (which I call \textbf{AllStateBan}), I compare states which have bans for all employers to those which have no bans.  

Using my most preferred specification I find that salary history bans significantly reduce gender gaps in hourly wages by 2 p.p. (s.e. 0.8) and in weekly earnings by 1.9 p.p. (s.e. 0.8).
When I estimate the effects of the ban separately for men and women, I find that this reduction in gender pay gap is driven almost entirely by increase in female wages and weekly earnings, respectively and not because of any adverse effects on male earnings. 

I also investigate whether SHBs weaken the link between past and present earnings by estimating the effect of SHBs on the auto-correlation in earnings. 
Indeed, I find that SHBs reduce the auto-correlation between these two earnings measures, especially for the sample of workers who can be credibly identified as having changed jobs. 

Next, I investigate heterogeneous treatment effects by race, education, age, and sector of employment. 
I find that SHBs increased earnings only for white female workers, thus widening both the white-black racial gap among women, as well as improving the gender pay gap only among white workers. 
Positive effects of SHB on earnings also seem restricted to young workers, particularly those with lower education levels. 
When I decompose the effects of SHBs on the gender pay gap by public versus private sectors, I find that SHBs reduced the pay gap only in the private sector, while the gap in the public sector seems to have increased by around 2-3 p.p. 
Finally, there is little evidence that SHBs changed other labor market outcomes like labor force participation, unemployment rate, and job turnover. 

It is important to emphasize here that the success of SHBs hinge on prospective employers not getting access to information about job applicants' previous salary. 
However, voluntary disclosure on part of applicants have the potential to unravel these effects. 
Therefore, for SHBs to have had any impact, average disclosure rates would have to decrease with these bans and the size of this reduction would have to differ by gender. 
Indeed I find, using a novel dataset from PayScale, that the bans reduced disclosure rates by 22 p.p. (s.e. 3.5) among men and by 24 p.p. (s.e. 4.7) among women, conditional on other observables. 
In fact, there is little difference in the disclosure rates between men and women after the ban.
Does this imply that it is the nudge from the employer that might have induced more job applicants to disclose before the ban, and more so for women? 

To check for this, I compare disclosure behavior among male and female job applicants who were asked and not asked for their salary history, even before the bans were put into place. 
I find that conditional on other observables, men were 59 p.p. (s.e. 1.4) and women were 66 p.p. (s.e. 0.8) more likely to disclose before the ban when asked versus when not asked. 
Moreover, conditional on being asked, women were 4.2 p.p. (s.e. 1.3) more likely than men to comply with information requests. 
This, coupled with the fact that women were 4 p.p. (s.e. 1.6) more likely than men to be asked about their salary history before the ban\footnote{Men were 27.4\% (s.e. 3.1) and women were 31.3\% (s.e. 3.0) likely to be asked about their salary history before the ban, conditional on other observables.}, led to a higher disclosure rate among women in pre-SHB periods by 3 p.p. (s.e. 1.2).  
This confirms my initial hypotheses that it is really the prompt from the prospective employer to reveal information, which induces more job applicants to comply. 
On further investigation, I also find that high-earners (i.e., those who earn above the occupation-specific median salary), among both men and women, were 6-8 p.p. more likely to disclose salary history in pre-SHB periods. 
Indeed, high-earners might find it beneficial to reveal information because their already-high salaries would help them negotiate better offers, while low-earners would withhold information because recruiters might offer them only a fixed percent raise on their already-low salaries. 
But financial incentives may not be the only drivers of disclosure decisions. 
For example, job applicants might feel more obligated to share information when prompted, because refusal might put off the prospective employer. 
To better understand these motivations, I conducted a nationally representative online survey of 5,700 US workers, where I asked respondents whether they would reveal their salary history and why they would make those choices. 
Evidence from this survey confirms my findings from the PayScale data, that applicants are more likely to reveal information only when asked\footnote{In the survey over 70\% of respondents said they would share information only when asked and withhold when not asked. Around 18\% said they would never disclose, and the rest said they would always disclose.}. 
While respondents are not necessarily put off by salary history questions during hiring, they also recognize these as strategies for low-balling offers. 
Men are more likely to say they would disclose because their salaries are already high, and women are more likely to state that they feel uncomfortable with disclosing as well as withholding information. 

While there could be many financial and behavioral incentives that make job applicants choose whether to disclose, it is still striking that 88\% of applicants do not volunteer information and 23\% refuse to comply with information requests. 
Does this imply that employers do not perceive non-disclosure as a sign of already-low salaries, and therefore do not statistically discriminate against those who withhold information? 
I argue that statistical discrimination is observationally consistent with an equilibrium in which a non-zero proportion of job applicants do not reveal salary history. 
However, statistical discrimination or its lack thereof, can still not explain why disclosure rates are higher among those who are asked about their salary, and why disclosure rates are not 100\% even among high-earners. 

To bridge this gap, I propose a theoretical framework of salary negotiation between a job applicant and a prospective employer, and I use this framework to rationalize my empirical findings. 
In this simple framework, the employer has information about the worker's productivity but not their current wage, which can take from one of two values - high or low. 
Only the distribution of wages conditional on gender, is public information and the worker can choose whether to reveal their actual wage. 
I incorporate two types of utility costs for the worker: (a) a psychic cost of disclosing private information like salary, which is invariant to gender and (b) a psychic cost of withholding information when asked, which is higher among women. 
The first type of utility cost implies that when an applicant discloses salary history, the prospective employer has to compensate them for their utility cost of sharing private information, if the applicant were to accept the offer. 
This allows us to sustain a separating equilibrium where only low-wage applicants withhold information if the proportion of low-earners is too high, and a pooling equilibrium where all applicants withhold information when the proportion of low-earners falls below a certain threshold. 
In a separating equilibrium the prospective employer can extract full rent from both high and low-wage applicants, while in a pooling equilibrium they can extract full rent only from high-wage applicants, while paying a premium to low-earners. 
Therefore, if the proportion of low-earners is too high, the employer is better off choosing a wage offer which incentivizes only high-wage applicants to share information. 
Moreover, applicants face a utility cost of withholding information when asked than when not asked. 
Therefore, if applicants are asked and a pooling equilibrium arises, employers have to pay non-disclosing low-earners for their cost of withholding information on top of a premium for pooling with high-earners. 
This implies that the threshold proportion of low-earners above which employers begin to offer separating wages, decreases when applicants are asked in comparison to when they are not.  
I use this to prove analytically that there exists gender-specific initial wage distributions, such that only high-earners disclose information when applicants are asked (pre-SHB) and all workers withhold information when they are not asked (post-SHB). 

Gender differences in the second type of utility cost then leads to different pre and post-SHB equilibria for men and women. 
In particular, I show that there exists initial wage distributions for men and women, with a more right-skewed distribution for women, such that women are offered separating wages in pre-SHB periods and pooling wages in post-SHB period, while men are offered pooling wages in both periods. 
This helps us capture four empirical findings on disclosure: (a) women are more likely to disclose when asked, (b) women are more likely to disclose in pre-SHB period, (c) high-earners are more likely to disclose than low-earners, and (d) SHB reduces average disclosure rates, more so for women. 
I then prove analytically how a ceiling on women's utility cost of withholding information when asked, helps us capture the effects of SHB on the gender pay gap. 
This upper bound ensures that average female wages are lower than the pooling wage men receive in pre-SHB periods, thus reflecting the baseline gender pay gap. 
In the simple framework, men and women receive the same pooling wages in post-SHB periods, thus capturing the improvement in gender pay gap with SHBs in place. 
This improvement in gender wage gap comes from low-earning women, who would have received a separating wage in the pre-SHB case and they are pooled with high-wage women in the post-SHB case. 

I then extend this stylized framework in two different ways: (a) gender-invariant heterogeneity in productivity and utility costs, while preserving the two-valued support of initial wage distributions, (b) gender-invariant heterogeneity along with continuous initial wage distributions for both men and women. 
Since analytical results for these two extensions are impossible without restrictive functional form assumptions, I numerically simulate these two frameworks. 
In particular I show, how for a given set of parameters, there exists initial wage distributions with a more right-skewed distribution for women, such that the model and its extensions capture my main empirical findings. 
The difference in skewness of initial wage distributions is a reasonable model outcome, in light of the actual earnings distribution in the CPS data.

My paper contributes to the literature on the effects of restricting employers' access to different kinds of information about job applicants. 
\citet{GoldinRouse2000} found that orchestras switching to blind auditions where the gender of the musician is not observable, increases the proportion of female musicians employed. 
Similarly, \citet{Wozniak2015} found that when employers require drug tests for workers, employment of low-skilled black men increases. 
However, restricting information also has the potential to generate unintended effects. 
For example, \citet{BartikNelson2016} show that banning employers from checking job applicants' credit histories, reduces job-finding rates for young black job seekers. 
Using an audit study, \citet{AganStarr2017} show that limiting employer access to criminal records increases the white-black gap in call-back rates for job applicants. 
In related work, \citet{DoleacHansen2018} study real world implications of ``ban-the-box movement'' (BTB) policies in the US, which prohibit employers from accessing criminal records and incarceration history of job applicants.
Consistent with statistical discrimination, they found that young, low-skilled black men were less likely to be employed after BTB than before.  
While statistical discrimination is a valid possibility in my context, salary history bans are quite different from some of the measures above, since salary history and expectations are often early and commonplace discussions during hiring, as opposed to more auxiliary information like criminal records and credit histories. 

Whilst ours is not the only paper to study the effects of SHB policies, it is fundamentally different from contemporary literature on this topic along multiple dimensions. 
\citet{BarachHorton2020} conduct a field experiment in an online labor market, where treated employers could not observe the compensation history of their applicants. 
However, in their setting, applicants themselves were unaware of this feature and therefore, had no control over disclosing information. 
\citet{AganCowgillGee2020b} conduct a small online survey of 500 respondents and ask participants whether they would choose to disclose their salary history in hypothetical scenarios.  
\citet{Sranetal2020} show SHBs increase the number of job postings and also the likelihood that postings carry salary information. 
\citet{HansenNichols2020} replicate my work using a synthetic control approach, focusing primarily on California, and confirm my results on the gender pay gap. 
\citet{Bessenetal2020} focus on job changers and find that earnings increase not just for women but also for non-white male workers. 
And finally, \citet{Mask2021} focus on the effects of SHB on previously `scarred' workers. 

In contrast to contemporary literature, my paper introduces three new components to this topic, besides providing robust empirical evidence on the effects of SHB on the gender pay gap. 
First, I bring in novel evidence on real-world disclosure behavior among job applicants from a large nationally representative survey. 
I use this evidence to show not only how voluntary disclosure does not unravel intended effects, but also how employers' prompt and gender differences to these nudges, are the primary drivers of underlying mechanisms. 
Therefore, I provide the first direct link between the success of SHB policies and conscious changes in interviewee strategies. 
Second, I generate new qualitative evidence through an online survey that both confirms my empirical findings on disclosure behavior as well as sheds light on the financial and behavioral motivations that drive these decisions among job applicants. 
To the best of my knowledge, my survey is the first to generate these insights on privacy and non-conformity costs, in the domain of salary negotiation. 
Third, my paper is the first to endogenize job applicants' disclosure choice in a theoretical framework. 
This is in contrast to \citet{AganCowgillGee2020b}, who consider workers as immutable `defiers' or `compliers', and \citet{MeliSpindler2019} or \citet{Bessenetal2020}, who assume away the possibility of non-disclosure before the ban and that of disclosure after the ban. 
This is also in contrast to standard search theoretic models (\citet{PostelVinayRobin2002}, \citet{CahucPostelVinayRobin2006}, \citet{Baggeretal2014}), where there is no asymmetric information between the worker and the prospective employer, and current wages are irrelevant to the bargaining outcome. 
And finally, I translate my empirical evidence on disclosure into notions of psychic costs and incorporate them in my model to bring together the two dimensions to this story - disclosure behavior and earnings.

The rest of this paper is structured as follows. 
In Section \ref{Laws} I provide an overview of salary history bans and their variations across the US. 
Section \ref{Data} discusses my data and in Section \ref{RDesign} I outline my main empirical strategy. 
I report my findings on earnings in Section \ref{ResultsEarningsAll} and in Section \ref{ResultsDisclosure} I show evidence on pre-SHB disclosure behavior and how disclosure rates have changed with SHB. 
In Section \ref{TheoreticalModel} I discuss my theoretical framework and show how this model helps rationalize my empirical findings using both analytical derivations and simulations. 
Section \ref{Conclusion} concludes. 

\section{Background on Salary History Ban and Geographic Variation \label{Laws}}

New York was the first to institute a state-wide salary history ban for state employers and agencies in January, 2017. 
Since then, other states have followed, including California, Massachusetts and Connecticut, along with many local governments at the city and county levels. 
As of December 2019, 18 states (including District of Columbia), 6 counties, and 10 cities have implemented different versions of salary history bans (See Fig $\ref{Fig01}$) with close to 40\% of the US adult population in the age group 22-64 subject to such legislations (See Fig $\ref{Fig02}$).

There is considerable variation across state and local bans in both the scope of these bans and their stringency\footnote{For more details on the effective dates, scope and provisions for each ban, see Table $\ref{Tab02}$}. 
While most states (eg, California, Massachusetts and Connecticut) have bans for all employers, others including New York and New Jersey\footnote{Both NY and NJ extended bans to all employers in early January, 2020. For more details, see Table \ref{Tab02}.} have bans only for state employers and agencies. 
Local bans also vary in their scope. 
While most local bans are applicable for city employers and agencies alone (eg, Chicago (IL), New Orleans (LA), Salt Lake City (UT), Kansas(MO)), others in New York city and counties like Westchester (NY) and Albany (NY) apply to all employers. 

The primary provision of these bans, which is common across jurisdictions, is to restrict employers from asking applicants their salary history at any stage in the hiring process, including application forms in most cases\footnote{Example of an exception to the ban on application form disclosure is New York city. 
Since New York state has a ban on application form disclosure for all state employers, all city agencies and private employers in New York city are exempt from this requirement.}. 
Employers cannot seek information on salary history from public records, background checks or ask applicants' current or former employers. 
In some states employers are also prohibited from asking hiring agencies for information and these agencies are required to have written consent from job applicants before they can disclose such information to prospective employers\footnote{San Francisco requires hiring agencies to seek consent from applicants before they disclose any salary information to prospective employers, while the state of California has no such restriction.}. 
However, an important caveat of all these bans is that applicants can voluntarily disclose information about pay history without being prompted. 
In some states, if applicants volunteer information on salary history or employers accidentally discover such information while conducting background checks, then they cannot use this information to either discriminate against applicants in hiring or decide compensation for a new position. 
Applicants cannot be discriminated against because of their refusal to disclose salary history on being prompted. 
Employers are allowed to ask about pay history only after they have already made an offer that includes compensation details. 
If applicants choose to disclose after they already know about their pay in the new position, employers can verify this information. 
These laws are not applicable for internal transfers, promotions, union bargaining and are subject to federal exemptions. 
In cases where there are conflicts between state-wide bans and local bans within that state, state restrictions override all local ambiguities. 
Any breach of these bans can be contested in a court of law; and if found guilty, the employer can incur monetary penalties. 
In some cases, the employer is not liable for any breach of laws by a hiring agency as long as the employer had explicitly directed the agency to not collect information on pay history.

In this paper, I focus only on the scope of the ban (i.e., the type of employer subject to these bans) and the main provision which restricts employers from asking pay history during job interviews. 
Since there is considerable variation in scope as well, I conduct multiple robustness checks in my analyses by both restricting the sample I include as well as by redefining the treatment I assign. 
The variation in scope also helps us test for spillover effects from bans in the private sector on the public sector and vice-versa. 
This is discussed in more detail in Section $\ref{ResultsSpillover}$. 
I believe that many of the other provisions, for example, where the employer cannot use voluntarily disclosed or accidentally discovered information, or hold an applicant's refusal to disclose against her, are difficult to prove in a court of law and would therefore have little effect in the labor market.

\section{Data \label{Data}}

The first dataset I use in this paper comes from the Current Population Survey (CPS) (\citet{CPSData}). 
I use two kinds of data from the CPS: Basic Monthly Files,
and the Outgoing Rotation Group supplement from January, 2010 to December, 2019\footnote{I have decided to limit this analyses through 2019 in order to avoid any potential bias from how the 2020 coronavirus pandemic affected the labor market}. 
The CPS is a nationally representative monthly survey of around 60,000 households who are interviewed for four consecutive months, go out of the sample for eight consecutive months, and then re-enter the survey for another four consecutive months before leaving the survey altogether. 
Besides the demographic characteristics for each individual in the household like gender, age, education, location, etc., the basic monthly files also record the individual's labor market outcomes, including labor force participation, employment status, industry and occupation. 
For six of those eight months in the sample, excluding the 1st and the 5th months, each employed individual is also asked whether they changed their employer from the previous month\footnote{There are two questions related to any change in employment: (1) Still working for the same employer, and (2) Still have the same work activities. I use the answer to the first question to detect any change in employer.}. 
This variable can be used to construct monthly transitions between jobs, sectors, between employment and unemployment for each of the two 3-month windows in the survey. 
The basic monthly files do not have any information on individual earnings. 
The earnings data comes from the Outgoing Rotation Group supplement, which is essentially the basic monthly file for the individual in the 4th and the 8th months in the sample. 
Two types of earnings are recorded for each of these 2 months: (a) hourly wages for those who are paid hourly, and (b) weekly earnings for all workers. 
While the earnings information is topcoded, only about 3\% of the sample is topcoded in their weekly earnings, and almost 0\% of the sample is topcoded in their hourly wages.

In this paper I restrict my sample to civilian non-institutionalized population between the ages 22 and 64. 
I focus on this subset of population because most individuals would have enough time to complete college and enter the workforce by the age of 22, and remain in the labor force before retiring. 
Since earnings is observed only for those who are employed at the time of the survey, the sample used for analyzing earnings is smaller with 
around 53.84\% of men and 59.12\% of women being hourly paid.

My second data comes from \citet{PayScaleData}\footnote{\href{https://www.payscale.com}{Data provided by \href{https://www.payscale.com}{\includegraphics[height=5.0mm]{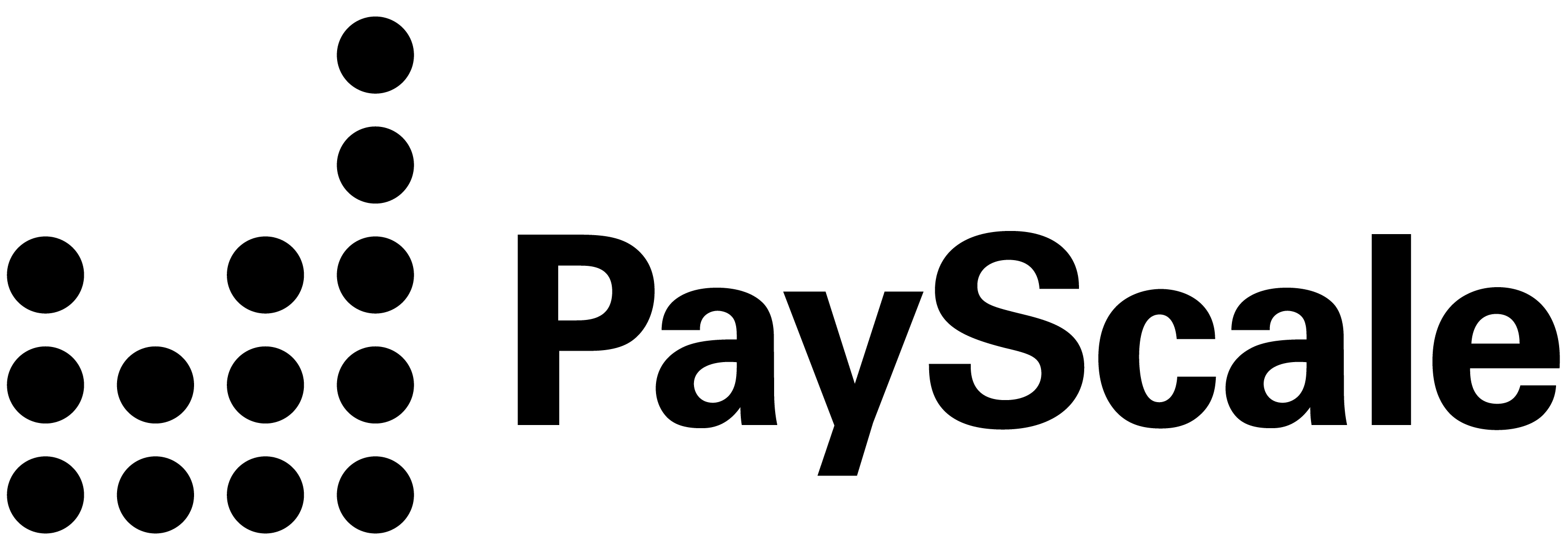}}, Last Updated: August, 2019.}}, which is a compensation software and data company that helps employers manage compensation schedules, and employees to evaluate their skills, jobs and offers. 
PayScale conducted a survey of its users who wanted to evaluate their job offers. Between April 2017 to July 2017, and then from February 2018 till August 2019, around 140,000 respondents were asked whether the employer had asked them about their previous salary at any point in the interview process and whether they had revealed information in response to a prompt or volunteered information\footnote{Users were asked the following question: \textit{At any point during the interview process, did you disclose your pay at previous jobs?} Users could select from the following five options: \textit{(a) No, and they did not ask, (b) No, but they asked, (c) Yes, they asked about my salary history, (d) Yes, I volunteered information about my salary history, and (e) I do not recall.} The data that I received from PayScale has responses recorded only for the first four options. The details of PayScale's own methodology and an overview of their findings can be found \href{https://www.payscale.com/data/salary-history/methodology}{\underline{here}} and \href{https://www.payscale.com/data/salary-history}{\underline{here}} respectively.}. 
The data also has information on worker characteristics like gender, education, race, occupation (SOC codes), industry, total annual earnings, age, geographical location like state, and a time stamp at the week level. 
I restrict this data to workers between the ages of 22 and 64 years to match my exclusion criteria for the CPS sample. This leaves me with a sample of 66,387 women and 57,052 men. 
I reweight the PayScale data to match distributions of worker characteristics as observed in the NY Fed's Survey of Consumer Expectations (SCE) Job Search Module. 
Details of my weighting strategy and sample comparisons are outlined in Appendix $\hyperref[C]{B}$. 
Figure $\ref{Fig03a}$ shows a breakdown of responses into four categories by gender. 
About 70\% of respondents (including those who report having volunteered information) are not asked about their salary history. 
This also includes post salary history ban periods when employers can no longer ask about pay history. 
When prompted, 75\% of men disclose pay history as opposed to 78\% of women. 
When not asked, 12.8\% of men and 10.3\% of women choose to volunteer. 
Over the entire sample period men are slightly more likely to be asked about pay history than women (34.6\% versus 33.1\%). 

I show a final piece of summary statistic before discussing my empirical strategy in the next section. 
Fig $\ref{Fig04}$ shows the pre-ban gender differences for a host of labor market outcomes, conditional on other observables, separately for states that have no salary history ban and others that have some version of a ban as of March, 2019\footnote{To generate these differences, I regress the earnings and other labor market outcomes on a bunch of individual characteristics, labor market characteristics like sector, industry and occupation, state and year fixed effects, and a treatment dummy. The treatment dummy takes the value 1 for all states that have bans later in my sample, and 0 for all others. Standard errors are clustered at the state level.}. 
Two of these outcomes are worth pointing out. 
States which have salary history bans already have a smaller gender gap in $log$(hourly wages) and $log$(weekly earnings) than states that have no bans. 
It is evident from these differences that adoption of salary history bans by states is not a random event. 
In fact, it could be the case that states that are more proactive towards reducing workplace gender disparities are more likely to adopt salary history bans. 
My empirical strategy does not require adoption to be random, only that there are no pre-ban differences in the outcome trends between the treated and control groups. 
However, these pre-ban differences suggest that by reducing gender pay gaps in treated states, salary history bans have the potential to increase spatial inequality in gender disparities across the country. 

\section{Empirical Strategy And Sample Construction \label{RDesign}}

I exploit the staggered roll-out of salary history bans by different states over time, to estimate the causal effects of SHB on gender pay gap using a staggered difference-in-differences design\footnote{I assume here that there are no heterogeneous treatment effects both across units for the same time and across time for the same unit. Under these assumptions, the two-way fixed effects model identifies the ATT, without any bias.}. 
My baseline specification is as follows:
\begin{multline}\label{Eq01}y_{ismt} = \alpha+\textcolor{red}{\beta_1}\mathbb{I}_{i}^{f}\mathbb{T}_{smt}+\beta_2\mathbb{I}_{i}^{f}+\beta_3\mathbb{T}_{smt}+\beta_4 \mathbb{I}_{i}^{f}EverTreat_{s}+\gamma X_{ismt}\\ +\lambda_s + \lambda_t+\delta_1\lambda_s t + \delta_2\lambda_I t + \delta_3\lambda_f t+\epsilon_{ismt},\end{multline}

where $y_{ismt}$ is the log-earnings measured for the individual $i$ in state $s$, in calendar month $m$ and calendar year $t$. 
$\mathbb{I}_{i}^{f}$ is a female dummy, $EverTreat_{s}$ is a dummy variable if state $s$ ever implements salary history ban, $X_{ismt}$ is a set of individual characteristics like race, education, a polynomial (cubic) in age, part/full time work status, sector, industry (2-digit NAICS), occupation (2-digit SOC 2010), and calendar month. $\lambda_s$ and $\lambda_t$ are state and year fixed-effects, $\lambda_s t$, $\lambda_I t$ and $\lambda_f t$ are state, industry and female-specific linear time trends. 
The outcome variables $y$ are $log$(hourly wage) or $log$(weekly earnings) (or a set of labor market outcomes like labor force participation, unemployment status, private sector employment, public sector employment, U2E, J2J or E2U transitions in later sections)\footnote{U2E: Unemployment to Employment, J2J: Job to Job, E2U: Employment to Unemployment}. 
When the outcome variable is $log$(weekly earnings) I also control for whether the worker is hourly paid, and the number of hours worked interacted with the part/full time work status. 
For outcomes like labor force participation (LFP), unemployment status (UR) and U2E I control only for individual characteristics, along with unemployment duration where applicable. 
For turnover outcomes, I control for the previous industry and occupation. 
I cluster standard errors in all specifications at the state level.

The wide variations across states and local jurisdictions in the scope of their salary history bans, as well as the differences in the scope between state-wide bans and local jurisdictions within those states, complicates my treatment and sample definitions. 
I define my treatment indicator $\mathbb{T}_{smt}$ in various ways depending on the sample and my analyses. 
For my main estimation results, I restrict the sample by dropping all states which have state-wide bans for only state employers and agencies (New Jersey, Pennsylvania, Illinois\footnote{I drop Illinois because IL implemented SHB for only state employers and agencies in Jan, 2019 but then extended it to all employers in September, 2019. }, Michigan, North Carolina), and ignore all local bans (at city levels). 
I only include states which have either no state-wide bans, or state-wide bans for all employers (Delaware, Oregon, California, Massachusetts, Vermont, Connecticut, Hawaii, Washington, Maine, Alabama). 
This also forces me to drop New York as a state, because while NY has a state-wide ban for state employers and agencies, New York city and multiple counties in NY have local bans for all employers. 
I call this main sample \textbf{AllStateBan}. The treatment indicator $\mathbb{T}_{smt}$ is then defined at the state-month-year level and takes the value 1 for all calendar months and years after the effective date for that state and 0 otherwise. 
All estimation results shown in subsequent sections come from this sample and treatment definition\footnote{In Appendix Section \ref{AppendixPScoreWeighting} I exploit the scope of the SHBs to construct other samples and use them to study spillover effects of SHB across sectors.}. 

I also estimate heterogeneous effects of the salary history by sector of employment, education levels and age using the following specification:
\begin{multline}\label{Eq02}y_{ismt} = \alpha+(\beta_0+\textcolor{red}{\beta_1}\mathbb{I}_{i}^{f}\mathbb{T}_{smt}+\beta_2\mathbb{I}_{i}^{f}+\beta_3\mathbb{T}_{smt})*(\gamma_0+\textcolor{red}{\gamma_1}X_{ismt}^{1})+\beta_{4}\mathbb{I}_{i}^{f}EverTreat_{s} \\ +\gamma_2 X_{ismt}^{2}+\lambda_s + \lambda_t+\delta\lambda_s t + \delta_2\lambda_f t + \epsilon_{ismt},\end{multline}
where $X_{ismt}\equiv\left[X_{ismt}^{1}\  X_{ismt}^{2}\right]$. $X_{ismt}^{1}$ is the set of education, polynomial in age and their interactions, whereas $X_{ismt}^{2}$ is the set of other individual characteristics like race, occupation, industry, part-time/full-time status and sector. 
I use a cubic in age for most specifications, five levels of education: (a) No High School, (b) High School, (c) Some college but no degree, (d) Associate Degree/Diploma, (e) College; and include only public and private sector workers. 
Depending on the sample, I distinguish between federal, state and local employees among all public sector workers. 

Finally, to test for parallel pre-trends in my outcomes of interest, I use the following specification:
\begin{multline}\label{Eq02a}y_{isk} = \alpha + \sum_{l=-M}^{M}\textcolor{red}{\beta_{1}^{l}}\mathbb{I}_{i}^{f}EverTreat_{s}\mathbb{I}_{k=l} + \sum_{l=-M}^{M}\beta_{2}^{l}EverTreat_{s}\mathbb{I}_{k=l} + \beta_{3}\mathbb{I}_{i}^{f}EverTreat_{s}\\ + \beta_{4}\mathbb{I}_{i}^{f} + \sum_{l=-M}^{M}\beta_{5}^{l}\mathbb{I}_{i}^{f}\mathbb{I}_{k=l} + \sum_{l=-M}^{M}\beta_{6}^{l}\mathbb{I}_{k=l}  + \beta_{7}X_{isk} + \lambda_{s} + \epsilon_{isk},\end{multline}
where $k$ denotes month-since-treatment, $\mathbb{I}_{k=l}$ is an indicator function, and all other variables have the same interpretation as in ($\ref{Eq01}$) and ($\ref{Eq02}$)\footnote{For control states which do not have salary history bans I randomly assign a hypothetical treatment start month drawn from the range of start months of all \textit{EverTreat} states.}.

\section{Effects of Salary History Bans on Earnings and Earnings Gaps \label{ResultsEarningsAll}}

\subsection{Parallel Trends in Pre-Treatment Gender Pay Gap \label{ResultsParallel}}

Before I discuss the estimation results from my specifications in the previous section, I first show evidence for parallel pre-trends in my outcome variables.  
Fig $\ref{Fig05}$ shows the time-varying effects on gender pay gap ($\beta_{1}^{l}$ in specification ($\ref{Eq02a}$) above), separately for hourly wages and weekly earnings. 
Time is re-centered as `month-since-treatment' such that for each treated state, the first month after the effective date is denoted as 0. 
All states which have no salary history bans are assumed to have a hypothetical effective date of September, 2018, which is the average effective date for all treated states. 
Visual inspection suggests that the gender pay gap (in both hourly wages and weekly earnings) followed reasonably parallel trends between treated and control states before SHB went into effect. 
More specifically, I can say with 95\% confidence that the gender pay gap was no different from the period just before SHB. 
In Appendix Figure $\ref{Fig07}$ I show similar parallel pre-trends for nine different labor market outcomes including turnover rates. Again, these graphs reasonably satisfy parallel trends. 

\subsection{Effects on Average Earnings \label{ResultsEarnings}}

I first show the effects of the salary history ban on the two earnings measures using a variant of the specification in ($\ref{Eq01}$), where I drop the interaction term $\mathbb{I}_{i}^{f}*T_{smt}$. 
The coefficient on $T_{smt}$ then denotes the average effects of SHB on pay.   
The results for hourly wages and weekly earnings are shown in Tables $\ref{Tab03new}$ and $\ref{Tab04new}$ respectively. 
Our coefficient of interest is $Treat$ and I find no evidence of any impact of SHB on weekly earnings (row 1 in Table $\ref{Tab04new}$), and only marginally significant (at 10\% significance) increase in hourly wages by 1 p.p. (row 1 of Table $\ref{Tab03new}$). 
These estimates are robust to inclusion of controls and different time trends. 
The lack of any effects on overall earnings could be explained by either no change in both male and female pay, or changes for both genders in the opposite direction. 
This therefore, has implications for individual earnings as well as the gender pay gap. 
I investigate this in the next section. 

\subsection{Effects on Gender Pay Gap \label{ResultsGPG}}

Table $\ref{Tab03}$ and Table $\ref{Tab04}$ show my main estimates of the effects of SHB on gender gap in hourly wages and weekly earnings respectively using my baseline specification in ($\ref{Eq01}$). 
Our main coefficient of interest is $TreatXFemale$ (row 1), which is the female premium. 
A positive value of this estimate implies a reduction or improvement in the gender pay gap. 
The coefficient on $Treat$ (row 2) is the effect of the ban on male pay, while the $Female$ indicator (row 3) denotes the baseline gender pay gap. 
Both tables show that the effects on gender gap for both earnings measures are positive and significant at the 95\% level. 
The most comprehensive specifications in Column (7) of either table shows a reduction in gender pay gap by 2 p.p. in hourly wages, and 1.9 p.p. in weekly earnings.  

In contrast, the effects on both male hourly wages and weekly earnings are small, precisely estimated, and statistically insignificant. 
Using my most preferred specification in Column (7), male wages increased by a mere 0.1\% (s.e. 0.006) and male weekly earnings decreased marginally by only 0.4\% (s.e. 0.004). 
Therefore, it is evident that the SHBs were effective in their first intended objective of decreasing the gender pay gap by increasing female earnings, and there is no evidence for adverse effects on male earnings. 

Next, I investigate whether the bans were effective in their second intended objective of reducing the path-dependence in earnings, in particular the link between current and past earnings. 

\subsection{Effects on Path-Dependence in Earnings \label{ResultsAutoCorrelation}}

The CPS basic monthly files provide earnings information for the 4$^{th}$ and the 8$^{th}$ months that the household (individual) is in the sample.  
I regress the demeaned earnings in the 8$^{th}$ month on the demeaned earnings in the 4$^{th}$ month interacted with the treatment indicator to estimate the effect of SHB on the auto-correlation between the two measures. 
In particular I use the following two specifications:

\begin{multline}\label{EQAutoCorrA} 
\tilde{y}_{ismt}^{(8)} = \alpha + \textcolor{red}{\beta_1}\mathbb{T}_{smt}\mathbb{I}_{i}^{f}\tilde{y}_{ism't'}^{(4)} + \textcolor{red}{\beta_{2}}\mathbb{T}_{smt}\tilde{y}_{ism't'}^{(4)} + \beta_{3}\mathbb{I}_{i}^{f}\tilde{y}_{ism't'}^{(4)} + \beta_{4}\tilde{y}_{ism't'}^{4} + \beta_{5}EverTreat_{s}\tilde{y}_{ism't'}^{(4)}\\ +\lambda_{s} + \lambda_{mt} + \epsilon_{ismt}\tag{5a} 
\end{multline}
\begin{multline}\label{EQAutoCorrB} 
\tilde{y}_{ismt}^{(8)} = \alpha + \textcolor{red}{\delta_1}\mathbb{T}_{smt}\tilde{y}_{ism't'}^{(4)} + \delta_{2}\tilde{y}_{ism't'}^{4} + \delta_{3}EverTreat_{s}\tilde{y}_{ism't'}^{(4)} +\lambda_{s} + \lambda_{mt} + \epsilon_{ismt}\tag{5b} 
\end{multline}

where $\tilde{y}_{ismt}^{(j)}$ is the demeaned\footnote{I residualize earnings by regressing hourly wages and weekly earnings on cubic polynomial of age, education, race, industry, occupation, sector, calendar month, calendar year, part/full-time status, hours worked, and gender interacted with \textit{EverTreat} variable.} earnings measured in the $j^{th}$ month of the survey, and all other variables have the same interpretation as in (\ref{Eq01}). 
In specification (\hyperref[EQAutoCorrA]{5a}), I estimate gender-specific effects while in (\hyperref[EQAutoCorrB]{5b}) I estimate overall effects. 
Our coefficients of interest in (\hyperref[EQAutoCorrA]{5a}) are $\beta_{2}$ which measures the effect of SHB on the auto-correlation between the two earnings measures for men and ($\beta_{2}+\beta_{1}$) which is the effect of SHB on the auto-correlation for women. 
$\beta_{4}$ denotes the baseline auto-correlation for men while ($\beta_{4}+\beta_{3}$) measures the baseline for women. 
Similarly, in (\hyperref[EQAutoCorrB]{5b}) $\delta_{1}$ measures the effect of SHB on overall auto-correlation in earnings and $\delta_{2}$ measures the baseline value. 
These results are shown in Table \ref{tab:AutocorrelationNEW}. 
Before I discuss these results, a caveat. 

SHBs can affect path-dependence in earnings only when a worker moves to a new job and their previous/current earnings is not observable to the new employer at the time of hiring. 
SHBs are legally not applicable to promotions and lateral movements with the same employer and it is reasonable to assume that earnings information is not hidden from the employer during these salary re-negotiations. 
Therefore, by including all workers in my analyses above, I might underestimate the effects of SHB on the auto-correlation in earnings, since path dependence will still apply to lateral moves and promotions. 
To address this problem, I run the above analyses separately for the group of workers who I can credibly identify as having changed jobs between the 4th and the 8th months in the sample\footnote{While the CPS tracks, for each month, whether the individual has changed their employer after the previous month, this information is not collected in the 5th month when the household or individual re-enters the sample after the 8 month absence window. Since earnings information is available only for the 4th and the 8th months in the sample, it is therefore not possible for me to pin down whether there's been a change in employer between the 4th and the 8th months for each individual in the sample. Instead, I denote as 'Job Changers' those, who report to have changed their employers in the 6th, 7th, or 8th months in the sample. This way, I lose some job changers who would have changed their jobs during the 8 month outside the sample. Since entry into the sample is random, it is reasonable to assume that individuals who would change their jobs during the 8 month gap are no different from those I denote as `Job Changers'.}. 
I call this restricted sample of workers as `Job Changers' and run the same analyses as with the full set of workers. 
The results for hourly wages and weekly earnings for this restricted sample are shown in the even-numbered columns of Table \ref{tab:AutocorrelationNEW} while the results for the full sample are shown in the odd-numbered columns. 

Our coefficients of interest are \textit{Treat*PastEarn}, \textit{Treat*Male*PastEarn}, and \textit{Treat*Female*PastEarn} which denote the effects of SHB on the auto-correlation in earnings for all workers, male, and female workers respectively. 
All these estimates are negative (although the effects on women are imprecisely estimated), implying that the auto-correlation in earnings decreases because of SHB. 
This is true for both Job Changers and the full sample. 
Finally, the effects for Job Changers are stronger than those for the full sample and this is true for both men and women. 
Overall, these results suggest that the SHBs were somewhat successful in their second intended objective of reducing path-dependence in wage setting\footnote{How do we reconcile the decrease in auto-correlation in earnings for both men and women as a result of SHB while the policy increases women's earnings without much effect on men's earnings? I re-estimate the effects of SHB on earnings for quintiles of $\tilde{y}_{ism't'}^{(4)}$, where I compute quintiles separately for men and women. These results are shown in Figure \ref{fig:SHbEffectbyIncQuintile}. For both men and women in the lowest income quintile, SHB increases earnings and marginally more for women. For men in higher income quintiles, earnings decreases after SHB. This is why there is little evidence for overall effects on men's earnings (as shown in Tables \ref{Tab03} and \ref{Tab04}) while the auto-correlation in earnings decreases. In contrast, for women in higher income quintiles the effects of SHB are still mostly positive, although smaller than that in the lowest quintile. Overall this results in an increase in women's earnings while still decreasing the auto-correlation. }. 

\subsection{Spillover Effects on Earnings Across Sectors \label{ResultsSpillover}}

In all my previous results, I have used the main sample (\textbf{AllStateBan}) where I compare states that have SHBs for all employers (both private and public) with states that have no bans for any employer. 
These sector-specific results are a composition of two effects: (a) effect of the sector-specific ban, and (b) spillover effect from SHB in the other sector. 
In this section I first decompose the effects of SHBs by sector. 
I then exploit the differences in SHB's scope across states to separate out sector-specific effects from spillover effects. 
To do this, I construct two additional samples by restricting the data in two other ways.  

The first sample, called \textbf{PublicStateBan}, includes states which have state-wide bans for only state agencies (treatment group) and also states with no bans for any employer (control group).\footnote{To construct this sample I drop all states that have statewide bans for all employers, and local jurisdictions at city and county levels that have local bans (e.g., New York). I also identify and drop the MSAs which correspond to these local areas. Since only 75\% of my sample is identifiable at the MSA level, I also summarily drop all non-identified MSAs.}. 
Then I estimate sector specific effects for this sample.  
Here, the treatment effect on the public sector is the direct sector-specific SHB effect on public sector workers, and the treatment effect on the private sector is the spillover effect from public sector SHB. 

Next, I construct another sample called \textbf{AllStateBan-PP} where I include only states which have statewide bans for either all employers (treatment group) or statewide bans only for state agencies\footnote{I drop New York because although it has a statewide ban for public employers, New York city has bans for all employers} (control group).  
Again I estimate sector-specific effects.
The treatment effect on public sector workers would be the spillover effects from bans in the private sector, because the public-sector specific effects would cancel out from the treatment and control group. 
And the treatment effect on private sector workers would be the direct sector-specific effect of SHB because the spillovers from the public sector SHBs would cancel out from treatment and control groups\footnote{Both of these statements are true under the assumption that there is no complementarity between sector-specific and spillover effects.}. 

To ensure that my results are not confounded by sample selection, I calculate propensity score-based weights\footnote{This method is described in more detail in Appendix Section \ref{AppendixPScoreWeighting}. In short, I split data across the three samples by gender and employment sector. Then for each subsample and each observation I predict the probabilities of belonging to each of the 3 samples, using a multinomial probit model. For this, I control for all worker covariates excluding time and state of residence and also include in other covariates which I do not use in my final analyses. Then I use the inverse of these predicted probabilities as weights after normalizing the weights within each group.} for each observation in each of the three samples, such that the weighted covariate distributions across the three samples are the same\footnote{The covariate balance is shown in Appendix Tables \ref{Tab-CovBalWeek} and \ref{Tab-CovBalHour} for all workers, and for hourly paid workers separately.}. 
I bootstrap this analyses to compute standard errors. 

Appendix Table \ref{Tab07} shows the effects of SHBs on public and private sector workers in the first sample \textbf{AllStateBan}. 
For both hourly wages and weekly earnings as outcomes, the gender pay gap actually increases in the public sector (not statistically significant for weekly earnings), driven almost entirely by an increase in male earnings by about 3\%.  
In contrast, the gender pay gap decreases by 2.7 p.p. in the private sector, again driven mostly by an increase in female wages and earnings.
Therefore, SHBs seem to have achieved their intended goals only in the private sector, but have actually widened the gap in the public sector. 

Next, in Appendix Table \ref{Tab08} I show the effects from samples \textbf{PublicStateBan} and \textbf{AllStateBan-PP} in the first and the last two columns respectively. I find that public sector-specific SHB worsens gender pay gaps in the public sector and has moderate positive effects on the gender pay gap in the private sector. 
In contrast, private sector-specific SHB has little effects on the public sector, while it improves the gender pay gap in the private sector.

\subsection{Heterogeneous Effects of SHB by Worker Subgroups \label{ResultsByWorkerSubgroups}}

In Appendix Table \ref{tab:GPGRace} I show the effects of SHB on earnings separately by race (rows 1 and 2) and the gender pay gap by race (rows 3 and 4), and in rows (5) and (6) I show the baseline gender pay gap by race. 
As with my main results, I find little evidence for any substantial effects on average wages and weekly earnings for both white and black workers. 
SHBs decrease the gender pay gap, but only for white workers, and this is driven again by an increase in women's earnings (row 2 in columns 5 and 6 of Appendix Table \ref{tab:RaceGap}). 
However, I do not find much evidence for SHBs decreasing the gender pay gap among black workers. 

In Appendix Table \ref{tab:RaceGap} I further show the effects of SHBs on the gap between white and black workers (denoted as RaceGap) in row 1. 
These results suggest that the white premium among women increases by around 3 p.p. (significant at 95\%), two-thirds of which is driven by an increase in white women's earnings. 
In contrast, the race gap between white and black men decreases by about 1.5 p.p. (not significant), half of which is driven by a marginal increase in black men's earnings.  

In Appendix Fig \ref{fig:SHBEffectOnHWByAgeXEduc} I plot the effects of SHB on hourly wages for men and women and gender wage gap, separately for each of five education levels (less than high school, high school, some college, associate degrees, and four year college degrees) and age.  
Similarly, in Appendix Figure \ref{fig:SHBEffectOnWEByAgeXEduc}, I do the same for weekly earnings instead. 
Together, these two sets of results suggest that SHB decreases the gender pay gap mostly for young workers with higher levels of education.

\subsection{Effects of SHB on Labor Force Participation and Job Turnover \label{ResultsPT}}

My results in Section $\ref{ResultsGPG}$ could be driven by changes in labor market turnover rates in response to the bans, as opposed to wage growth conditional on turnover. 
For example, if salary history bans result in higher (lower) job-to-job (J2J) transitions for women (men), then the gender pay gap will decrease in repeated cross-sections, even if there is no change in wage growth conditional on job change. 
Again, if more women select into the labor force, especially in high-income jobs, the gender pay gap will decrease even if wage growth does not change conditional on J2J transitions. 
To check whether my results are driven by these channels, I use linear probability models similar to the baseline specification in ($\ref{Eq01}$) to estimate the effects of salary history bans on nine outcomes: labor force participation (LFP), unemployment rate (UR), employment in the private sector, employment in the public sector, monthly unemployment-to-employment (U2E), monthly job-to-job (J2J), monthly employment-to-unemployment (E2U) transitions, monthly transitions from the Private sector to the Public sector (Pr2Pu), and monthly transitions from public sector to private sector (Pu2Pr). 

These results are shown in Appendix Table $\ref{Tab10}$. 
I fail to find any large or significant effects on these outcomes, with the exception of small and positive (0.4\%) effects on male unemployment rate (Row 2 of Column 2).
These results show that my estimates on pay and gender pay gap are not subject to any large effects from either selection into work or selection into specific types of jobs. 


\section{Pre-Ban Disclosure Behavior and Effects of Ban on Disclosure \label{ResultsDisclosure}} 

SHBs do not restrict job applicants from voluntarily disclosing their earnings information. 
Therefore, for SHBs to have any effect on earnings, average disclosure rates would have to change after the bans. 
More specifically, for SHBs to have different effects by gender, disclosure rates would have to change differently for men and women. 
This line of reasoning suggests that it is the prospective employers' nudge for information that might have induced job applicants to disclose salary history more frequently before the ban, and when SHBs restrict enquiry, average disclosure rates go down. 
Therefore, it must follow that a significant proportion of job applicants were being asked about their salary history before the ban and they would have been less likely to disclose, if only they were not asked. 
Following the same reasoning, it must also be the case that before the ban, disclosure rates were higher among job applicants who were asked about their salary history than those who were not. 
Is this line of argument valid and do the associated hypotheses about disclosure behavior hold true empirically? 
This is precisely what I investigate in this section using the PayScale data on job interviewees' disclosure behavior. 
To do this, I construct a sample using the same restrictions as used for my main earnings sample \textbf{AllStateBan}.

\subsection{Do Salary History Bans reduce overall disclosure rates? \label{ResultsPayScaleSHBEffects}}

To check whether SHBs reduce disclosure rates when employers can no longer nudge for information, I use the following specification:

\begin{multline}\label{EQSHBEffectsDisclosure}
\mathbb{I}_{ismt}^{\text{Disclosed}} = \alpha+\beta_1\mathbb{I}_{i}^{f}\mathbb{T}_{smt}+\beta_2\mathbb{I}_{i}^{f}+\beta_3\mathbb{T}_{smt}+\beta_4 \mathbb{I}_{i}^{f}EverTreat_{s}+\gamma X_{ismt}\\ +\lambda_s + \lambda_m + \lambda_t+\delta\lambda_f t+\epsilon_{ismt} 
\end{multline}

These results are shown in Table \ref{tab:EffectSHBDisclosure}. 
Rows 1 and 2 show that SHBs reduce disclosure rates for both women and men by 24 p.p. and 22 p.p. respectively. 
These results are statistically significant at 99\% level and robust across models and sample weighting. 
Moreover, women reduce their disclosure rates by 2 p.p. more than men in response to SHBs. 
Therefore, my initial hypothesis about SHBs reducing disclosure rates holds true. 
These results suggest that it is really the nudge from the employer that induces most applicants to disclose information; and when SHBs restrict that nudge, candidates are much less likely to volunteer salary information. 

Row 3 in Table \ref{tab:EffectSHBDisclosure} shows that pre-SHB women were about 2 p.p. more likely than men to disclose salary information and row 4 shows that there is little evidence for gender gaps in disclosure rates after SHB.  
What drives this higher pre-SHB disclosure rates among women? 
Is it the case that women were asked about their salary history more frequently than men, before the ban? 
Or is the case that conditional on being asked, women were more likely to disclose than their male counterparts? 
I examine these questions in the next two sections. 

\subsection{Who gets asked about salary history when there is no SHB? \label{ResultsPayScaleEnquiry}}

To check whether women were more likely to get asked about their salary history before the ban, I use the following specification:
\begin{equation}\label{EQAsked}\mathbb{I}_{ismt}^{\text{Asked}} = \alpha + \textcolor{red}{\beta}\mathbb{I}_{i}^{f} + \gamma X_{it} + \lambda_{s} + \lambda_{m} + \lambda_{t} + \delta\lambda_{f}t + \epsilon_{ismt},\end{equation}

where the outcome $\mathbb{I}_{ismt}^{\text{Asked}}$ is an indicator of whether the individual $i$ residing in state $s$ on calendar month $m$ and year $t$ was asked about salary history, $\mathbb{I}_{i}^{f}$ is a female dummy, $X_{it}$ are time varying and time-invariant individual covariates, $\lambda$ are fixed effects, and $f(t)$ is a linear time trend. 
The coefficient $\beta$ measures the conditional gender gap in enquiry rates and is shown in Table \ref{tab:PreBanEnquiryGenderDiff}. 
Besides this linear probability model, I also use logit and probit models and find the marginal effects of the female dummy. 
These results show, that conditional on other observables, women were significantly (at the 95\% confidence level) more likely by 2.8-3.9 p.p. than men to be asked about their salary history before the ban. 

One possible explanation is that these candidates work in positions where there is higher variation in earnings and recruiters ask for salary history as a way to set their offer ranges. 
To test this hypotheses, in Figure \ref{fig:CorrOccVarOcc} I plot the occupation fixed effects from (\ref{EQAsked}) against within-occupation earnings variance relative to the base occupation (Administrative Support) which is used to normalize the occupation fixed effects\footnote{I control for calendar year and calendar month when I estimate the occupation fixed effects. When computing the within-occupation earnings variance I pool data across the relatively short time horizon in the PayScale data - second quarter of 2017 to third quarter of 2019.}. 
In Panel A I show this for men and women together and in panel B I estimate gender-specific effects and relative variance. 
Both these panels show that the occupation-specific likelihood of being asked about salary history is positively correlated with occupation-specific earnings variance. 
Therefore, it appears that employers are more likely to ask for information when it is more difficult for them to narrow down the applicants' current salary range, conditional on all other observables. 

\subsection{Pre-Ban Disclosure Behavior among Job Applicants \label{ResultsPayScaleDisclosure}}
 
In the previous section I showed that women were indeed about 3-4 p.p. more likely than men to be asked, conditional on all other observables. 
Is it also the case that they were more likely than men to disclose, conditional on being asked? In order to investigate how disclosure rates differ by enquiry and gender I run the following three regressions:
\begin{align}
\text{Panel A: } & \mathbb{I}_{ismt}^{\text{Disclosed}} &=& \alpha + \beta_{1}\mathbb{I}_{ismt}^{\text{Asked}} + \beta_{2}\mathbb{I}_{i}^{f} + \gamma X_{it} + \lambda_{s} + \lambda_{m} + \lambda_{t} + \delta\lambda_{f}t + \epsilon_{ismt} \label{EQPreBanDiscPanelA} \\
\text{Panel B: } & \mathbb{I}_{ismt}^{\text{Disclosed}} &=& \alpha + \beta_{1}\mathbb{I}_{ismt}^{\text{Asked}} + \beta_{2}\mathbb{I}_{i}^{f} + \beta_{3}\mathbb{I}_{ismt}^{\text{Asked}}*\mathbb{I}_{i}^{f} + \gamma X_{it} + \lambda_{s} + \lambda_{m} + \lambda_{t} + \delta\lambda_{f}t + \epsilon_{ismt} \label{EQPreBanDiscPanelB} \\
\text{Panel C: } & \mathbb{I}_{ismt}^{\text{Disclosed}} &=& \alpha + \beta_{1}\mathbb{I}_{ismt}^{\text{Asked}} + \gamma X_{it} + \lambda_{s} + \lambda_{m} + \lambda_{t} + \delta\lambda_{f}t + \epsilon_{ismt}, \label{EQPreBanDiscPanelC}
\end{align}
where $\mathbb{I}_{ismt}^{\text{Disclosed}}$ is a dummy for whether an individual $i$ residing in state $s$ at calendar month $m$ and calendar year $t$ had disclosed salary information, $\mathbb{I}_{ismt}^{\text{Asked}}$ is a dummy for whether they were asked, $\mathbb{I}_{i}^{f}$ is a female dummy, and all other variables have the same interpretation as in equations before.  
The results are shown in the three panels of Table \ref{tab:PreBanDisclosure}.  

Panel A shows that controlling for gender and other covariates, candidates are roughly 62-64 p.p. (significant at 99\% confidence level) more likely to disclose if they are asked in comparison to when they are not asked (row 1).
If I look at the unadjusted proportions, 77.07\% of those who are asked disclose information in contrast to only 12.28\% who volunteer. 
Does this imply that both men and women are both more likely to disclose when asked versus when not asked? 
Moreover, are men and women equally likely to disclose regardless of whether they are asked? 
That is what I analyze using the specification in (\ref{EQPreBanDiscPanelB}) and show these results in Panel B of Table \ref{tab:PreBanDisclosure}. 

In rows 1 and 2 of Panel B, I show that women are roughly 66 p.p. more likely and men are about 60 p.p. more likely to disclose when asked versus when not asked. 
These effects are precisely estimated, statistically significant at 99\% confidence level, and robust across models and sample weighting.  
Row 3 shows that among candidates who were asked about salary history, women were 4-5 p.p. (significant at 99\% confidence level) more likely than men to disclose information. 
In contrast, row 4 shows that among candidates who were not asked, women were about 2 p.p. less likely than men to disclose information, although this result is not statistically significant. 
Overall, these results show that both men and women are more likely to disclose salary history when nudged and especially among those who are asked, women are more likely to share information. 
Together with the fact that women are about 3-4 p.p. more likely to be asked than men, average disclosure rates among women were 2-3 p.p. (significant at 95\% confidence levels) higher than men before the ban (Panel C of Table \ref{tab:PreBanDisclosure}). 

Why don't most job applicants volunteer information and a prompt from the employer incentivize more candidates to share salary history?
These are precisely the questions that I explore in more detail in Section \ref{disclosurereasons} 
However, before that, I investigate whether I can identify in more detail the type of job applicants who disclose and those who don't. 

Why do some job applicants choose to disclose when asked or volunteer information even when not asked? 
Is it the case that high-earners are more likely to disclose, because they can use their already high salaries to bargain for even higher offers? 
To test for this, I estimate the correlation between earnings and disclosure rates, conditional on observables. 
More specifically, I show whether there are income gaps in disclosure rates for the same gender and gender differences in disclosure rates for the same income level. 
These results are shown in Table \ref{tab:DiscByIncomeGender}. 

Rows 1 and 2 of show that among both women and men, high earners were 6-8 p.p. (significant at 99\% level) more likely to disclose than low earners\footnote{In Appendix Table \ref{tab:AppendixDiscByIncomeGender} I show these results separately for those who were asked (Panel B) and those who were not asked (Panel C). These additional results suggest that when asked, low earning women were more likely to disclose than low earning men. But when not asked, high earning men were more likely to disclose than high earning women.}. 
The income gap in disclosure rates for men vanish after the ban. 
In contrast, high-earning women are still more likely (by around 2 p.p.) to disclose than low-earning women. 

Rows 3 and 4 show that before SHB, among candidates who earned less than their occupation-specific median salary, women were roughly 3 p.p. (significant at 99\% level) more likely than men to disclose salary information. 
In contrast, there is little evidence for gender differences in disclosure rates among those who earned more than their occupation-specific median salary. 
After SHB, there is still little evidence for gender differences in disclosure rates among high earners. 
However, among low earners, men appear to disclose slightly more than women (although this effect is not significant in some specifications). 


\subsection{Motivations behind Job Applicants' Disclosure Behavior: Survey Evidence \label{disclosurereasons}}

Applicants could have different motivations for disclosing or withholding salary information from prospective employers. 
These incentives could be both financial and behavioral, and correlated with applicant demographics. 
For example, high-earners could be more likely to disclose information than low-earners because revealing already high salaries offers them a bargaining advantage. 
Applicants could also be apprehensive that if they do not disclose salary history, recruiters might think they earn less and offer them lower salaries. 
Beyond these financial incentives, behavioral factors could drive disclosure decisions as well. 
Applicants could withhold information either because they actively dislike sharing confidential information like salary, or because they feel that salary offers should not depend on what they earn, but on the skills and experience that they bring to the job. 
On the flip side, applicants might feel uncomfortable about refusing information especially when asked, because they don't want to come off as disagreeable to recruiters. 

To better understand motivations that drive disclosure decisions, I conducted a representative survey\footnote{I designed the survey in Qualtrics and administered the survey online through the sample survey platform \href{https://luc.id/}{\underline{Lucid}}. I targeted my survey to those who were employed and within the age range of 25-50 years. An equal number of men and women were surveyed from 4 census regions across the US. Conditional on these variables, survey participants were representative of the US population.} of $\sim$5,700 US respondents, where I asked survey participants whether they would disclose salary history. 
Conditional on their response, I then asked them why they would or would not disclose information\footnote{The full set of survey questions is available in Appendix Section \ref{JobInterviewBehaviorSurvey}}. 

Figure \ref{fig:survey_wouldyoudisclose} shows that among both men and women, over 70\% of survey participants stated that they would disclose information only when asked and withhold information when not asked. 
Around 18\% of respondents said that they would refuse to disclose, and only around 11\% of respondents said they would volunteer information. 
This evidence lines up with my previous findings from the PayScale disclosure data, where I had discovered that most job applicants do not volunteer information and are more likely to disclose information only when asked. 

Figure \ref{fig:survey_perceivesalaryquestions} shows how respondents perceive questions about salary history during job interviews and decompose these responses by gender. 
Although around 50\% of survey participants (among both men and women) state that they are likely to think the recruiter is trying to drive down the offer, 50\% of participants also state that they wouldn't think too much of these questions since it's only a bargaining strategy. 
Furthermore, only around of 30\% of respondents believe that questions about salary history imply the prospective employer does not care about the candidates' skills or experience. 
Overall, I do not find much evidence for gender differences in these responses and it appears that candidates are not necessarily put off by enquiries.

In Figure \ref{fig:survey_reasonfordisclosing} I show what proportion of respondents agree with several stated reasons for choosing to disclose salary history. 
Men are about 8 p.p. more likely than women to state that their current salary is already high and therefore they can bargain better if they reveal information. 
In contrast, women are about 10 p.p. more likely to state that they would disclose because they are uncomfortable negotiating salary. 
Men and women are almost equally likely to state that they would reveal salary information because otherwise the recruiter might think their current salary is low or even dislike them. 
In Figure \ref{fig:survey_reasonfornotdisclosing} I show what proportion of respondents agree with several stated reasons for not disclosing salary history. 
Over 70\% of both men and women state that they believe new offers should depend on their skills and experience; over 50\% of respondents state their previous salary should not matter and because withholding information helps them negotiate better. 
About 30\% of respondents state they feel uncomfortable talking about salary, and women are about 10\% more likely than men to say so. 

Overall, these findings suggest that while job applicants are not necessarily put off by salary history questions from recruiters, they feel obligated to reveal information, especially when asked. 

\subsection{How do I explain refusal to disclose or withholding of information? \label{whydontdisclose}}

The results in previous sections show that before the ban 88\% of workers do not volunteer information when not asked and 23\% of workers refuse to disclose when asked. 
This, coupled with the fact that only 36\% of workers were asked, implies that around 65\% of workers did not share salary history with their prospective employers before the ban. 
A natural question at this point is why such a large proportion of workers do not share information.
More specifically, how do I reconcile these disclosure decisions with concerns about statistical discrimination where information withholding might send negative signals about the worker's pay and productivity to the prospective employer? 
Does this imply that statistical discrimination is not a concern in this setting? 

I first argue that statistical discrimination is not observationally inconsistent with a positive non-disclosure rate, if the underlying wage distribution has two supports. 
To see this, think of a framework where workers' current salary can take from only two discrete values and these workers are all observationally similar otherwise. 
If everybody withholds information, then the prospective employer would offer a wage that matches the outside option of the `average' worker and this would incentivize high-earners to disclose salary\footnote{This is similar to the unravelling of the lemon's market in \citet{Akerlof1970}. In fact, even with a continuous initial wage distribution, statistical discrimination would ensure that all but the lowest earning workers have an incentive to disclose salary.}. 
Remaining workers are then credibly identified as low-earners and they are indifferent between disclosing and withholding information.  

However, this framework still does not capture the following two empirical observations: (1) not all high wage earners disclose information, (2) candidates are much more likely to disclose information when asked versus when not asked. 
Nor does it help explain how the SHB improves the gender pay gap. 
Therefore, statistical discrimination alone cannot explain reduction in disclosure rates after the ban and the accompanying reduction in the gender pay gap. 
How do I then explain the disclosure results in the previous sections? 

A possible starting point is the empirical observation that candidates are more likely to disclose when asked in comparison to when not asked. 
In fact, when not asked an overwhelming majority (88\%) of candidates do not volunteer information. 
This alludes to two facts: (a) job applicants might face a higher `penalty' (financial or otherwise) when they refuse information in comparison to when they simply do not volunteer information, (b) job applicants might actively `dislike' sharing confidential information like salary, in general. 
I incorporate these two ideas in a comprehensive framework in the next section, and show how this helps me predict pre-SHB disclosure behavior and SHB's effects on the gender pay gap. 

\section{Model \label{TheoreticalModel}}

In this section I discuss a theoretical framework that helps reconcile my empirical findings on pre-SHB disclosure behavior, pre-SHB gender pay gap, and the effects of SHB on disclosure rates and gender pay gap. 
To ensure analytical tractability, I first introduce a stylized model of salary negotiation and use this simple framework to prove my main results. 
Then in Sections \ref{discretewage} and \ref{continuouswage}, I generalize the stylized model, and using numerical simulations I show that my main results still go through.  

\subsection{Model Set-Up \label{modelsetup}}

Time is discrete and there are 2 time periods: $t\in\left\{1,2\right\}$. 
For simplicity, let's assume that there is one female worker (F) and one male worker (M) in the economy. 
In time period $t=1$ each worker $i$ is employed with a match-specific output $z$ (invariant across gender) and they are paid a wage $w^{i}\in\left\{w_{L}, w_{H}\right\}$, where $w_{L}<w_{H}\leq z$. 
For the female (male) worker, the probability that they are a low-earner (i.e., earns $w_{L}$) is $f_{L}^{F}$ ($f_{L}^{M}$). 
Given a wage $w_{i}$ and output $z$, the worker's flow utility from the match is given by $u(w_{i}, z)$. 
At the end of time period $t=1$, both workers receive the same additive shock $c$ which decreases their flow utility of $t=1$. 

At the beginning of time period $t=2$, both workers get matched to new risk-neutral firms which observe their initial output $z$, their utility shock $c$, type-specific wage distribution $f_{L}^{k},\  k\in\left\{F,M\right\}$, but not their actual wage $w^{i}$. 
At the beginning of $t=2$, both worker and new firm also observe a new match-specific output $z'$ (gender-invariant), an enquiry shock $e^{i}\in\left\{0,1\right\}$ and an array of gender-specific \textit{psychic costs}: $\left\{c_{ed}^{i}|d^{i}\in\left\{0,1\right\}\right\}$, where $d^{i}$ is the worker's disclosure decision. 
Then the worker decides whether to disclose ($d^{i}$) initial wage ($w^{i}$), and the new firm simultaneously commits to a new wage offer ($w_{i}^{'}$) which depends on $d^{i}$ and $w^{i}\Big|_{d^{i}=1}$. 

In this setting, I capture the pre-SHB interaction between the worker and the prospective employer through the enquiry shock $e^{i}$. 
In particular, I assume that in the pre-SHB case, all workers receive a positive enquiry shock ($e^{i}=1,\  \forall\  i\in\left\{F,M\right\}$) and in the post-SHB case all workers receive a zero enquiry shock ($e^{i}=0$)\footnote{In the simulations of Sections \ref{discretewage} and \ref{continuouswage} I relax this assumption to $0<\text{Pr}(e^{i}=1|k)<1\  \forall\  k\in\left\{F,M\right\}$.}. 

The timing of the decision game between the worker and the new firm is as follows:
\begin{itemize}
    \item[1.] Worker $i$ gets matched with new firm $j'$. 
    \item[2.] $j'$ observes $z$, $c$, $\left\{f_{L}^{k}|i\in k\right\}$. 
    \item[3.] $i$ and $j'$ both observe $z'$ and an enquiry shock $e^{i}\in\left\{0,1\right\}$.  
    \item[4.] $i$ and $j'$ both observe an array of additive \textit{psychic costs}: $\left\{c_{ed}^{i}|d^{i}\in\left\{0,1\right\}\right\}$.
    \item[5.] $j'$ commits to a non-disclosure wage ($w_{i}^{'}(X, d^{i}=0)$) and a disclosure wage schedule ($w_{i}^{'}(X, d^{i}=1, w_{i})_{w_{i}\in\left\{w_{L}, w_{H}\right\}}$), where $X$ refers to $(z,z',c,e,c_{e1}^{i},c_{e0}^{i})$. 
    Simultaneously, worker $i$ chooses whether to disclose ($d^{i}\in\left\{0,1\right\}$) and whether to accept the new offer ($a_{i}\in\left\{0,1\right\}$)\footnote{I do not allow the incumbent firm to make counter offers to the worker. 
    This is because it is difficult to reconcile counter-offers with the notion of one-shot disclosure or withholding of information.
    More specifically, it is unclear how to interpret the worker not disclosing their own previous salary but credibly conveying a counter offer to the prospective employer.}. 
    \item[6.] The true wage $w_{i}$ of the worker is either revealed or not revealed to $j'$ depending on the worker's decision $d^{i}$\footnote{I do not allow the worker to lie about their initial wage when they decide to disclose ($d^{i}=1$). 
    This is because prospective employers are not prohibited from verifying salary information once the job applicant has disclosed this information. 
    An equivalent way to ensure that in equilibrium the worker does not lie would be to include costless salary verification once disclosed and an infinite penalty on the worker if they are found to have lied.}. New output, new wages, and turnover are realized.   
\end{itemize} 
I impose the following assumptions on my model:
\begin{assumption}{1}{}\label{Assumputil}
$\frac{\partial u}{\partial w}(w, z)\Bigg{|}_{z}>0$, $\frac{{\partial}^{2}u}{\partial{w}^{2}}(w, z)\Bigg{|}_{z}<0$,  $\frac{\partial u}{\partial z}(w, z)\Bigg{|}_{w}<0$.
\end{assumption}
\begin{assumption}{2}{}{\label{Assumpprivacycost}}
$c_{e1}^{i}>c_{e0}^{i}$ $\forall$ $i\in\left\{F,M\right\}$, $e\in\left\{0,1\right\}$. \end{assumption}
\begin{assumption}{3}{}\label{Assumpnonconformity}
$c_{10}^{F}>c_{10}^{M}>c_{00}^{F}=c_{00}^{M}$
\end{assumption}
\begin{assumption}{4}{}\label{Assumpequalgender}
$c_{e1}^{F}=c_{e1}^{M}$ $\forall$ $e\in\left\{0,1\right\}$
\end{assumption} 

Before I write down the worker and the firm's optimization problems in Section \ref{optimalproblems}, I make a small detour in the next section to explain my assumptions and in particular, the idea of \textit{psychic costs} in disclosure behavior. 

\subsection{Psychic Costs of Disclosing and Withholding Salary \label{psychiccostidea}}

Assumption \ref{Assumputil} captures the fact that a worker's flow utility increases with wages and there is diminishing marginal utility from wage increases.
However, if output increases without any increase in the worker's wage, then the worker might fell undervalued and their flow utility decreases. This is motivated by a body of literature which shows how workers' job satisfaction decreases when they find out that they are underpaid relative to their co-workers (\citet{ClarkOswald1996}, \citet{Luttmer2005}, \citet{Brownetal2008}, \citet{Clarketal2009}, \citet{GodechotSenik2015}, \citet{CullenPerezTruglia2021}). 
If higher salaries among co-workers, especially those who do similar work, is indicative of the output that a job generates, then these findings would parallel the essence of this Assumption \ref{Assumputil}. 

Assumption \ref{Assumpprivacycost} captures the fact that workers value privacy of confidential information like salary and therefore, they face psychic costs of having to disclose private information. 
This is motivated by existing survey evidence (\citet{Cranoretal1999}, \citet{Phelpsetal2000}, \citet{Singeretal2001}, \citet{Waldoetal2007}) that suggests individuals value privacy of their personal information and are reluctant to share this information in online platforms and in surveys. 
In the field of human-computer interaction, survey design, consumer research, and marketing, researchers have tried to valuate information privacy and designed mechanisms to elicit information (\citet{Kleinbergetal2001},  \citet{Hannetal2002},  \citet{Hubermanetal2005}, \citet{WathieuFreidman2007},  \citet{GhoshRoth2011},  \citet{LigettRoth2012},  \citet{RothSchoenbeck2012},  \citet{RothSchoenbeck2012}). 
In a related sense, the final assumption captures the fact that men and women are equally private about sharing confidential information about salary. 

Assumption \ref{Assumpnonconformity} captures the notion that both men and women face a utility cost of refusing to disclose salary history when asked. 
In particular, women face higher utility costs than men if they refuse to disclose when asked. 
This relates to the fact that women are more likely to face higher non-conformity costs than men. 
It is motivated by existing literature on the `psychic cost/stress' faced by economic agents when they do not conform to accepted standards (\citet{Bernheim1994}, \citet{ErardFeinstein1994}, \citet{Dullecketal2016}) and gender differences in these psychic costs (\citet{DAttoma2017}, \citet{GroschHolger2016}, \citet{TomGranie2011}). 
Much of this literature pertains to tax compliance, criminal behaviour and the main thesis of this literature is that individual psychic costs are often more salient than legal frameworks, in inducing compliance among agents.

\subsection{Worker and Firm Optimization Problems \label{optimalproblems}}

Given a match, i.e., $X:=(k,z,c,f_{L}^{k},z',e^{i},c_{e1}^{k},c_{e0}^{k})$ the worker decides on the optimal disclosure strategy  $d^{i}\in\left\{0,1\right\}$ and acceptance $a^{i}\in\left\{0,1\right\}$, and the prospective employer simultaneously commits to a non-disclosure wage ($w_{i}^{'}(X, d^{i}=0)$) and a disclosure wage schedule $(w_{i}^{'}(X, d^{i}=1, w_{i}))_{w_{i}\in\left\{w_{L},w_{H}\right\}}$. 
I abbreviate the new wage offers as $w_{i0}^{'}, w_{i1}^{'}(w_{L}), w_{i1}^{'}(w_{H})$. 
I can therefore write down the worker and the new firm's optimization problems as follows:
\begin{flushleft}\textbf{Worker's Optimization Problem:}\end{flushleft}
\begin{equation}\label{optimworker}\max_{d\in\left\{0,1\right\},a\in\left\{0,1\right\}}\  \Bigg(a\Big[du(\mathbf{w_{i1}^{'}(w)},z')+(1-d)u(\mathbf{w_{i0}^{'}},z')-dc_{e1}^{i}-(1-d)c_{e0}^{i})\Big]+(1-a)\Big[u(z,w)-c\Big]\Bigg{|}X,w\Bigg)\end{equation}
\begin{flushleft}\textbf{Firm's Optimization Problem:}\end{flushleft}
\begin{equation}\label{optimfirm}\max_{w_{i0}^{'}\leq z',w_{i1}^{'}(w_{L})\leq z',w_{i1}^{'}(w_{H})\leq z'}\Bigg(\mathbf{a}z'-\mathbf{a}\Big[w_{i0}^{'}\text{Pr}(\mathbf{d=0})+w_{i1}^{'}(w_{L})\text{Pr}(\mathbf{d=1}, w=w_{L})+w_{i1}^{'}(w_{H})\text{Pr}(\mathbf{d=1}, w=w_{H})\Big]\Bigg{|}X\Bigg)\end{equation} 

\subsection{Model Solutions and Discussion \label{modelsolutions}}

In this section I show that there exists an equilibrium set of parameter values, such that the solutions capture the following set of empirical observations:
\begin{itemize}
    \item[1.] A non-zero proportion of workers do not disclose their salary history before the ban. 
    \item[2.] Disclosure rates are higher among workers who are asked about salary history than among those who are not asked. 
    \item[3.] When asked, disclosure rates are higher among women than men. 
    \item[4.] Disclosure rates are higher among high-earners than among low-earners. 
    \item[5.] SHB reduces disclosure rates on an average. 
    \item[6.] SHB reduces the gender wage gap. 
\end{itemize}

To do this, I proceed in three steps. 
First, I prove that for given a type of worker and other parameter values, there exists an initial wage distribution such that some workers do not disclose salary history in equilibrium. 
Then, I contrast the cases when the same type of worker is asked and not asked for salary history. 
Here, I show that there exists an initial distribution of wages, for which there are differences in both offered wages and disclosure rates when asked versus when not asked. 
Finally, in the third step I show that there exists different initial wage distributions for men and women such that disclosure rates are higher among women than men when asked, and SHB reduces the gender wage gap. 

\begin{flushleft}\textit{Step 1: There exists an initial wage distribution for a given type of worker, such that a non-zero proportion of workers of that type do not disclose salary history, regardless of enquiry.}\end{flushleft}

\begin{proposition}{1}{}\label{prop:existnondisc}
Given $(z,c,z',e,c_{e1},c_{e0},w_{L},w_{H})$ $\exists\  \tilde{f}_{e}\in\left(0,1\right)$ such that:
\begin{itemize}
    \item[1.] $\forall\  f_{L}>\tilde{f}_{e}$ there is a \textbf{Separating Equilibrium} where the firm commits to offering $\hat{w}_{e1}$ if the worker discloses and is found to be a high-earner, and $\tilde{w}_{e0}$ in all other cases. 
    Low-earners (i.e., $w=w_{L}$) optimally choose to not disclose ($d=0$) and high-earners (i.e., $w=w_{H}$) choose to disclose ($d=1$).  
    \item[2.] $\forall\  f_{L}\leq\tilde{f}_{e}$ there is a \textbf{Pooling Equilibrium} where the firm offers a flat wage $\hat{w}_{e0}$ regardless of disclosure $d$ and initial wage $w$ (if disclosed).
    Both high and low-earners optimally choose to not disclose ($d=0$).  
    \item[3.] $\tilde{f}_{1}<\tilde{f}_{0}$. 
\end{itemize}
\end{proposition} 
The proof of this proposition can be found in Appendix \ref{ProofsOfPropositions}.
The intuition behind this result is as follows. 

First, when workers disclose their salary history, they face a \textit{psychic cost} of disclosing information. 
Therefore the new firm has to compensate them for this additional utility cost to make them indifferent between accepting and rejecting the new offer. 
Second, when the new firm can distinguish between high and low-earners it can extract full surplus from both, while in the pooling equilibrium it can extract full surplus only from high-earners while paying some rent to low-earners. 
Therefore, when the number of low-earners is high ($f_{L}>\tilde{f}$), instead of having to pay rent to a large number of low-earners, the firm finds it optimal to extract full surplus from both types of workers (i.e., separating equilibrium) and then use this surplus to compensate the small number of high earners for their utility cost of disclosing information. 
In contrast, when the number of high-earners is high ($f_{L}\leq\tilde{f}$), the firm cannot compensate for the utility costs of disclosing for such a large number of high-earners even by extracting full surplus from everyone. 
Instead, the firm decides to offer a flat wage.  
It's important to note here that the first two results in the proposition hold both when workers are asked $e=1$ and when they are not asked $e=0$, which mimic pre and post-SHB cases respectively. 

The final result in the proposition says that the threshold proportion of low-earners ($\tilde{f}_{e}$), below which the firm offers pooling wages, decreases when workers are asked (i.e., pre-SHB) in comparison to when workers are not asked. 
This is because workers face an additional cost of withholding information when asked. 
Therefore, in both pooling and separating equilibrium when asked, the firm has to compensate non-disclosing low-earners for their cost of non-compliance with information requests. 
However, in a pooling equilibrium the firm also has to pay some rent to low-earners. 
This is why the firm has a lower tolerance threshold for low-earners when workers are asked. 
In particular, the threshold of low-earners ($f_{L}$) above which the firm begins to offer separating wages, decreases when workers are asked in comparison to when they are not asked. 

In the next step I show how the outcomes for a given type of worker vary when they are asked and when not asked. 

\begin{flushleft}\textit{Step 2: There exists an initial wage distribution for a given type of worker, such that when workers are asked, the firm offers separating wages and when workers are not asked the firm offers a pooling wage.}\end{flushleft}

\begin{proposition}{2}{}\label{prop:SHBchangedisc}
Given $(z,c,z',e,c_{11}=c_{01}>c_{00},w_{L},w_{H})$ $\exists\  \bar{c},\tilde{f}_{1},\tilde{f}_{0}$ such that $\forall\  c_{10}\in(c_{00},\bar{c}),\  f_{L}\in(\tilde{f}_{1},\tilde{f}_{0})$:
\begin{itemize}
    \item[1.] The new firm offers separating wages when $e=1$ and pooling wages when $e=0$. 
    \item[2.] $\text{Pr}(d=1|e=1)>\text{Pr}(d=1|e=0)$. 
    \item[3.] $\mathbb{E}(w^{'}|e=1, w=w_{H})>\mathbb{E}(w^{'}|e=0, w=w_{H})$.
    \item[4.] $\mathbb{E}(w^{'}|e=1, w=w_{L})<\mathbb{E}(w^{'}|e=0, w=w_{L})$. 
\end{itemize}
\end{proposition}

The proof of this proposition and other cases ($c_{10}>\bar{c},\  f_{L}<\tilde{f}_{1},\  f_{L}>\tilde{f}_{0}$) are shown in Appendix \ref{ProofsOfPropositions}. 
These results naturally follow from Proposition \ref{prop:existnondisc}.

If the fraction of low-earners ($f_{L}$) is too low (i.e., $f_{L}<\tilde{f}_{1}<\tilde{f}_{0}$), then the firm offers pooling wages both when workers are asked and when they are not asked, and therefore there are no differences in disclosure rates. 
In contrast, when the fraction of low-earners is too high (i.e., $f_{L}>\tilde{f}_{0}>\tilde{f}_{1}$), the firm offers separating wages in both cases and again there are no differences in disclosure rates. 
However, when $f_{L}$ lies between these two thresholds ($\tilde{f}_{1}, \tilde{f}_{0}$), the firm offers pooling wages when workers are not asked and separating wages when workers are asked. 
Only high-earners disclose in a separating equilibrium while no worker discloses in a pooling equilibrium. 
This implies a reduction in expected disclosure rates from pre to post-SHB periods\footnote{In Section \ref{discretewage} and \ref{continuouswage}, I assume $0<\text{Pr}(e^{i}=1|k)<1\  \forall\  k\in\left\{F,M\right\}$ in pre-SHB period and $\text{Pr}(e^{i}=1|k)=0\  \forall\  k\in\left\{F,M\right\}$ in post-SHB period. Therefore, Proposition \ref{prop:SHBchangedisc} would suggest that the disclosure rate in pre-SHB period lies strictly between 0 and 1, and disclosure rates are higher among those who are asked. Furthermore, it would suggest that disclosure rates decrease in post-SHB period.}. 

In the separating equilibrium high-earners are being compensated for their utility cost of disclosing information ($c_{11}$), which is higher than their utility cost of withholding information ($c_{10}$) when asked, which in turn is higher than the cost of not volunteering information when not asked ($c_{00}$). 
Therefore, when $e=0$ and high-earners pool in a non-disclosing equilibrium, they would see their wage offers decrease. 
This is captured in point (3) in the proposition above. 

Furthermore, the upper bound ($\bar{c}$) on $c_{10}$ implies that when $e=1$ and low-earners do not disclose in a separating equilibrium, the additional compensation that the firm has to pay them for non-compliance, is not too high. 
This ensures that when $e=0$ and low-earners pool with high-earners, the pooling wage is higher than the separating wage they would have received when $e=1$. 
In other words, low-earners make a better wage when they are not asked in comparison to when they are asked.  

This proposition therefore captures the following empirical observations: (1) high-earners disclose more than low-earners when asked and (2) disclosure rates are higher among those who are asked. 
Since asking workers mimics the pre-SHB case and not asking mimics the post-SHB case, the results in Proposition \ref{prop:SHBchangedisc} would suggest that (3) SHB decreases disclosure rates. 

In the final step I distinguish between initial wage distributions by gender and show how I capture the pre-SHB disclosure differences between men and women, and SHB's effects on the gender wage gap. 

\begin{flushleft}\textit{Step 3: Different initial wage distributions for men and women result in different equilibria, and therefore different disclosure rates and wages, when asked versus when not asked.}\end{flushleft}

\begin{proposition}{3}{}\label{prop:diffgenderdist}
Given ($z,c,z',c_{11}=c_{01}>c_{10}^{M},c_{00},w_{L},w_{H}$) $\exists\  \underline{c}_{F}$ ($>c_{10}^{M}$) such that \\ 
$\forall\  c_{10}^{F}\in(c_{10}^{M},\underline{c}_{F})$, $\exists\  \tilde{f}_{1M},\bar{f},\tilde{f}_{0F}$ such that $\forall\  f_{L}^{M}<\tilde{f}_{1M}<\bar{f}<f_{L}^{F}<\tilde{f}_{0F}$:
\begin{itemize} 
    \item[1.] Firm offers separating wages to female workers and pooling wages to male workers when asked, and pooling wages to all workers when not asked. 
    \item[2.] $\text{Pr}(d=1|e=1,k=F)>\text{Pr}(d=1|e=1,k=M)$.
    \item[3.] $\text{Pr}(d=1|e=1)>\text{Pr}(d=1|e=0)=0$.
    \item[4.] $\mathbb{E}(w'|e=1,w=w_{L},k=F)<\mathbb{E}(w'|e=0,w=w_{L},k=F)$, \\ $\mathbb{E}(w'|e=1,w=w_{L},k=M)>\mathbb{E}(w'|e=0,w=w_{L},k=M)$. 
    \item[5.] $\mathbb{E}(w'|e=1,w=w_{H},k)>\mathbb{E}(w'|e=0,w=w_{H},k)\  \forall\  k\in\left\{F,M\right\}$.
    \item[6.] $\mathbb{E}(w'|e=0,k=F)-\mathbb{E}(w'|e=0,k=M)>\mathbb{E}(w'|e=1,k=F)-\mathbb{E}(w'|e=1,k=M)$.
\end{itemize}
\end{proposition}
The proof of this proposition is shown in Appendix \ref{ProofsOfPropositions}. 
Again, these results follow naturally from Propositions \ref{prop:existnondisc} and \ref{prop:SHBchangedisc}. 

Here, the fraction of low-earners among men is low enough such the firm offers pooling wages to men both when male workers are asked and when they are not asked. 
In contrast, the proportion of low-earners among women is high enough (but not too high) such that the firm offers separating wages when female workers are asked and pooling wages when they're not. 
This captures two empirical observations: (1) disclosure rates among women are higher than men when asked and (2) disclosure rates are higher on average when workers are asked. 

Both types of men are being compensated for their cost of not disclosing information when $e=1$.
Hence, when $e=0$ their wage offers decrease even when they do not volunteer information ($\because c_{00}<c_{10}^{M}$).
High-earning women are being compensated for their cost of disclosing information when $e=1$, and when they do not disclose for $e=0$, their wage offers decrease ($\because c_{10}^{F}<c_{11}$). 
In contrast, low-earning women were being compensated for withholding information when $e=1$, but this additional compensation was not too high because of the upper bound ($\underline{c}_{F}$) on $c_{10}^{F}$. 
So, when they pool with high-earners for $e=0$, their pooling wages are higher than their separating wages for $e=1$. 

Finally, when $e=0$, pooling wages are same for men and women since neither face any cost of refusing information and there is no other heterogeneity between the genders. 
So there is zero female premium in wages in the post-SHB case. 
When $e=1$, the low cost of withholding information among women ($c_{10}^{F}<\underline{c}_{F}$) ensures that average female wages are lower than average male wages. 
In other words there is a positive male premium in the pre-SHB case. 
Therefore, SHB narrows the gender wage gap. 

In this section I have shown how a stylized model can rationalize my empirical findings using a single female and a single male worker. 
In the immediate next section, I introduce a mass of workers and within-gender heterogeneity along some dimensions in both time periods. 
However, I stick to the gender-invariant, two-valued initial wage distribution. 
I relax this assumption further in Section \ref{continuouswage}. 

\subsection{Worker Heterogeneity with Discrete Initial Wage Distribution \label{discretewage}} 

I now extend the stylized model in the previous section by considering a unit mass of workers, instead of a single worker, for each type/gender $k$. 
I assume that they are all employed in $t=1$. 
Both female and male workers have the same initial wage support $\left\{w_{L},w_{H}\right\}$, but the fraction of low earners ($f_{L}^{k}$) can differ by type. 
Match-specific productivity for worker $i$ is drawn from the same distribution for both men and women, i.e., $z^{i}\sim F(.)$. 
At the end of $t=1$, both men and women receive additive utility shocks drawn from the same distribution, i.e., $c^{i}\sim C(.)$. 
This shock reduces the utility of remaining in the current job. 

At the beginning of $t=2$, worker $i$ gets matched to a new firm $j'$ and a new match-specific productivity $z^{i'}\sim F(.)$ is drawn. 
$j'$ observes $z^{i},c^{i}, z^{i'}$ but not the worker's actual wage $w^{i}$; only the type-specific fraction of low-earners $f_{L}^{k},\  i\in k$. 
Also, at the beginning of $t=2$, worker $i$ receives a publicly observable enquiry shock $e^{i}\sim \text{Binomial}(p_{k})$, where $p_{k}$ is the probability with which workers of type $k$ are asked about salary history. 
Depending on $e^{i}$ both observe an array of utility costs $\left\{c_{e1}^{i},c_{e0}^{i}\right\}$ where $c_{ed}^{i}\sim \mathscr{C}(.|e,d,k)$. 
In particular, I assume $\mathscr{C}(.|e=1,d=1,k)\stackrel{d}{=}\mathscr{C}(.|e=0,d=1,k)\stackrel{d}{=}\mathscr{C}(.|d=1)$ and $\mathscr{C}(.|e=0,d=0,k)\stackrel{d}{=}\mathscr{C}(.|e=0,d=0)$. 
These restrictions capture Assumption \ref{Assumpequalgender} in Section \ref{modelsetup}. 
However, the distributions for the utility cost of refusing information is gender-specific, i.e., $\mathscr{C}(.|e=1,d=0,F)\stackrel{d}{\neq}\mathscr{C}(.|e=1,d=0,M)$. 
To capture the relative shifts in this utility cost between men and women, I impose the following restriction:
\begin{equation}\label{FOSDConditions}\mathscr{C}(.|d=1)\succ_{FOSD} \mathscr{C}(.|e=1,d=0,F)\succ_{FOSD}\mathscr{C}(.|e=1,d=0,M)\succ_{FOSD} \mathscr{C}(.|e=0,d=0),\end{equation} Where $FOSD$ implies first-order stochastic dominance. 
However, for an individual worker $i$, I ensure that $c_{11}^{i}=c_{01}^{i}>c_{10}^{ik}>c_{00}^{i}$.

\subsection*{Setting up Numerical Simulations \label{discretewagesimu}}

To simulate outcomes for a given set of distributions, I consider two different cases: (a) pre-SHB case where $1<p_{F}=p_{M}<0$, and (b) post-SHB case where $p_{k}=0,\forall\  k\in\left\{F,M\right\}$. 
For both cases I use the same distribution of workers, except for the draws of $c_{e0}$. 
For simulation purposes, I further assume that all these distributions are normal, while still satisfying the distributional restrictions above. 
I also assume that for a given type, the distributions of underlying variables are independent of each other\footnote{I relax this assumption in Appendix Section \ref{AppendixSimResultsDiscreteFrameworkCorrwoldzold} where I allow for correlations between initial $w$ and $z$, separately for men and women, and allow this correlation to vary from negative to positive.}. 
For each case, I separately simulate the model and generate predictions for wages, disclosure decisions, job turnover, and gender gap outcomes in $t=2$. 
I then compare the two cases to find the effects of SHB on wages, disclosure rates, and gender gaps. 
I show how my results change as I vary the values of $f_{L}^{F}\text{ and }f_{L}^{M}$.

\subsection*{Results and Discussion \label{discretewageresults}}

In Figure \ref{fig:fLMfLF_PreBanDisclosure} I show the pre-SHB gender gap (female-male) in disclosure rates (left panel) and the pre-SHB difference in disclosure rates between those who are asked and those who are not (right panel). 
The plot on the left shows that disclosure rates among women are higher than men when the fraction of female low-earners is higher than the fraction of male low-earners, but not too high. 
This is equivalent to the condition $f_{L}^{M}<\tilde{f}_{1M}<\bar{f}<f_{L}^{F}<\tilde{f}_{0F}$ in Proposition \ref{prop:diffgenderdist} above. 
For a given $f_{L}^{M}$, gender differences in pre-SHB disclosure rates decrease as $f_{L}^{F}$ increases because only high-earners disclose and the proportion of high-earners decreases as $f_{L}^{F}$ increases. 
Furthermore, for a given $f_{L}^{F}$, gender gap in disclosure rates initially decrease as $f_{L}^{M}$ increases, because the firm begins to offer separating wages to men when asked. 
However, beyond a certain threshold value of $f_{L}^{M}$ ($\tilde{f}_{1M}$), the firm begins to offer pooling wages to men, and hence the gender gap in disclosure rates increases again. 
The plot on the right in Figure \ref{fig:fLMfLF_PreBanDisclosure} shows that when the fraction of low-earners is either not too low or too high, workers are more likely to disclose when asked than when they are not asked. 
This is because the firm offers separating wages to those who are asked and pooling wages to those who are not. 
Pooling and separating wages are offered to all when $f_{L}$ is too high or too low respectively. 
Therefore, there is little difference in enquiry rates between those who are asked and those who are not in these cases. 

In Figure \ref{fig:fLMfLF_SHBEffectOnDisclosure} I show that SHB reduces disclosure rates among both men and women when the fraction of low-earners is either not too high or too low. 
If the fraction of low-earners is too small (large) to begin with, then the firm would offer separating (pooling) wages both before and after SHB.  
In that case, SHB would have zero effect on disclosure rates. 

Figure \ref{fig:fLMfLF_GenderPayGap} shows the pre-SHB gender wage gap (left panel) and the SHB's effects on the wage gap (right panel) as $f_{L}^{F}$ and $f_{L}^{M}$ vary. 
Again, I find that when $f_{L}^{M}$ is low and $f_{L}^{F}$ is higher, women on average earn less than men in pre-SHB case (i.e., female premium is negative). 
This is driven by the fact that women, in contrast to men, are being offered separating wages where the large number of low-earning women earn their non-disclosure reservation wages, instead of earning rent through pooling wages. 
Moreover, for a given $f_{L}^{F}$, as the value of $f_{L}^{M}$ increases, the pre-SHB gender wage gap (female premium) increases because the firm begins to offer separating wages to men, which pushes down an increasingly larger number of low-earning men to their reservation wages.  
The figure on the right shows the effects of SHB on the gender wage gap. 
Again, when the fraction of low-earners is not too high or too low, then the gender wage gap (female premium) improves after SHB. 

Figure \ref{fig:fLMfLF_EffectOnSDWages} shows that standard deviation in wages decreases because of wage-pooling when the fraction of low-earners is not too high or too low. 
Finally, in Figure \ref{fig:fLMfLF_EffectOnWages} I show the effect of SHB on wages. 
Following Proposition \ref{prop:diffgenderdist}, wages decrease for the average worker. 
However, if $f_{L}^{F}$ or $f_{L}^{M}$ are not too high or too low, then wages increase for low-earners. 

While these simulations substantiate my empirical findings in a piecemeal fashion, a natural question at this point is whether there are initial wage distributions for men and women, such that all my empirical findings can be jointly validated by simulation. 
Since I have not calibrated the parameters for this numerical exercise, I care only about the existence of $f_{L}^{F}\in\left(0,1\right)$ and $f_{L}^{M}\in\left(0,1\right)$, which characterize initial wage distributions, given the values of other parameters. 
In Appendix Figure \ref{fig:AppendixfLMfLF_exist} I show which combinations of $(f_{L}^{F},f_{L}^{M})$ generate simulation results that parallel my empirical findings, for the given set of parameter values.  
Indeed there exists $f_{L}^{M}<f_{L}^{F}$, such that simulations validate my empirical results. 
This is not an unreasonable outcome of the model, since Appendix Figure \ref{fig:AppendixEarningsDist} shows that the female wage distribution is indeed more right-skewed than the male wage distribution in the CPS data. 

Finally, in Appendix \ref{AppendixSimResultsDiscreteFrameworkCorrwoldzold}, I account for correlation between initial wages ($w$) and initial productivity ($z$) separately for men and women. 
I show how model predictions change as I vary the degree of correlation between these two variables, separately for men and women. 
These simulation results are shown in Appendix Figures \ref{fig:corrwoldzold_preshbgdgaskeffect}-\ref{fig:corrwoldzold_effectwages}. 
Overall, they capture the general trends from the baseline model where there was no correlation between $w$ and $z$. 
My empirical findings are validated qualitatively even when there is positive correlation between $w$ and $z$ for both men and women.

\subsection{Worker Heterogeneity with Continuous Initial Wage Distribution \label{continuouswage}} 

Finally, I extend the model and simulations in the previous section by starting out with a continuous initial wage distribution instead of just two values. 
I still restrict initial wage distributions of men and women to have the same support, but allow them to be differently skewed. 
As in the previous section, I allow workers to be heterogeneous along all other dimensions ($z, z', c, c_{e1}$), and draw from the same distributions for both men and women, except for the utility cost of not disclosing when asked ($c_{10}$). 
Here, I stick to the stochastic dominance condition in (\ref{FOSDConditions}). 
At the individual worker level, I still impose the Assumptions in (\ref{Assumputil}), (\ref{Assumpprivacycost}), and (\ref{Assumpnonconformity}). 
For the baseline analyses, I assume that all these distributions are independent conditional on the worker's type $k$. 
For a given set of distributions, I first compute the optimal wage offer schedule for each worker ($i$) given $X:=(k,z,z',c,e,c_{e1},c_{e0})$ and $G(w|X)$, where $G(w|X)$ is the prior of initial wage distribution. 
This helps me pin down the pre-SHB disclosure behavior, gender wage gap, and the effects of SHB on disclosure, wages, and wage gap, for one set of male and female distributions. 
I then vary only the skewness ($\mu$) of the initial wage distributions ($G(w)$) independently for men and women, to see how the pre-SHB outcomes and the effects of SHB change.

These results are shown in Appendix Section \ref{SimResultsContinuousFramework}. 
Broadly, these results mirror what I know from the propositions and simulations in Section \ref{discretewage}. 
I find that there exists initial wage distributions for men and women, with a more positively skewed distribution for women than men, such that simulations match my empirical results qualitatively. 
Appendix Figures \ref{fig:OldWageDist_PreSHBGDG}-\ref{fig:OldWageDist_PreSHBIncEffect} show the pre-SHB disclosure differences by gender, enquiry status, and income respectively, while \ref{fig:OldWageDist_SHBEffectonDiscRates}-\ref{fig:OldWageDist_PostSHBGDG} show the effects of SHB on disclosure behavior. 
Appendix Figures \ref{fig:OldWageDist_PreEffectSHBGPG}-\ref{fig:OldWageDist_SHBEffectonMeanWage} show the effects of SHB on wages, gender pay gap, auto-correlation in wages, and standard deviation in wages. 

\section{Conclusion \label{Conclusion}}

Salary history bans continue to gain momentum in the US as states and local governments strive to reduce gender pay gaps. 
By restricting employers from asking job applicants their pay history, these policies intend to reduce history dependence in pay setting and narrow the gender pay gap in the process. 
However, since job applicants are allowed to voluntarily disclose their salary, there are plausible concerns about the success of these policies.   

Using data from the Current Population Survey and a difference-in-differences design, I show that SHBs reduced gender gap in both hourly wages and weekly earnings by 2 p.p.. 
This is driven almost entirely by an increase in female earnings and not because of any adverse effects on male earnings. 
Consistent with motivations, the bans were also successful in weakening the link between past and present earnings for those who changed jobs. 
Therefore, SHBs have the potential to prevent wage growth depression over time, particularly for workers who might have experienced wage losses uncorrelated with productivity, earlier in their careers. 

Does the success of salary history bans then suggest that job applicants were more likely to withhold pay information after SHB? 
Furthermore, does the differential effect on women imply that disclosure rates changed differently by gender?
Using individual level data from PayScale, I show that this is indeed the case. 
After the ban, both men and women became less likely to share salary information with prospective employers and by 2 p.p. more among women. 
Moreover, applicants who were asked about their pay history before the ban were over 60 p.p. more likely to comply than those who were not asked. 
These two empirical observations suggest that it is really the prospective employer's prompt for information which might have induced most job applicants to reveal pay history before SHB and when SHBs restricted such nudges, applicants became less likely to volunteer information. 
Furthermore, pre-SHB disclosure rates among women were higher than among men by 2 p.p., both because women were more likely to be asked about their salary history, but also because women were more likely than men to comply with information requests. 

Why do so many applicants either not volunteer information or refuse to disclose when asked? 
To better understand these motivations, I conduct a nationwide survey where I ask job applicants why they would or would not share salary information. 
Overall, evidence from this survey suggests that while candidates are not necessarily put off by salary history questions, most respondents would rather that their salary offer not depend on their previous pay, but on their skills and experiences. 
Over 50\% of respondents stated that they would not disclose salary history because it helps them negotiate better. 
Around 30\% of men and 40\% of women also stated that they would not disclose because they feel uncomfortable discussing salary. 

Therefore, it is evident that behavioral motivations are equally important as financial incentives for applicants' decisions to share information. 
But how do we rationalize this evidence against the backdrop of statistical discrimination which might induce job applicants to share, lest withholding sends a negative signal to the prospective employer about current pay and therefore productivity? 
I argue that while statistical discrimination is consistent with some workers withholding salary history, by itself it fails to explain why disclosure rates are higher among those who are asked and therefore, the effects of the policy. 

To bridge this gap, I build a theoretical framework of salary negotiation where the prospective employer does not have access to the applicant's pay history and the candidate can choose whether to reveal information. 
I bring in two types of utility cost - privacy cost of having to disclose salary history which is gender-invariant, and non-conformity cost of refusing information when asked, which is higher for women. 
Privacy cost allows me to sustain an equilibrium where all workers withhold salary history if the proportion of previously low-earning workers is low, and high-earners begin to disclose information as the initial wage distribution becomes more right-skewed. 
I show that for appropriate initial wage distributions, the presence of non-conformity costs can capture the reduction in disclosure rates by enquiry status as well as between pre and post-SHB cases. 
Finally, I show that a more right skewed initial wage distribution for women in comparison to men, combined with higher non-conformity costs among women can capture pre-SHB disclosure differences between men and women, as well as reduction in disclosure rates and gender pay gap after SHB.

\newpage
\printbibliography

\newpage
\input{Tables_and_Figures}

\newpage 
\appendix 
\input{Appendix}

\end{document}

%% file: Title_Page_and_Author_Details
\author{Sourav Sinha\footnote{ Yale University, sourav.sinha@yale.edu} }
\title{US Salary History Bans\\
Strategic Disclosure by Job Applicants and the Gender Pay Gap\footnote{The first version of this paper was circulated \href{https://ssrn.com/abstract=3458194}{\underline{here}} on October 2, 2019 under the title "Salary History Ban: Gender Pay Gap and Spillover Effects".
I thank Joseph Altonji, Costas Meghir, Cormac O'Dea, Johanna Rickne, John Eric Humphries, and participants at the Yale labor/public finance workshops and Yale ISPS Policy Fellows workshops, for their helpful comments. 
Research funding from the Yale Institute of Social and Policy Studies, the Cowles Foundation, and the Department of Economics at Yale, is gratefully acknowledged. 
I thank PayScale for providing access to their confidential survey data, and Pamela O'Donnell and Dorothy Ovelar for their administrative support.}
}

\maketitle

%% file: Abstract
\begin{abstract} 

I study the effects of US salary history bans which restrict employers from inquiring about job applicants' pay history during the hiring process, but allow candidates to voluntarily share information. 
Using a difference-in-differences design, I show that these policies narrowed the gender pay gap significantly by 2 p.p., driven almost entirely by an increase in female earnings. 
The bans were also successful in weakening the auto-correlation between current and future earnings, especially among job-changers. 
I provide novel evidence\footnote{\href{https://payscale.com}{\includegraphics[height=5.0mm]{payscale_logo_black.png}}, Last Updated: August, 2019.} showing that when employers could no longer nudge candidates for information, the likelihood of voluntarily disclosing salary history decreased among job applicants and by 2 p.p. more among women.  
I then develop a salary negotiation model with asymmetric information, where I allow job applicants to choose whether to reveal pay history, and use this framework to explain my empirical findings on disclosure behavior and gender pay gap. 
\end{abstract}
\addcontentsline{toc}{chapter}{Abstract} 
Journal Classifications: J08, J31, J38, J78 \\
Keywords: Salary History Ban, Gender Pay Gap, Asymmetric Information, Information Disclosure 

%% file: Tables_and_Figures
\section*{Tables and Figures \label{TablesFigures}}
\addcontentsline{toc}{chapter}{Tables and Figures}

\begin{figure}[H]
\centering
\caption{\small{US states with state-wide salary history bans} \label{Fig01}}
\begin{minipage}{12cm}
\emph{\footnotesize{\newline\newline The states in red have state-wide bans for state employers alone, while those in blue have state-wide bans for all employers, as of December 2019. For more details on effective dates and other local (city and county) bans, see Table $\ref{Tab02}$.}}
\end{minipage}
\includegraphics[width=13cm, height=8.5cm]{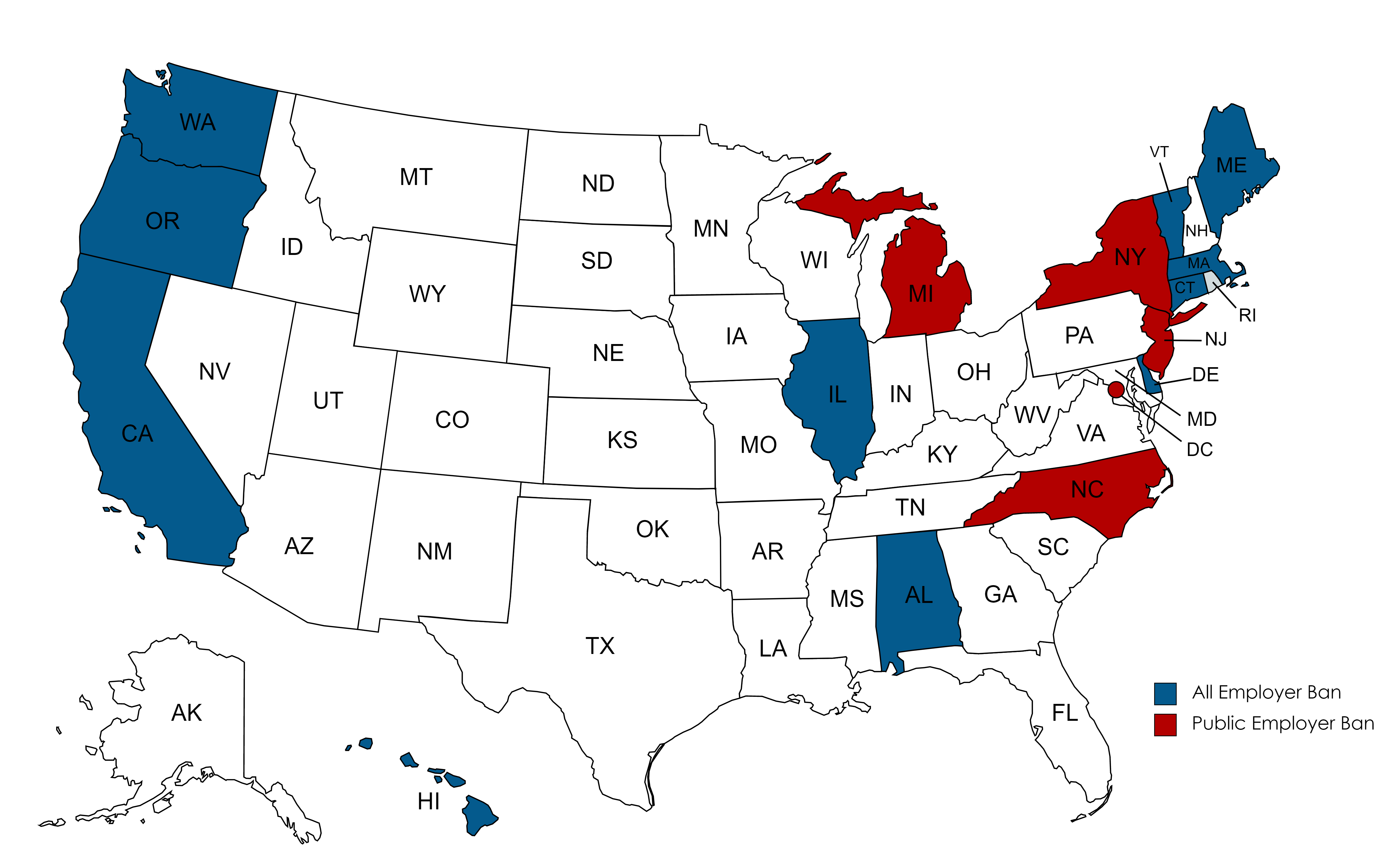}
\end{figure}

\begin{figure}[H]
\centering
\caption{\small{Fraction of US Adult Population (22-64 years) Exposed to Salary History Bans} \label{Fig02}}
\begin{minipage}{15cm}
\emph{\footnotesize{\newline The figure below shows the proportion of US employed workers in the age group of 22-64 years, who are subject to salary history bans. Data Source: Current Population Survey. }}
\end{minipage}
\includegraphics[width=11cm, height=7.5cm]{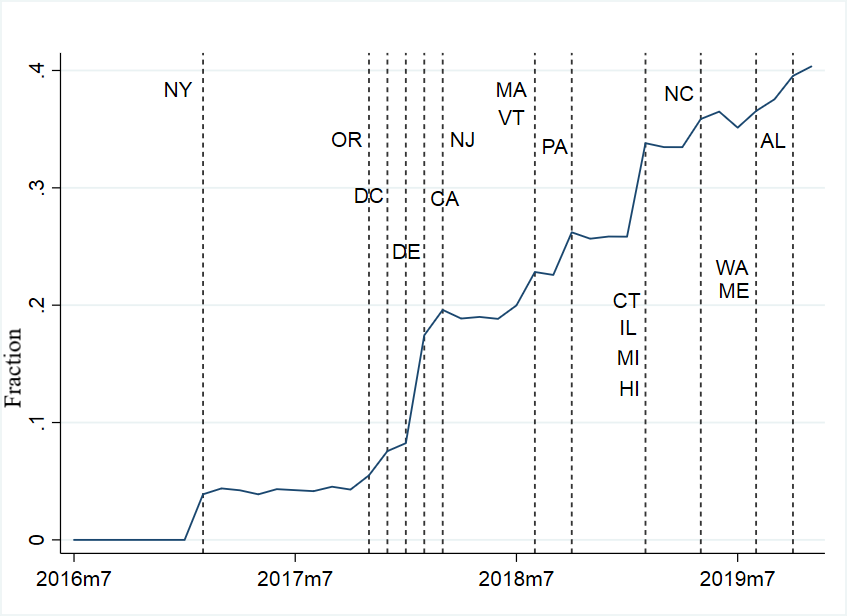}
\end{figure}

\begin{figure}[H]
\centering
\caption{\small{Pre-Trends in Hourly Wages and Weekly Earnings by Gender} \label{Fig05}}
\includegraphics[width=16cm, height=9cm]{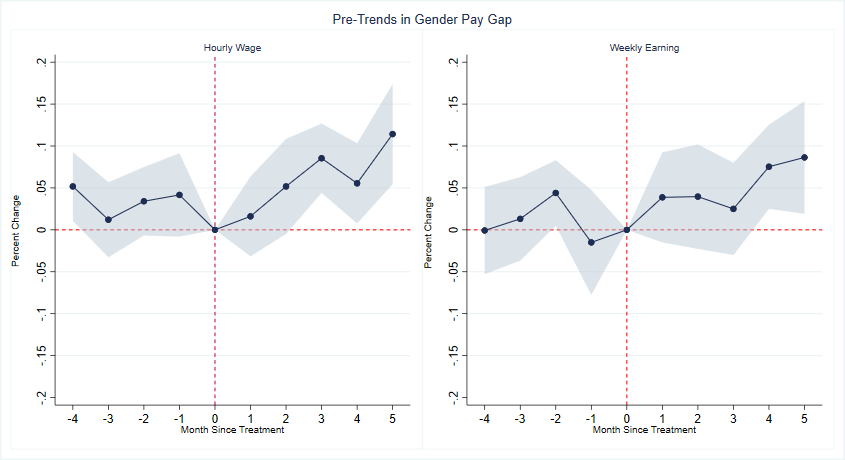}
\end{figure}

\begin{sidewaystable}
\begin{table}[H]
\centering
\scriptsize
\caption{\small{Timeline and Legislation Details of US Salary History Bans (updated as of December, 2021)}\label{Tab02}}
\begin{minipage}{24cm}
\emph{\newline\newline The table below lists states, counties, and cities which have some version of salary history ban in place. It includes details on implementation, coverage, and provisions.\\ \\ 
-- ``$\checkmark$'' implies that the action is allowed under the law. \\
-- ``X'' implies that the action is prohibited under the law. \\ 
-- ``U'' implies that the law is either silent or ambiguous on these details. \\ \\ 
Information is drawn from the following sources -- 
Running List:[\href{https://www.hrdive.com/news/salary-history-ban-states-list/516662/}{\underline{1}}], 
DC:[\href{https://dchr.dc.gov/sites/default/files/dc/sites/dchr/publication/attachments/edpm_11B_92_salary_history_instruction.pdf}{\underline{1}}], 
Delaware:[\href{https://legis.delaware.gov/json/BillDetail/GenerateHtmlDocument?legislationId=25664&legislationTypeId=6&docTypeId=2&legislationName=HS1}{\underline{1}}], 
Louisiana:[\href{https://ogletree.com/insights/new-orleans-mayor-signs-executive-order-prohibiting-wage-history-inquiries/}{\underline{1}}], 
California:[\href{https://leginfo.legislature.ca.gov/faces/billTextClient.xhtml?bill_id=201720180AB2282}{\underline{1}}, \href{https://www.hrdive.com/news/joining-a-national-trend-san-francisco-bans-salary-history-questions/447094/}{\underline{2}}], 
Kentucky:[\href{https://d12v9rtnomnebu.cloudfront.net/diveimages/Kentucky.pdf}{\underline{1}}], 
Connecticut:[\href{https://www.cga.ct.gov/2018/ACT/pa/2018PA-00008-R00HB-05386-PA.htm}{\underline{1}}], 
Hawaii:[\href{https://www.capitol.hawaii.gov/session2018/bills/SB2351_CD1_.PDF}{\underline{1}}], 
Maine:[\href{https://mainelegislature.org/legis/bills/getPDF.asp?paper=SP0090&item=3&snum=129}{\underline{1}}], 
Alabama:[\href{https://www.littler.com/publication-press/publication/alabama-enacts-pay-equity-law}{\underline{1}}], 
Maryland:[\href{http://mgaleg.maryland.gov/2020RS/bills/hb/hb0123t.pdf}{\underline{1}},\href{https://www.montgomerycountymd.gov/council/resources/files/lims/bill/2019/Enacted/pdf/7130_2600_Enacted_05102019.pdf}{\underline{2}}], 
Colorado:[\href{https://leg.colorado.gov/sites/default/files/2019a_085_signed.pdf}{\underline{1}}], 
Georgia:[\href{https://www.atlantaga.gov/Home/Components/News/News/11942/672}{\underline{1}}], 
Massachusetts[\href{https://malegislature.gov/Laws/SessionLaws/Acts/2016/Chapter177}{\underline{1}}], 
Michigan:[\href{https://www.michigan.gov/whitmer/0,9309,7-387-90499_90704-486962--,00.html}{\underline{1}}], 
Mississippi:[\href{https://apnews.com/article/8aa20d4205074ab48f2da8c0e2a1708c}{\underline{1}}], 
Missouri:[\href{https://www.documentcloud.org/documents/6789627-St-Louis-71095-Combined.html}{\underline{1}}, \href{https://d12v9rtnomnebu.cloudfront.net/diveimages/190380_Sub_1.pdf}{\underline{2}}], 
Nevada:[\href{https://www.leg.state.nv.us/App/NELIS/REL/81st2021/Bill/7896/Text}{\underline{1}}], 
New Jersey:[\href{http://nj.gov/infobank/eo/056murphy/pdf/EO-1.pdf}{\underline{1}}, \href{https://www.njsba.org/news-publications/school-leader/march-april-2020-vol-50-no-5/legally-speaking-new-jersey-bans-salary-history-inquiries/}{\underline{2}}], 
New York:[\href{https://www.governor.ny.gov/sites/governor.ny.gov/files/atoms/files/EO_161.pdf}{\underline{1}}, \href{https://legislation.nysenate.gov/pdf/bills/2019/S6549}{\underline{2}}, \href{https://legistar.council.nyc.gov/LegislationDetail.aspx?ID=2813507&GUID=938399E5-6608-42F5-9C83-9D2665D9496F&Options=ID|Text|&Search=salary+history}{\underline{3}}, \href{http://app.albanycounty.com/legislature/resolutions/2017/10/16-LL_P.pdf}{\underline{4}}, \href{https://www.scnylegislature.us/DocumentCenter/View/57524/Introductory-Resolution-1856-18-PDF?bidId=}{\underline{5}}, \href{https://d12v9rtnomnebu.cloudfront.net/diveimages/Westchester_county.pdf}{\underline{6}}], 
North Carolina:[\href{https://files.nc.gov/governor/documents/files/EO93-_Prohibiting_the_Use_of_Salary_History_in_the_State_Hiring_Process.pdf}{\underline{1}}], 
Ohio:[\href{https://d12v9rtnomnebu.cloudfront.net/diveimages/Legislation_Text.pdf}{\underline{1}}, \href{https://www.kmklaw.com/labor-employment/cincinnati-salary-history-ban-takes-effect\#:~:text=The\%20city\%20of\%20Cincinnati's\%20salary,Friday\%2C\%20March\%2013\%2C\%202020.&text=Refuse\%20to\%20hire\%20or\%20otherwise,his\%20or\%20her\%20salary\%20history.}{\underline{2}}], 
Oregon:[\href{https://www.jacksonlewis.com/publication/oregon-enacts-expansive-pay-equity-law}{\underline{1}}], 
Pennsylvania:[\href{https://phila.legistar.com/LegislationDetail.aspx?ID=2849975\&GUID=239C1DF9-8FDF-4D32-BACC-296B6EBF726C}{\underline{1}}, \href{https://pittsburgh.legistar.com/LegislationDetail.aspx?ID=2931161&GUID=E45D1721-68E5-4BEC-9989-59C275B74AA7\&Options=ID\%7cText\%7c\&Search=salary+history\&FullText=1}{\underline{2}}], 
Puerto Rico:[\href{https://www.jacksonlewis.com/publication/puerto-rico-enacts-equal-pay-law-prohibits-employers-inquiring-about-past-salary-history}{\underline{1}}], 
Illinois:[\href{https://www2.illinois.gov/Pages/news-item.aspx?ReleaseID=19609}{\underline{1}}, \href{https://www2.illinois.gov/idol/FAQs/Pages/Equal-Pay-Act-Salary-History-Ban-FAQ.aspx}{\underline{2}}, \href{https://d12v9rtnomnebu.cloudfront.net/diveimages/EXEC_ORD_NO._2018-1_gen_20180410095817.pdf}{\underline{3}}], 
Rhode Island:[\href{}{\underline{1}}], 
South Carolina:[\href{https://d12v9rtnomnebu.cloudfront.net/diveimages/Attachment-7494_3.pdf}{\underline{1}}, \href{https://d12v9rtnomnebu.cloudfront.net/diveimages/Pages_from_Richland_County_Council_Regular_Session_-_June_4_2019_1.pdf}{\underline{2}}], 
Utah:[\href{https://www.slcinfobase.com/PPAREO/\#!WordDocuments/30110genderpayequity.htm}{\underline{1}}], 
Vermont:[\href{https://www.shrm.org/resourcesandtools/legal-and-compliance/state-and-local-updates/pages/vermont-enacts-salary-history-inquiry-law.aspx}{\underline{1}}], 
Virginia:[\href{https://lis.virginia.gov/cgi-bin/legp604.exe?201+sum+HB416}{\underline{1}}], 
Washington:[\href{http://lawfilesext.leg.wa.gov/biennium/2019-20/Pdf/Bills/House\%20Passed\%20Legislature/1696-S.PL.pdf\#page=1}{\underline{1}}]\newline\newline}
\end{minipage}
\begin{tabular}{| l | p{2.9cm} | p{1.5cm} | p{1.5cm} | p{1.5cm} | p{1.5cm} | p{1.5cm} | p{1.5cm} | p{1.5cm} | p{1.5cm} | p{1.5cm} | p{1.5cm}|}
\hline & Region & Effective Date (yyyy-mm-dd) & Employer Type & Application Form Enquiry & Interview Enquiry & Voluntary Disclosure & Info search in public/private records, current/former employers & Exclude based on reporting or refusal & Ask and Verify after offer & Rely on voluntary reporting or accidental discovery\footnote{In most cases, reliance on salary history to determine offer or new salary is allowed only if the applicant has voluntarily shared information, and not when the information was either accidentally discovered while conducting background checks or available from public records.} & Exemptions for internal transfers, promotion, union bargaining \\
\hline &&&&&&&&&&& \\
\multirow{8}{*}{States} & Puerto Rico & 2017-03-08 & All & X & X & $\checkmark$ & U & X & $\checkmark$ & U & $\checkmark$ \\
& Oregon & 2017-10-06 & All & X & X & $\checkmark$ & X & X &$\checkmark$ & X & $\checkmark$ \\
& District of Columbia & 2017-11-01 & State & U & X & $\checkmark$ & U & U & $\checkmark$ & $\checkmark$ & $\checkmark$ \\
& Delaware & 2017-12-14 & All & X & X & $\checkmark$ & U & X &$\checkmark$ & U & $\checkmark$\\
& California & 2018-01-01 & All & X&X & $\checkmark$ & U &X  & $\checkmark$& $\checkmark$ & $\checkmark$ \\
& Massachusetts & 2018-07-01 & All & X& X& $\checkmark$ & X & X & $\checkmark$& U & $\checkmark$ \\
& Vermont & 2018-07-01 & All & X & X& $\checkmark$ & X & X & $\checkmark$& U & $\checkmark$\\
& Connecticut & 2019-01-01 & All & X & X& $\checkmark$ & U & X & $\checkmark$& U & $\checkmark$ \\ 
& Hawaii & 2019-01-01 & All & X & X & $\checkmark$ & X & U & $\checkmark$ & X & $\checkmark$ \\ 
& Michigan & 2019-01-15 & State & X & X & $\checkmark$ & U & X & $\checkmark$& U & $\checkmark$ \\
& North Carolina & 2019-04-02 & State & X& X& $\checkmark$ & X & X & $\checkmark$& X & $\checkmark$ \\
& Washington & 2019-07-28 & All & X & X & $\checkmark$ & X & X & $\checkmark$ & U & $\checkmark$ \\
& Maine & 2019-09-17 & All & X & X & $\checkmark$ & X & U & $\checkmark$ & U & $\checkmark$ \\
& Alabama & 2019-09-01 & All & U & X & $\checkmark$ & U & X & $\checkmark$ & U & U \\
& Illinois\footnote{Illinois had banned salary history questions for only state employers and agencies on Jan 15, 2019. This was extended to all employers in September, 2019. } & 2019-09-29 & All & X & X & $\checkmark$ & X & X & $\checkmark$& X & $\checkmark$ \\
& New Jersey\footnote{\footnotesize{New Jersey implemented SHB for only state employers and agencies on Feb 1, 2018. It was extended to cover all employers from Jan 1, 2020.}} & 2020-01-01 & All & X & X & $\checkmark$ & $\checkmark$ & X & $\checkmark$ & $\checkmark$ & $\checkmark$ \\
& New York\footnote{New York implemented a state-wide SHB for only state employers and agencies on Jan 9, 2017. This was extended to all employers on Jan 6, 2020.} & 2020-01-06 & State & X & X & $\checkmark$ & X & X & $\checkmark$ & X & $\checkmark$\\
& Maryland & 2020-10-01 & All & X & X & $\checkmark$ & X & X & $\checkmark$ & $\checkmark$ & $\checkmark$ \\ 
& Colorado & 2021-01-01 & All & X & X & $\checkmark$ & U & X & $\checkmark$ & X & U \\ 
& Nevada & 2021-10-01 & All & X & X & $\checkmark$ & X & X & $\checkmark$ & X & $\checkmark$ \\
& Rhode Island & 2022-01-01 & All & X & X & $\checkmark$ & X & X & $\checkmark$ & X & $\checkmark$ \\
\hline 
\end{tabular}
\begin{minipage}{24cm}
\emph{\footnotesize{\newline Table \ref{Tab02} continues on next page.}}
\end{minipage}
\end{table}
\end{sidewaystable}

\begin{sidewaystable}
\begin{table}[H]
\centering
\scriptsize
\caption*{\small{Table \ref{Tab02} Contd: Timeline and Legislation Details of US Salary History Bans (updated as of December, 2021)}}
\begin{minipage}{24cm}
\emph{\newline\newline The table below lists states, counties, and cities which have some version of salary history ban in place. It includes details on implementation, coverage, and provisions.\\ \\ 
-- ``$\checkmark$'' implies that the action is allowed under the law. \\
-- ``X'' implies that the action is prohibited under the law. \\ 
-- ``U'' implies that the law is either silent or ambiguous on these details. \\ \\ 
Information is drawn from the following sources -- 
Running List:[\href{https://www.hrdive.com/news/salary-history-ban-states-list/516662/}{\underline{1}}], DC:[\href{https://dchr.dc.gov/sites/default/files/dc/sites/dchr/publication/attachments/edpm_11B_92_salary_history_instruction.pdf}{\underline{1}}], Delaware:[\href{https://legis.delaware.gov/json/BillDetail/GenerateHtmlDocument?legislationId=25664&legislationTypeId=6&docTypeId=2&legislationName=HS1}{\underline{1}}], Louisiana:[\href{https://ogletree.com/insights/new-orleans-mayor-signs-executive-order-prohibiting-wage-history-inquiries/}{\underline{1}}], California:[\href{https://leginfo.legislature.ca.gov/faces/billTextClient.xhtml?bill_id=201720180AB2282}{\underline{1}}, \href{https://www.hrdive.com/news/joining-a-national-trend-san-francisco-bans-salary-history-questions/447094/}{\underline{2}}], Kentucky:[\href{https://d12v9rtnomnebu.cloudfront.net/diveimages/Kentucky.pdf}{\underline{1}}], Connecticut:[\href{https://www.cga.ct.gov/2018/ACT/pa/2018PA-00008-R00HB-05386-PA.htm}{\underline{1}}], Hawaii:[\href{https://www.capitol.hawaii.gov/session2018/bills/SB2351_CD1_.PDF}{\underline{1}}], Maine:[\href{https://mainelegislature.org/legis/bills/getPDF.asp?paper=SP0090&item=3&snum=129}{\underline{1}}], Alabama:[\href{https://www.littler.com/publication-press/publication/alabama-enacts-pay-equity-law}{\underline{1}}], Maryland:[\href{http://mgaleg.maryland.gov/2020RS/bills/hb/hb0123t.pdf}{\underline{1}},\href{https://www.montgomerycountymd.gov/council/resources/files/lims/bill/2019/Enacted/pdf/7130_2600_Enacted_05102019.pdf}{\underline{2}}], Colorado:[\href{https://leg.colorado.gov/sites/default/files/2019a_085_signed.pdf}{\underline{1}}], Georgia:[\href{https://www.atlantaga.gov/Home/Components/News/News/11942/672}{\underline{1}}], Massachusetts[\href{https://malegislature.gov/Laws/SessionLaws/Acts/2016/Chapter177}{\underline{1}}], Michigan:[\href{https://www.michigan.gov/whitmer/0,9309,7-387-90499_90704-486962--,00.html}{\underline{1}}], Mississippi:[\href{https://apnews.com/article/8aa20d4205074ab48f2da8c0e2a1708c}{\underline{1}}], Missouri:[\href{https://www.documentcloud.org/documents/6789627-St-Louis-71095-Combined.html}{\underline{1}}, \href{https://d12v9rtnomnebu.cloudfront.net/diveimages/190380_Sub_1.pdf}{\underline{2}}], Nevada:[\href{https://www.leg.state.nv.us/App/NELIS/REL/81st2021/Bill/7896/Text}{\underline{1}}], New Jersey:[\href{http://nj.gov/infobank/eo/056murphy/pdf/EO-1.pdf}{\underline{1}}, \href{https://www.njsba.org/news-publications/school-leader/march-april-2020-vol-50-no-5/legally-speaking-new-jersey-bans-salary-history-inquiries/}{\underline{2}}], New York:[\href{https://www.governor.ny.gov/sites/governor.ny.gov/files/atoms/files/EO_161.pdf}{\underline{1}}, \href{https://legislation.nysenate.gov/pdf/bills/2019/S6549}{\underline{2}}, \href{https://legistar.council.nyc.gov/LegislationDetail.aspx?ID=2813507&GUID=938399E5-6608-42F5-9C83-9D2665D9496F&Options=ID|Text|&Search=salary+history}{\underline{3}}, \href{http://app.albanycounty.com/legislature/resolutions/2017/10/16-LL_P.pdf}{\underline{4}}, \href{https://www.scnylegislature.us/DocumentCenter/View/57524/Introductory-Resolution-1856-18-PDF?bidId=}{\underline{5}}, \href{https://d12v9rtnomnebu.cloudfront.net/diveimages/Westchester_county.pdf}{\underline{6}}], North Carolina:[\href{https://files.nc.gov/governor/documents/files/EO93-_Prohibiting_the_Use_of_Salary_History_in_the_State_Hiring_Process.pdf}{\underline{1}}], Ohio:[\href{https://d12v9rtnomnebu.cloudfront.net/diveimages/Legislation_Text.pdf}{\underline{1}}, \href{https://www.kmklaw.com/labor-employment/cincinnati-salary-history-ban-takes-effect\#:~:text=The\%20city\%20of\%20Cincinnati's\%20salary,Friday\%2C\%20March\%2013\%2C\%202020.&text=Refuse\%20to\%20hire\%20or\%20otherwise,his\%20or\%20her\%20salary\%20history.}{\underline{2}}], Oregon:[\href{https://www.jacksonlewis.com/publication/oregon-enacts-expansive-pay-equity-law}{\underline{1}}], Pennsylvania:[\href{https://phila.legistar.com/LegislationDetail.aspx?ID=2849975\&GUID=239C1DF9-8FDF-4D32-BACC-296B6EBF726C}{\underline{1}}, \href{https://pittsburgh.legistar.com/LegislationDetail.aspx?ID=2931161&GUID=E45D1721-68E5-4BEC-9989-59C275B74AA7\&Options=ID\%7cText\%7c\&Search=salary+history\&FullText=1}{\underline{2}}], Puerto Rico:[\href{https://www.jacksonlewis.com/publication/puerto-rico-enacts-equal-pay-law-prohibits-employers-inquiring-about-past-salary-history}{\underline{1}}], Illinois:[\href{https://www2.illinois.gov/Pages/news-item.aspx?ReleaseID=19609}{\underline{1}}, \href{https://www2.illinois.gov/idol/FAQs/Pages/Equal-Pay-Act-Salary-History-Ban-FAQ.aspx}{\underline{2}}, \href{https://d12v9rtnomnebu.cloudfront.net/diveimages/EXEC_ORD_NO._2018-1_gen_20180410095817.pdf}{\underline{3}}], Rhode Island:[\href{}{\underline{1}}], South Carolina:[\href{https://d12v9rtnomnebu.cloudfront.net/diveimages/Attachment-7494_3.pdf}{\underline{1}}, \href{https://d12v9rtnomnebu.cloudfront.net/diveimages/Pages_from_Richland_County_Council_Regular_Session_-_June_4_2019_1.pdf}{\underline{2}}], Utah:[\href{https://www.slcinfobase.com/PPAREO/\#!WordDocuments/30110genderpayequity.htm}{\underline{1}}], Vermont:[\href{https://www.shrm.org/resourcesandtools/legal-and-compliance/state-and-local-updates/pages/vermont-enacts-salary-history-inquiry-law.aspx}{\underline{1}}], Virginia:[\href{https://lis.virginia.gov/cgi-bin/legp604.exe?201+sum+HB416}{\underline{1}}], Washington:[\href{http://lawfilesext.leg.wa.gov/biennium/2019-20/Pdf/Bills/House\%20Passed\%20Legislature/1696-S.PL.pdf\#page=1}{\underline{1}}]\newline\newline}
\end{minipage}
\begin{tabular}{| l | p{2.9cm} | p{1.5cm} | p{1.5cm} | p{1.5cm} | p{1.5cm} | p{1.5cm} | p{1.5cm} | p{1.5cm} | p{1.5cm} | p{1.5cm} | p{1.5cm}|}
\hline & Region & Effective Date (yyyy-mm-dd) & Employer Type & Application Form Enquiry & Interview Enquiry & Voluntary Disclosure & Info search in public/private records, current/former employers & Exclude based on reporting or refusal & Ask and Verify after offer & Rely on voluntary reporting or accidental discovery\footnote{In most cases, reliance on salary history to determine offer or new salary is allowed only if the applicant has voluntarily shared information, and not when the information was either accidentally discovered while conducting background checks or available from public records.} & Exemptions for internal transfers, promotion, union bargaining \\
\hline &&&&&&&&&&& \\
\multirow{4}{*}{Counties} & Albany (NY) & 2017-12-17 & All & X& X & $\checkmark$ & U & X &$\checkmark$ & U & $\checkmark$\\
& Louisville (KY)  & 2018-05-17 & Metro & X& X & $\checkmark$ & X & X & $\checkmark$& X & $\checkmark$\\
& Westchester (NY) & 2018-07-09 & All & X& X & $\checkmark$ & X & X &$\checkmark$ & U &$\checkmark$ \\
& Richmond (SC) & 2019-05-23 & County & X & X & $\checkmark$ & X & U & $\checkmark$ & U & $\checkmark$ \\
& Montgomery (MD) & 2019-08-14 & County & X & X & $\checkmark$ & X & X & $\checkmark$ & $\checkmark$ & $\checkmark$ \\
& Suffolk (NY) & 2019-06-30 & All & X& X & $\checkmark$ & X & U &$\checkmark$ & X &$\checkmark$ \\
& & & & & & & & & & & \\
\hline &&&&&&&&&&& \\
\multirow{6}{*}{Cities} & New Orleans (LA) & 2017-01-25& City & X& X & $\checkmark$ & U & X &$\checkmark$ & U &$\checkmark$ \\
& Pittsburgh (PA) & 2017-01-30 & City & X & X & $\checkmark$ & X & X &$\checkmark$ & U &$\checkmark$ \\
& New York City (NY) & 2017-10-31 & All & X & X & $\checkmark$ & X & U &$\checkmark$ & X & $\checkmark$\\
& Chicago (IL) & 2018-04-10 & City & X & X&  $\checkmark$ & X & X &$\checkmark$ & U &$\checkmark$ \\
& San Francisco (CA) & 2018-07-01 & All & X & X & $\checkmark$ & U & X &$\checkmark$ & X &$\checkmark$ \\
& Salt Lake (UT) & 2018-03-01 & City & X & X & $\checkmark$ & X & X &$\checkmark$ & X & $\checkmark$\\
& Atlanta (GA) & 2019-02-18 & City & X & X & $\checkmark$ & U & X &$\checkmark$ & U & $\checkmark$\\
& Jackson (MS) & 2019-06-13 & City & X & X & $\checkmark$ & X & U & $\checkmark$ & U & $\checkmark$ \\
& Columbia (SC) & 2019-08-06 & City & X & X & $\checkmark$ & X & X & $\checkmark$ & X & $\checkmark$ \\ 
& Kansas (MO) & 2019-10-31 & All & X & X & $\checkmark$ & X & X &$\checkmark$ & X & $\checkmark$\\
& St. Louis (MO) & 2020-03-11 & City & X & X & $\checkmark$ & X & X & $\checkmark$ & X & $\checkmark$ \\
& Cincinnati (OH) & 2020-03-01 & All & X & X & $\checkmark$ & X & X & $\checkmark$ & X & $\checkmark$ \\
& Toledo (OH) & 2020-06-25 & All & X & X & $\checkmark$ & X & X & $\checkmark$ & U & $\checkmark$ \\
& Philadelphia (PA) & 2020-09-01 & All & X & X & $\checkmark$ & X & X & $\checkmark$ & U & $\checkmark$ \\
& & & & & & & & & & & \\
\hline
\end{tabular}
\end{table}
\end{sidewaystable}

\newpage
\begin{table}[H]
\centering
\scriptsize
\caption{\small{Gender Pay Gap in Pre-Salary History Ban Period \label{Tab02a}}}
\begin{minipage}{17cm}
\emph{\newline \newline\footnotesize{The table below shows conditional gender gap in $log$(hourly wage) [Row 1] and $log$(weekly earnings) [Row 2] for all states pooled together across years in pre-ban periods. The results are regression coefficient on the ``\textnormal{Female}'' dummy and so a negative value implies that women earn less than men. All columns control for year and state fixed effects. All specifications with $log$(weekly earnings) as outcome control for an hourly paid indicator. Except for the one in column (1) all specifications with $log$(weekly earnings) as outcome also control for number of hours worked interacted with the Full/Part Time dummy. Data Source: Current Population Survey.}}
\end{minipage}
\begin{tabular}{l >{\centering\arraybackslash} m{1.3cm} >{\centering\arraybackslash} m{1.3cm} >{\centering\arraybackslash} m{1.3cm} >{\centering\arraybackslash} m{1.3cm} >{\centering\arraybackslash} m{1.3cm} >{\centering\arraybackslash} m{1.3cm} >{\centering\arraybackslash} m{1.3cm} >{\centering\arraybackslash} m{1.3cm} >{\centering\arraybackslash} m{1.3cm}} \\
\hline \hline 
& (1) & (2) & (3) & (4)  & (5) & (6) & (7) & (8) \\
\hline \\ \\
log(hourly wage)  & -0.136$^{***}$&      -0.105$^{***}$&      -0.103$^{***}$&      -0.144$^{***}$&      -0.156$^{***}$&      -0.157$^{***}$&      -0.135$^{***}$&      -0.115$^{***}$ \\
N$=$652,987          &     (0.003)         &     (0.003)         &     (0.003)         &     (0.002)         &     (0.002)         &     (0.002)         &     (0.002)         &     (0.002)         \\ \\
log(weekly earnings) & -0.263$^{***}$&      -0.115$^{***}$&      -0.112$^{***}$&      -0.154$^{***}$&      -0.164$^{***}$&      -0.165$^{***}$&      -0.143$^{***}$&      -0.132$^{***}$ \\  
N$=$1,097,396 &     (0.003)         &     (0.003)         &     (0.002)         &     (0.002)         &     (0.002)         &     (0.002)         &     (0.002)         &     (0.002)   \\ \\
\hline 
Full/Part Time & & $\checkmark$ & $\checkmark$ & $\checkmark$ & $\checkmark$ & $\checkmark$ & $\checkmark$ & $\checkmark$ \\
Race & & & $\checkmark$ & $\checkmark$ & $\checkmark$ & $\checkmark$ & $\checkmark$ & $\checkmark$ \\
Education & & & & $\checkmark$ & $\checkmark$ & $\checkmark$ & $\checkmark$ & $\checkmark$ \\
f(Age) & & & & & $\checkmark$ & $\checkmark$ & $\checkmark$ & $\checkmark$  \\
Sector & & & & & & $\checkmark$ & $\checkmark$ & $\checkmark$  \\
Industry & & & & & &  & $\checkmark$ & $\checkmark$  \\
Occupation & & & & & & &  & $\checkmark$ \\
\hline \hline \\
\multicolumn{9}{l}{* p$<$0.10 ** p$<$0.05 *** p$<$0.01} \\
\multicolumn{9}{l}{Standard errors in parentheses}
\end{tabular}
\end{table}

\newpage

\begin{table}[H]
\centering
\scriptsize
\caption{\small{Effects of Salary History Ban on Log(Hourly Wage)} \label{Tab03new}}
\begin{minipage}{15.5cm}
\emph{\footnotesize{\newline\newline The table below shows the effects of salary history ban on hourly wages of hourly-paid workers for the `\textbf{AllStateBan}' sample. Our coefficient of interest is ``\textnormal{Treat}''. Data Source: Current Population Survey.}}
\end{minipage}
\begin{tabular}{l >{\centering\arraybackslash} m{1.5cm}  >{\centering\arraybackslash} m{1.5cm}  >{\centering\arraybackslash} m{1.5cm}  >{\centering\arraybackslash} m{1.5cm}  >{\centering\arraybackslash} m{1.5cm}  >{\centering\arraybackslash} m{1.5cm}  >{\centering\arraybackslash} m{1.5cm}}\\
\hline \hline 
& (1) & (2) & (3) & (4) & (5) & (6) & (7) \\
\hline \\
Treat & 0.013$^{*}$ & 0.008 & 0.009 & 0.009 & 0.011$^{*}$ & 0.011$^{*}$ & 0.011$^{*}$ \\
& (0.006) & (0.006) & (0.005) & (0.005) & (0.005) & (0.005) & (0.005) \\ \\
Female &  -0.134$^{***}$ & -0.156$^{***}$ & -0.135$^{***}$& -0.115$^{***}$  & -0.115$^{***}$ & -0.115$^{***}$ & -0.108$^{***}$ \\
& (0.003) & (0.002) & (0.002) & (0.002) & (0.002) & (0.002) & (0.003) \\ \\
Cons & 2.733$^{***}$ & 0.518$^{***}$ & 0.810$^{***}$ & 1.053$^{***}$ & 1.034$^{***}$ & 1.019$^{***}$ & 1.015$^{***}$ \\
& (0.005) & (0.027) & (0.026) & (0.024) & (0.024) & (0.026) & (0.026) \\ \\
\hline 
R$^{2}$           &       0.051         &       0.245         &       0.323         &       0.406         &       0.406         &       0.406         &       0.406         \\ 
Controls  & & $\checkmark$ & $\checkmark$ & $\checkmark$ & $\checkmark$ & $\checkmark$ & $\checkmark$ \\
Industry & &  & $\checkmark$ & $\checkmark$ & $\checkmark$ & $\checkmark$ & $\checkmark$ \\
Occupation & & &  & $\checkmark$ & $\checkmark$ & $\checkmark$ & $\checkmark$ \\
$\lambda_s f(t)$ & & & & & $\checkmark$ & $\checkmark$ & $\checkmark$ \\
$\lambda_I f(t)$ & & & & & & $\checkmark$ & $\checkmark$ \\
$\lambda_f f(t)$ & & & & & & & $\checkmark$\\
N        &      675287         &      675287         &     675287      &     675287     &   675287      &     675287      &    675287       \\
\hline \hline\\
\multicolumn{8}{l}{\footnotesize{* p$<$0.10 ** p$<$0.05 *** p$<$0.01}}\\
\multicolumn{8}{l}{\footnotesize{Standard errors in parentheses}}
\end{tabular}
\end{table}
\medskip
\begin{table}[H]
\centering
\scriptsize
\caption{\small{Effects of Salary History Ban on Log(Weekly Earnings)} \label{Tab04new}}
\begin{minipage}{15.5cm}
\emph{\footnotesize{\newline\newline The table below shows the effects of salary history ban on weekly earnings for the `\textbf{AllStateBan}' sample. Our coefficient of interest is ``\textnormal{Treat}''. Data Source: Current Population Survey.}}
\end{minipage}
\begin{tabular}{l >{\centering\arraybackslash} m{1.5cm}  >{\centering\arraybackslash} m{1.5cm}  >{\centering\arraybackslash} m{1.5cm}  >{\centering\arraybackslash} m{1.5cm}  >{\centering\arraybackslash} m{1.5cm}  >{\centering\arraybackslash} m{1.5cm}  >{\centering\arraybackslash} m{1.5cm}}\\
\hline \hline 
& (1) & (2) & (3) & (4) & (5) & (6) & (7) \\
\hline \\
Treat            &           0.010    &       0.001  &         0.001  &         0.002    &       0.006    &       0.006   &        0.006    \\
                     &     (0.007)       &  (0.005)  &       (0.004)&         (0.005)  &       (0.005)   &      (0.005)     &    (0.005)    \\ \\
Female             &        -0.283$^{***}$    &   -0.164$^{***}$ & -0.143$^{***}$  &   -0.131$^{***}$   &  -0.131$^{***}$  &    -0.131$^{***}$    &   -0.131$^{***}$ \\
                       &   (0.003)      &   (0.002)    &     (0.002)  &       (0.002)    &     (0.002)     &    (0.002)      &   (0.004)   \\ \\
Constant            &        6.563$^{***}$      & 2.329$^{***}$    &   2.533$^{***}$ &     2.765$^{***}$  &     2.747$^{***}$&       2.729$^{***}$  &    2.730$^{***}$\\
                         & (0.005)       &  (0.031)      &   (0.033)    &     (0.032)      &   (0.031)       &  (0.032)       &  (0.033)     \\ \\
\hline 
R$^{2}$      &     0.052    &       0.484    &       0.503      &     0.531    &       0.531  &         0.531  &         0.531             \\
Controls  & & $\checkmark$ & $\checkmark$ & $\checkmark$ & $\checkmark$ & $\checkmark$ & $\checkmark$ \\
Industry & &  & $\checkmark$ & $\checkmark$ & $\checkmark$ & $\checkmark$ & $\checkmark$ \\
Occupation & & &  & $\checkmark$ & $\checkmark$ & $\checkmark$ & $\checkmark$ \\
$\lambda_s f(t)$ & & & & & $\checkmark$ & $\checkmark$ & $\checkmark$ \\
$\lambda_I f(t)$ & & & & & & $\checkmark$ & $\checkmark$ \\
$\lambda_f f(t)$ & & & & & & & $\checkmark$\\
N        &   1134788     &    1075187     &    1075187   &      1075187      &   1075187   &      1075187    &     1075187   \\
\hline \hline\\
\multicolumn{8}{l}{\footnotesize{* p$<$0.10 ** p$<$0.05 *** p$<$0.01}}\\
\multicolumn{8}{l}{\footnotesize{Standard errors in parentheses}}
\end{tabular}
\end{table}

\newpage

\begin{table}[H]
\centering
\scriptsize
\caption{\small{Baseline Specification I: Effects of Salary History Ban on Gender Gap in Log(Hourly Wage)} \label{Tab03}}
\begin{minipage}{16.8cm}
\emph{\footnotesize{\newline\newline The table below shows the effects of salary history ban on gender gap in $log$(hourly wage) for hourly-paid workers using my baseline specification ($\ref{Eq01}$) and my main sample `AllStateBan'. The coefficient on ``\textnormal{Treat}''  denotes the effect on men, and that on ``\textnormal{TreatXFemale}'' is the effect on gender pay gap. The coefficient on ``\textnormal{Female}'' is the baseline gender pay gap. ``\textnormal{FemaleXEverTreat}'' shows the baseline difference in gender pay gap between states which have salary history bans and other that do not. Data Source: Current Population Survey.}}
\end{minipage}
\begin{tabular}{l >{\centering\arraybackslash} m{1.5cm}  >{\centering\arraybackslash} m{1.5cm}  >{\centering\arraybackslash} m{1.5cm}  >{\centering\arraybackslash} m{1.5cm}  >{\centering\arraybackslash} m{1.5cm}  >{\centering\arraybackslash} m{1.5cm}  >{\centering\arraybackslash} m{1.5cm}}\\
\hline \hline 
& (1) & (2) & (3) & (4) & (5) & (6) & (7) \\
\hline \\
TreatXFemale      &          0.007       &    0.007   &        0.008   &        0.009   &        0.010    &       0.014$^{**}$     &      0.020$^{**}$   \\
                     &     (0.008)    &     (0.008)    &     (0.006)   &      (0.007) &        (0.007)   &      (0.007)  &       (0.008)   \\ \\
Treat                &       0.010    &       0.005   &        0.005   &        0.004   &        0.005    &       0.004    &       0.001   \\
                         & (0.007)  &    (0.005) &     (0.006) &     (0.006) &     (0.006) &     (0.006)  &    (0.006)   \\ \\
Female                  &   -0.145$^{***}$ &      -0.167$^{***}$ &      -0.142$^{***}$ &     -0.123$^{***}$ &     -0.123$^{***}$ &     -0.123$^{***}$ &  -0.115$^{***}$  \\
                          &(0.006)  &    (0.006)   &   (0.005)    &  (0.005)     & (0.005)  &    (0.005)  &    (0.005)    \\ \\
FemaleXEverTreat    &            0.040$^{**}$ &     0.039$^{***}$ &     0.026$^{**}$ &     0.028$^{**}$ &       0.028$^{**}$ &      0.027$^{**}$ &     0.026$^{*}$  \\
                     &     (0.016)   &   (0.013)   &   (0.012)   &   (0.013)  &    (0.013)    &  (0.013)   &   (0.013)     \\ \\
Constant           &         2.718$^{***}$ &      0.502$^{***}$ &    0.798$^{***}$ &   1.042$^{***}$ &     1.022$^{***}$ &    1.007$^{***}$ &   1.003$^{***}$  \\
                         & (0.007)  &    (0.039)    &  (0.047) &     (0.040)   &   (0.040)    &  (0.041)  &    (0.041)   \\ \\
\hline 
R$^{2}$           &       0.051     &      0.245     &      0.323      &     0.406      &     0.406    &       0.406       &    0.406          \\ 
Controls  & & $\checkmark$ & $\checkmark$ & $\checkmark$ & $\checkmark$ & $\checkmark$ & $\checkmark$ \\
Industry & &  & $\checkmark$ & $\checkmark$ & $\checkmark$ & $\checkmark$ & $\checkmark$ \\
Occupation & & &  & $\checkmark$ & $\checkmark$ & $\checkmark$ & $\checkmark$ \\
$\lambda_s f(t)$ & & & & & $\checkmark$ & $\checkmark$ & $\checkmark$ \\
$\lambda_I f(t)$ & & & & & & $\checkmark$ & $\checkmark$ \\
$\lambda_f f(t)$ & & & & & & & $\checkmark$\\
N        &      675287  &     675287       &    675287   &      675287      &     675287       &    675287    &      675287     \\
\hline \hline\\
\multicolumn{8}{l}{\footnotesize{* p$<$0.10 ** p$<$0.05 *** p$<$0.01}}\\
\multicolumn{8}{l}{\footnotesize{Standard errors in parentheses}}
\end{tabular}
\end{table}
\medskip
\begin{table}[H]
\centering
\scriptsize
\caption{\small{Baseline Specification II: Effects of Salary History Ban on Gender Gap in Log(Weekly Earnings)} \label{Tab04}}
\begin{minipage}{16.8cm}
\emph{\footnotesize{\newline\newline The table below shows the effects of salary history ban on gender gap in $log$(weekly earnings) using my baseline specification ($\ref{Eq01}$) and my main sample `AllStateBan'. The coefficient on ``\textnormal{Treat}''  denotes the effect on men, and that on ``\textnormal{TreatXFemale}'' is the effect on gender pay gap. The coefficient on ``\textnormal{Female}'' is the baseline gender pay gap. ``\textnormal{FemaleXEverTreat}'' shows the baseline difference in gender pay gap between states which have salary history bans and other that do not. Data Source: Current Population Survey. }}
\end{minipage}
\begin{tabular}{l >{\centering\arraybackslash} m{1.5cm}  >{\centering\arraybackslash} m{1.5cm}  >{\centering\arraybackslash} m{1.5cm}  >{\centering\arraybackslash} m{1.5cm}  >{\centering\arraybackslash} m{1.5cm}  >{\centering\arraybackslash} m{1.5cm}  >{\centering\arraybackslash} m{1.5cm}} \\
\hline \hline
& (1) & (2) & (3) & (4) & (5) & (6) & (7) \\
\hline \\
FemaleXTreat  &   0.042$^{***}$  &     0.014$^{*}$  &       0.014$^{*}$  &       0.014$^{**}$  &      0.015$^{**}$  &      0.018$^{**}$  &      0.019$^{**}$   \\ 
         &    (0.011)&      (0.008) &     (0.007)  &    (0.007) &     (0.007)  &    (0.007) &     (0.008)   \\ \\
Treat   & -0.011  &     -0.006  &     -0.006 &      -0.005  &     -0.002  &     -0.003   &    -0.004   \\
           &  (0.009)  &    (0.005)   &   (0.005)   &   (0.005)  &    (0.004) &     (0.004)  &    (0.004)   \\ \\
Female    &     -0.294$^{***}$  &   -0.172$^{***}$  &    -0.150$^{***}$  &  -0.138$^{***}$  &  -0.138$^{***}$  &   -0.138$^{***}$  &  -0.137$^{***}$   \\
            & (0.009)    &  (0.006)   &   (0.004)    &  (0.004) &     (0.004)  &    (0.004)   &   (0.005)   \\ \\
FemaleXEverTreat  &   0.035$^{*}$  &      0.031$^{**}$  &     0.024$^{**}$  &   0.026$^{**}$  & 0.026$^{**}$  &  0.026$^{**}$  &    0.026$^{**}$  \\ 
        &     (0.020)    &  (0.013)    &  (0.011)    &  (0.012)  &    (0.012)   &   (0.012)   &   (0.012)   \\ \\
Constant &         6.551$^{***}$  &   2.318$^{***}$  &     2.524$^{***}$  &    2.755$^{***}$  &     2.737$^{***}$  &   2.720$^{***}$  &     2.719$^{***}$   \\
        &     (0.009)    &  (0.057)    &  (0.067)    &  (0.063)   &   (0.063)   &   (0.064)    &  (0.065)  \\ \\             
             \hline 
R$^{2}$           &       0.052  &      0.484 &       0.503    &    0.531 &       0.531 &       0.531   &     0.531   \\
Controls  & & $\checkmark$ & $\checkmark$ & $\checkmark$ & $\checkmark$ & $\checkmark$ & $\checkmark$ \\
Industry & &  & $\checkmark$ & $\checkmark$ & $\checkmark$ & $\checkmark$ & $\checkmark$ \\
Occupation & & &  & $\checkmark$ & $\checkmark$ & $\checkmark$ & $\checkmark$ \\
$\lambda_s f(t)$ & & & & & $\checkmark$ & $\checkmark$ & $\checkmark$ \\
$\lambda_I f(t)$ & & & & & & $\checkmark$ & $\checkmark$ \\
$\lambda_f f(t)$ & & & & & & & $\checkmark$\\
N        &     1134788   &   1075187     & 1075187    &  1075187  &    1075187 &     1075187  &    1075187        \\
\hline \hline\\
\multicolumn{8}{l}{\footnotesize{* p$<$0.10 ** p$<$0.05 *** p$<$0.01}}\\
\multicolumn{8}{l}{\footnotesize{Standard errors in parentheses}}
\end{tabular}
\end{table}

\begin{table}[H]
\centering
\scriptsize
\caption{Effect of SHB on Correlation with Previous Earnings \label{tab:Autocorrelation}}
\begin{minipage}{17cm}
\emph{\small{\newline\newline The table below shows the effect of SHB on the correlation between current and previous pay for both hourly wages (Columns 1 and 2) and weekly earnings (Columns 3 and 4). Since, the Outgoing Rotation Group data records earnings only in the 4th and 8th months of the survey, the estimates below capture the effect of SHB on the correlation between these two data points. The analyses is conducted on two different samples: the first is entire sample of workers regardless of their change of job status between the 4th and 8th months, and the second is the subsample of workers who can be credibly identified to have changed jobs between these two data points. The results for the two samples are shown in Columns 1 and 3, and columns 2 and 4 respectively. Data Source: Current Population Survey. \\}}
\end{minipage}
\begin{tabular}{l >{\centering\arraybackslash} m{2cm} >{\centering\arraybackslash} m{2cm} >{\centering\arraybackslash} m{2cm} >{\centering\arraybackslash} m{2cm}} \\
\hline \hline \\
& \multicolumn{2}{c}{Hourly Wage} & \multicolumn{2}{c}{Weekly Earn} \\ \\
& (1) & (2) & (3) & (4) \\
& Full Sample & Job Changers & Full Sample & Job Changer \\
\hline \\
TreatXOverall & -0.047$^{***}$ & -0.066$^{**}$  & -0.022 & -0.049$^{***}$ \\
& (0.015) & (0.028) & (0.015) & (0.019) \\ \\
OverallBase & 0.384$^{***}$ & 0.331$^{***}$ & 0.261$^{***}$ & 0.190$^{***}$ \\
& (0.066) & (0.015) & (0.005) & (0.012) \\ \\
TreatXMale & -0.059$^{***}$ & -0.099$^{**}$ & -0.033$^{**}$ & -0.092$^{***}$ \\ 
& (0.014) & (0.046) & (0.014) & (0.031) \\ \\
MaleBase & 0.385$^{***}$ & 0.331$^{***}$ & 0.261$^{***}$ & 0.192$^{***}$ \\ 
& (0.006) & (0.015) & (0.005) & (0.012) \\ \\
TreatXFemale & -0.031 & -0.008 & -0.008 & -0.005 \\ 
& (0.027) & (0.047) & (0.019) & (0.030) \\ \\
FemaleBase & 0.386$^{***}$ & 0.323$^{***}$ & 0.248$^{***}$ & 0.181$^{***}$ \\ 
& (0.006) & (0.015) & (0.005) & (0.011) \\ \\
\hline \\
N & 159332 & 15305 & 324620 & 26714 \\   
\hline \hline \\
\end{tabular}
\end{table}

\begin{table}[H]
\centering
\scriptsize
\caption{Effect of SHB on Correlation with Previous Earnings \label{tab:AutocorrelationNEW}}
\begin{minipage}{17cm}
\emph{\small{\newline\newline The table below shows the effect of SHB on the auto-correlation between current and previous pay for both hourly wages (Columns 1-4) and weekly earnings (Columns 5-8). Since, the Outgoing Rotation Group data records earnings only in the 4th and 8th months of the survey, the estimates below capture the effect of SHB on the auto-correlation between these two data points. The analyses is conducted on two different samples: `Full Sample' which contains the entire sample of workers regardless of whether they changed jobs between the 4th and 8th months, and `Job Changers' which is the subsample of workers who can be credibly identified as having changed jobs between these two time points. The results for `Full Sample' are shown in odd-numbered columns and those for the `Job Changers' are shown in even-numbered columns. Data Source: Current Population Survey. \\}}
\end{minipage}
\begin{tabular}{l >{\centering\arraybackslash} m{1.3cm} >{\centering\arraybackslash} m{1.3cm} >{\centering\arraybackslash} m{1.3cm} >{\centering\arraybackslash} m{1.3cm} 
>{\centering\arraybackslash} m{1.3cm} >{\centering\arraybackslash} m{1.3cm} >{\centering\arraybackslash} m{1.3cm} >{\centering\arraybackslash} m{1.3cm}} \\
\hline \hline \\
& \multicolumn{4}{c}{Hourly Wage} & \multicolumn{4}{c}{Weekly Earn} \\ 
& \multicolumn{2}{c}{Full Sample} & \multicolumn{2}{c}{Job Changers} & \multicolumn{2}{c}{Full Sample} & \multicolumn{2}{c}{Job Changers} \\ \\
& (1) & (2) & (3) & (4) & (5) & (6) & (7) & (8) \\
\hline \\
TreatXMaleXPastEarn & -0.047$^{**}$ & & -0.106$^{**}$ & & -0.036$^{*}$ & & -0.119$^{**}$ & \\ 
& (0.019) && (0.053) && (0.018) && (0.051) & \\ \\
MaleXPastEarn & 0.577$^{***}$ && 0.506$^{***}$ && 0.553$^{***}$ && 0.465$^{***}$ & \\
& (0.007) && (0.015) && (0.007) && (0.018) & \\ \\ 
TreatXFemaleXPastEarn & -0.045 && -0.072 && -0.024 && -0.082 & \\ 
& (0.028) && (0.060) && (0.027) && (0.082) & \\ \\ 
FemaleXPastEarn & 0.633$^{***}$ && 0.569$^{***}$ && 0.603$^{***}$ && 0.565$^{***}$ & \\ 
& (0.007) && (0.016) && (0.008) && (0.014) & \\ \\ 
TreatXPastEarn && -0.048$^{***}$ && -0.092$^{**}$ && -0.030 && -0.097$^{**}$ \\
&& (0.017) && (0.037) && (0.020) && (0.048) \\ \\ 
PastEarn && 0.607$^{***}$ && 0.541$^{***}$ && 0.580$^{***}$ && 0.522$^{***}$ \\ 
&& (0.006) && (0.012) && (0.006) && (0.013) \\ \\
\hline \\ 
N & 159332 & 159332 & 15305 & 15305 & 339340 & 339340 & 28644 & 28644 \\ 
\hline\hline 
\end{tabular} 
\end{table} 

\begin{figure}[H]
\centering 
\scriptsize
\caption{\small{Effects of SHB on Earnings by Income Quintile}\label{fig:SHbEffectbyIncQuintile}}
\begin{minipage}{17cm}
\emph{\footnotesize{\newline\newline The figure below shows the quintile-specific effects of SHB on log(hourly wage) (Panel A) and log(weekly earnings) (Panel B) for men and women separately. To compute these effects, I first residualize  earnings in the 4th month of the survey ($\tilde{y}_{ism't'}^{(4)}$) and then compute quintile ranks based on this residualized earnings measure, separately by gender, and for hourly wages and weekly earnings. I then compute quintile specific effects of SHB using non-residualized earning in the 8th month ($y_{ismt}^{(8)}$) as an outcome variable, using a specification similar to (\ref{Eq01})). Since I can only use earnings in the 8th month, the sample size decreases to 159332 for hourly wages and 324620 for weekly earnings. The standard error bars represent 95\% confidence levels. Data Source: Current Population Survey. }\\}
\end{minipage}
\includegraphics[width=0.99\textwidth]{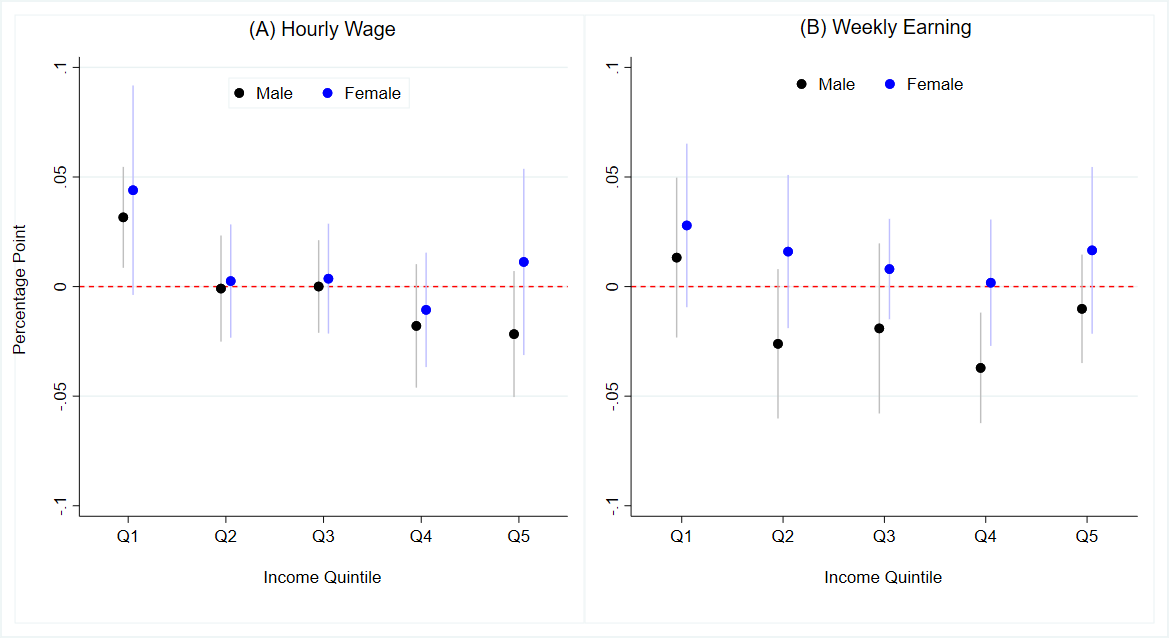}
\end{figure}

\begin{table}[H]
\centering 
\scriptsize 
\caption{\small{Effect of Salary History Ban on Disclosure Behavior} \label{tab:EffectSHBDisclosure}}
\begin{minipage}{18cm}
\emph{\footnotesize{\newline\newline The table below shows the effects of disclosure rates of men and women, and both the pre and post ban gender differences in disclosure rates. Results are shown for three models: Linear Probability (LPM), Logit, and Probit, and for both weighted and unweighted PayScale data. The Logit and Probit columns show the marginal effects on probability and not log odds. All regressions control for covariates including state of residence, calendar time, and female specific time trends. \textit{(PostBan - PreBan)} refers to the effect of SHB. Data Source: PayScale. \\}}
\end{minipage} 
\begin{tabular}{l >{\centering\arraybackslash} m{2cm} >{\centering\arraybackslash} m{2cm} >{\centering\arraybackslash} m{2cm} >{\centering\arraybackslash} m{2cm} >{\centering\arraybackslash} m{2cm} >{\centering\arraybackslash} m{2cm}} \\
\hline\hline \\
& \multicolumn{2}{c}{LPM} & \multicolumn{2}{c}{Probit} & \multicolumn{2}{c}{Logit} \\ \\
& (1) & (2) & (3) & (4) & (5) & (6) \\
& Weighted & Unweighted & Weighted & Unweighted & Weighted & Unweighted \\
\hline \\
FemaleX(PostBan - PreBan) & -0.241$^{***}$ & -0.234$^{***}$ & -0.242$^{***}$ & -0.243$^{***}$ & -0.243$^{***}$ & -0.246$^{***}$ \\
& (0.047) & (0.045) & (0.026) & (0.023) & (0.023) & (0.019) \\ \\
MaleX(PostBan - PreBan) & -0.219$^{***}$ & -0.208$^{***}$ & -0.230$^{***}$ & -0.221$^{***}$ & -0.234$^{***}$ & -0.226$^{***}$ \\
& (0.035) & (0.040) & (0.018) & (0.022) & (0.015) & (0.018) \\ \\
(Female - Male)XPreBan & 0.024 & 0.024$^{**}$ & 0.021$^{***}$ & 0.020$^{***}$ & 0.021$^{***}$ & 0.020$^{***}$ \\
& (0.016) & (0.012) & (0.007) & (0.006) & (0.007) & (0.006) \\ \\ 
(Female - Male)XPostBan & 0.002 & -0.002 & 0.003 & -0.007 & 0.002 & -0.008$^{*}$\\ 
& (0.029) & (0.020) & (0.010) & (0.004) & (0.010) & (0.004) \\ \\
\hline\\ 
N & 31818 & 31818 & 31818 & 31818 & 31818 & 31818 \\
Controls & $\checkmark$ & $\checkmark$ & $\checkmark$ & $\checkmark$ & $\checkmark$ & $\checkmark$ \\
\hline \hline \\
\multicolumn{7}{l}{* p$<$0.10 ** p$<$0.05 *** p$<$0.01} \\
\multicolumn{7}{l}{Standard errors in parentheses}
\end{tabular}
\end{table} 

\begin{table}[H]
\centering 
\scriptsize 
\caption{\small{Gender Differences in Pre-Ban Enquiry Rates} \label{tab:PreBanEnquiryGenderDiff}}
\begin{minipage}{16cm}
\emph{\footnotesize{\newline\newline The table below shows the differences between enquiry rates for women versus men. The estimates are coefficients on the female dummy in regressions where the outcome variable is an indicator of whether the applicant was asked about salary information and covariates are demographic variables and job characteristics. I show results for Linear Probability Model (LPM), Logistic and Probit regressions. For each model, I use the unweighted data or weight the data as explained in Appendix Section \ref{AppendixPayScaleWeighting}. Standard errors are clustered at the state level. Data Source: PayScale. \\}}
\end{minipage} 
\begin{tabular}{l >{\centering\arraybackslash} m{2cm} >{\centering\arraybackslash} m{2cm} >{\centering\arraybackslash} m{2cm} >{\centering\arraybackslash} m{2cm} >{\centering\arraybackslash} m{2cm} >{\centering\arraybackslash} m{2cm}} \\
\hline\hline \\
& \multicolumn{2}{c}{LPM} & \multicolumn{2}{c}{Logit} & \multicolumn{2}{c}{Probit} \\ \\
& (1) & (2) & (3) & (4) & (5) & (6) \\ 
& Weighted & Unweighted & Weighted & Unweighted & Weighted & Unweighted \\
\hline \\
Female-Male & 0.039$^{**}$ & 0.034$^{**}$ & 0.032$^{***}$ & 0.028$^{***}$ & 0.032$^{***}$ & 0.029$^{***}$ \\ 
& (0.016) & (0.014) & (0.007) & (0.006) & (0.007) & (0.006) \\
\hline \\
N & 26705 & 26705 & 26705 & 26705 & 26705 & 26705 \\
Controls & $\checkmark$ & $\checkmark$ & $\checkmark$ & $\checkmark$ & $\checkmark$ & $\checkmark$ \\
\hline \hline \\
\multicolumn{7}{l}{* p$<$0.10 ** p$<$0.05 *** p$<$0.01} \\
\multicolumn{7}{l}{Standard errors in parentheses}
\end{tabular}
\end{table}

\begin{figure}[H]
\centering 
\scriptsize 
\caption{\small{Occupation Effects on Prob(Asked) v/s Within-Occupation Variance in Earnings} \label{fig:CorrOccVarOcc}}
\begin{minipage}{17cm}
\emph{\footnotesize{\newline\newline The figures below show the correlation between the effects of occupation (2-digit SOC 2010) dummies on the probability of being asked about salary history, with the within-occupation variance in earnings. The earnings variance is computed relative to the same base state-occupation level as used in the regression. Panel (A) shows the average occupation effects for all workers, while Panel (B) separates out occupation effects by gender. Data Source: PayScale. \\}}
\end{minipage}
\begin{subfigure} 
\centering
\caption*{\small{(A)}}
\includegraphics[width = 16cm, height = 9cm]{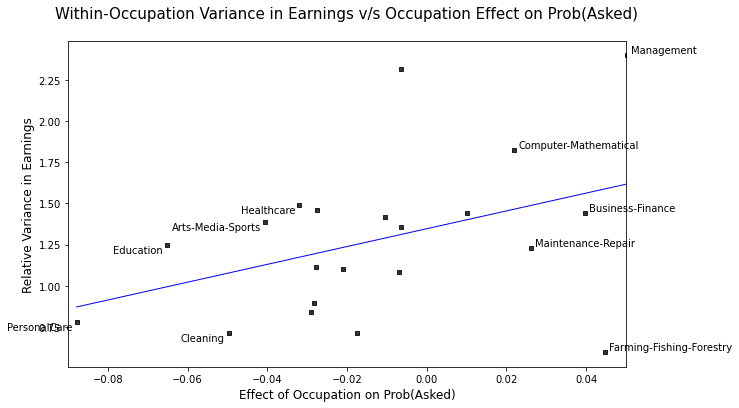}
\end{subfigure}
\begin{subfigure} 
\centering
\caption*{\small{(B)}}
\includegraphics[width = 14cm, height = 10cm]{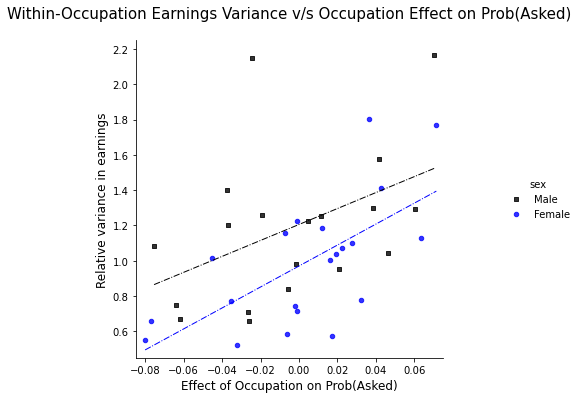}
\end{subfigure}
\end{figure}

\begin{table}[H]
\centering
\scriptsize  
\caption{\small{Pre-Ban Disclosure Differences Upon Enquiry and By Gender} \label{tab:PreBanDisclosure}}
\begin{minipage}{18cm}
\emph{\footnotesize{\newline\newline The table below shows the effects of enquiry and gender and their interaction on disclosure rates before the ban. In \textbf{PANEL A} I control for both a dummy for enquiry (\textit{Asked}) and a female. In \textbf{PANEL C}  I include only a dummy for female, and in \textbf{PANEL B} I include dummies for both enquiry, female, and their interaction. All three regressions are run using 3 different models: Linear Probability (LPM), Logit, and Probit for both weighted and unweighted PayScale data. All regressions control for other covariates including state, time, and female specific time trends. \textit{(Asked - NotAsked)} refers to the effects of enquiry, while \textit{(Female - Male)} refers to the gender gap. Data Source: PayScale.  \\}}
\end{minipage}
\begin{tabular}{l >{\centering\arraybackslash} m{2cm} >{\centering\arraybackslash} m{2cm} >{\centering\arraybackslash} m{2cm} >{\centering\arraybackslash} m{2cm} >{\centering\arraybackslash} m{2cm} >{\centering\arraybackslash} m{2cm}} \\
\hline\hline \\
& \multicolumn{2}{c}{LPM} & \multicolumn{2}{c}{Probit} & \multicolumn{2}{c}{Logit} \\ \\ 
& (1) & (2) & (3) & (4) & (5) & (6) \\
& Weighted & Unweighted & Weighted & Unweighted & Weighted & UnWeighted \\
\hline \\ \\
\textbf{PANEL A} &&&&&& \\ \\ \\
Asked - NotAsked & 0.624$^{***}$ & 0.644$^{***}$ & 0.623$^{***}$  & 0.643$^{***}$ & 0.623$^{***}$ & 0.643$^{***}$ \\
& (0.009) & (0.007) & (0.009) & (0.006) & (0.009) & (0.006) \\ \\
Female - Male & 0.004 & 0.006 & 0.000 & 0.001 & 0.001 & 0.002 \\
& (0.013) & (0.011) & (0.005) & (0.004) & (0.005) & (0.004) \\ \\
\hline \\ \\ \\
\textbf{PANEL B} &&&&&& \\ \\ \\
FemaleX(Asked - NotAsked) & 0.659$^{***}$ & 0.672$^{***}$ & 0.658$^{***}$ & 0.671$^{***}$ & 0.659$^{***}$ & 0.672$^{***}$ \\
& (0.008) & (0.006) & (0.008) & (0.007) & (0.008) & (0.007) \\ \\
MaleX(Asked - NotAsked) & 0.591$^{***}$ & 0.612$^{***}$ & 0.591$^{***}$ & 0.611$^{***}$ & 0.591$^{***}$ & 0.611$^{***}$ \\ 
& (0.014) & (0.010) & (0.013) & (0.010) & (0.013) & (0.009) \\ \\
(Female - Male)XAsked & 0.042$^{***}$ & 0.040$^{***}$ & 0.047$^{***}$ & 0.043$^{***}$ & 0.047$^{***}$ & 0.043$^{***}$ \\ 
& (0.013) & (0.012) & (0.012) & (0.009) & (0.007) & (0.009) \\ \\
(Female - Male)XNotAsked & -0.025 & -0.020$^{*}$ & -0.024$^{***}$ & -0.020$^{***}$ & -0.025$^{***}$ & -0.020$^{***}$\\ 
& (0.016) & (0.011) & (0.005) & (0.005) & (0.005) & (0.004) \\ \\
\hline \\ \\ \\
\textbf{PANEL C} &&&&&& \\ \\ \\
Female - Male & 0.028$^{*}$ & 0.028$^{**}$ & 0.021$^{***}$ & 0.020$^{***}$ & 0.021$^{***}$ &  0.020$^{***}$ \\ 
& (0.016) & (0.012) & (0.007) & (0.006) & (0.007) &  (0.006)\\ \\ 
\hline \\
N & 26705 & 26705 & 26705 & 26705 & 26705 & 26705 \\
Controls & $\checkmark$ & $\checkmark$ & $\checkmark$ & $\checkmark$ & $\checkmark$ & $\checkmark$ \\
\hline \hline \\
\multicolumn{7}{l}{* p$<$0.10 ** p$<$0.05 *** p$<$0.01} \\
\multicolumn{7}{l}{Standard errors in parentheses}
\end{tabular}
\end{table}

\begin{table}[H]
\centering 
\scriptsize 
\caption{\small{Disclosure Behavior by Income and Gender} \label{tab:DiscByIncomeGender}}
\begin{minipage}{18cm}
\emph{\footnotesize{\newline\newline The table below shows the gender gap in disclosure rates by income and income gap in disclosure rates by gender, both before and after the ban. \textit{(Female - Male)} refers to gender gap and \textit{(AboveMedian - BelowMedian)} refers to the gap between those earnings higher and lower than the nationwide occupation-year specific median income. Results are shown for the Linear Probability Model and using both weighted and unweighted PayScale data. \\}}
\end{minipage} 
\begin{tabular}{l >{\centering\arraybackslash} m{2.5cm} >{\centering\arraybackslash} m{2.5cm}  >{\centering\arraybackslash} m{2.5cm}  >{\centering\arraybackslash} m{2.5cm}} \\
\hline\hline \\
& \multicolumn{2}{c}{Pre Ban} & \multicolumn{2}{c}{\textbf{Post Ban}} \\ \\
& (1) & (2) & (3) & (4) \\
& Weighted & UnWeighted & Weighted & UnWeighted \\ 
\hline \\ 
FemaleX(AboveMedian - BelowMedian) & 0.055$^{***}$ & 0.055$^{***}$ & 0.022$^{***}$ & 0.020$^{***}$ \\ 
& (0.012) & (0.011) & (0.006) & (0.005) \\ \\
MaleX(AboveMedian - BelowMedian) & 0.078$^{***}$ & 0.084$^{***}$ & -0.004 & -0.005 \\ 
& (0.013) & (0.011) & (0.008) & (0.012) \\ \\
(Female - Male)XBelowMedian & 0.026$^{**}$ & 0.029$^{***}$ & -0.014 & -0.023$^{***}$ \\ 
& (0.011) & (0.009) & (0.016) & (0.006) \\ \\
(Female - Male)XAboveMedian & 0.003 & -0.000 & 0.012 & 0.001 \\ 
& (0.013) & (0.011) & (0.013) & (0.010) \\ \\ 
\hline \\ 
Controls & $\checkmark$ & $\checkmark$ & $\checkmark$ & $\checkmark$ \\
\hline\hline \\
\multicolumn{5}{l}{* p$<$0.10 ** p$<$0.05 *** p$<$0.01} \\
\multicolumn{5}{l}{Standard errors in parentheses}
\end{tabular}
\end{table}

\begin{figure}[H]
\centering 
\scriptsize 
\caption{\small{Disclosure Behavior Among Survey Respondents} \label{fig:survey_wouldyoudisclose}}
\begin{minipage}{18cm}
\emph{\footnotesize{The figure below shows the proportion of men and women who choose from among the three possible disclosure behaviors. Data Source: Salary Disclosure Motivations Survey.} \\}
\end{minipage} 
\includegraphics[width=0.7\textwidth]{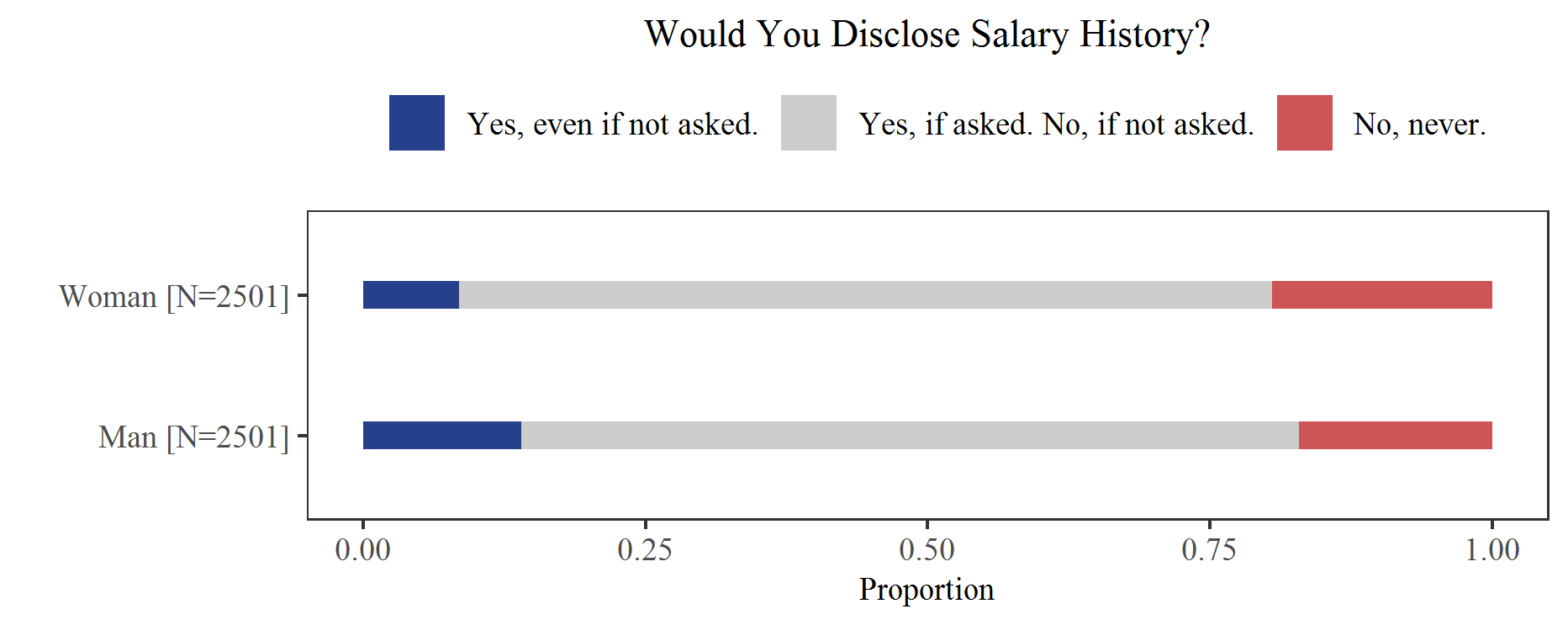}
\end{figure} 

\begin{figure}[H]
\centering\scriptsize 
\caption{\small{Perceptions On Salary History Questions During Job Interviews}\label{fig:survey_perceivesalaryquestions}}
\begin{minipage}{18cm}
\emph{\footnotesize{The figure below shows the proportion of men and women who state how likely they are to think of the following when asked about salary history during job interviews. Each survey participant was required to choose a response to each of the perceptions stated below. The number of male and female participants are shown at the bottom left of the figure. Data Source: Salary Disclosure Motivations Survey. }\\}
\end{minipage}
\includegraphics[width=0.8\textwidth]{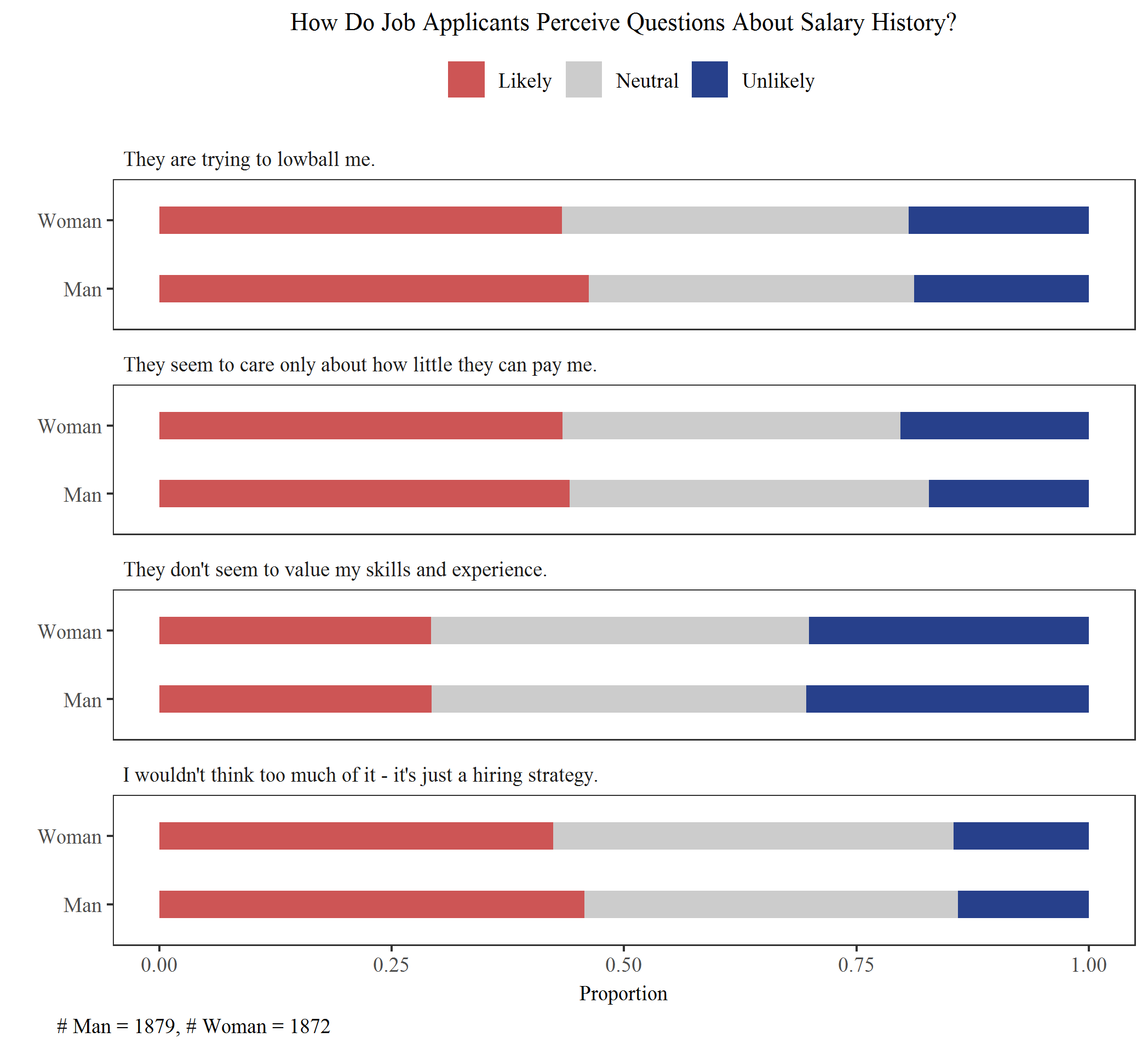}
\end{figure}

\begin{figure}[H]
\centering\scriptsize 
\caption{\small{Reasons For Disclosing Salary History} \label{fig:survey_reasonfordisclosing}}
\begin{minipage}{18cm}
\emph{\footnotesize{The figure below shows the proportion of respondents who agree or disagree with four stated reasons for disclosing salary. 
}\\}
\end{minipage} 
\includegraphics[width=0.56\textwidth]{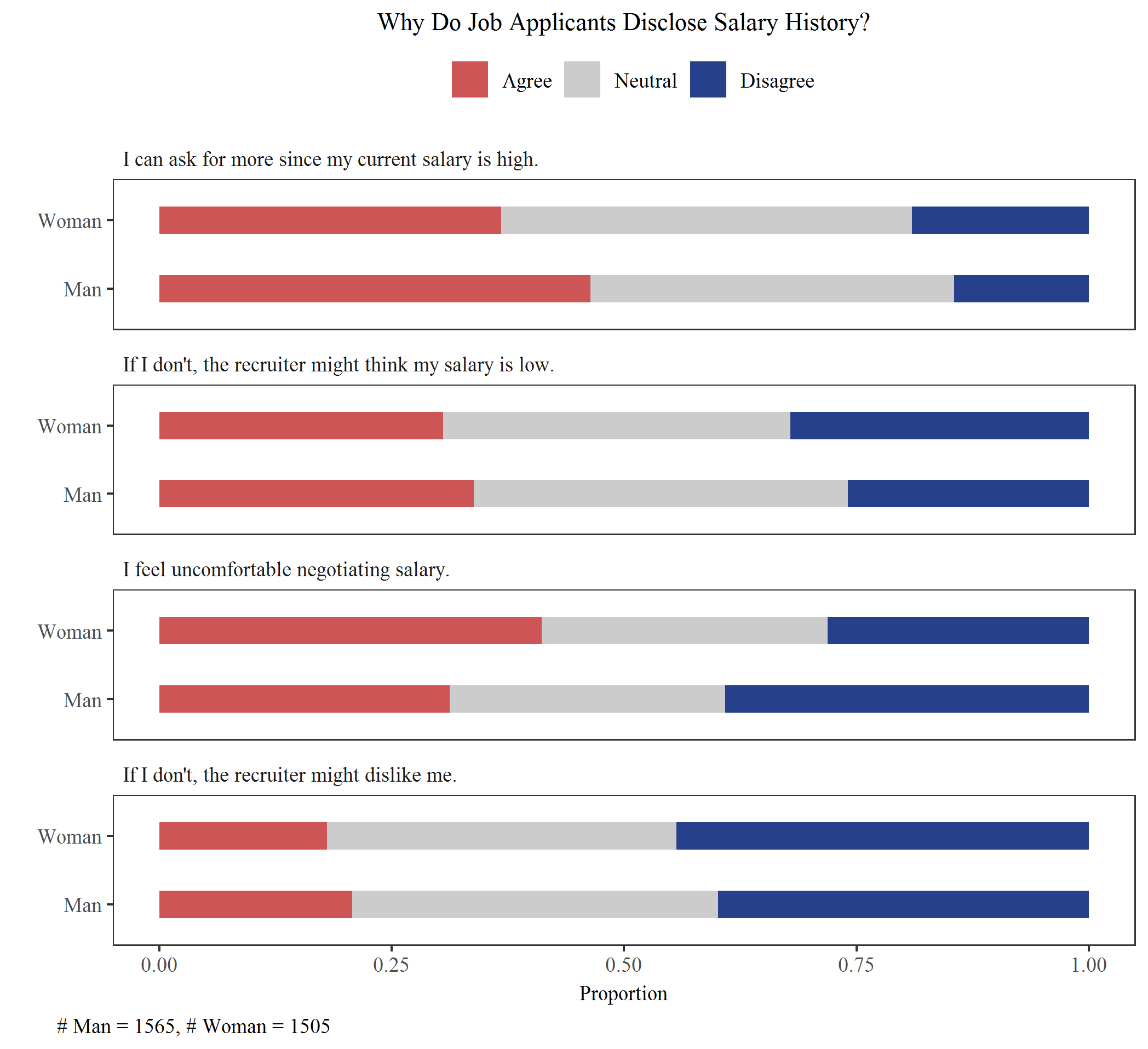}
\end{figure}

\begin{figure}[H]
\centering\scriptsize 
\caption{\small{Reasons For Not Disclosing Salary History} \label{fig:survey_reasonfornotdisclosing}}
\begin{minipage}{18cm}
\emph{\footnotesize{The figure below shows the proportion of respondents who agree or disagree with five stated reasons for not disclosing salary.
}\\}
\end{minipage} 
\includegraphics[width=0.66\textwidth]{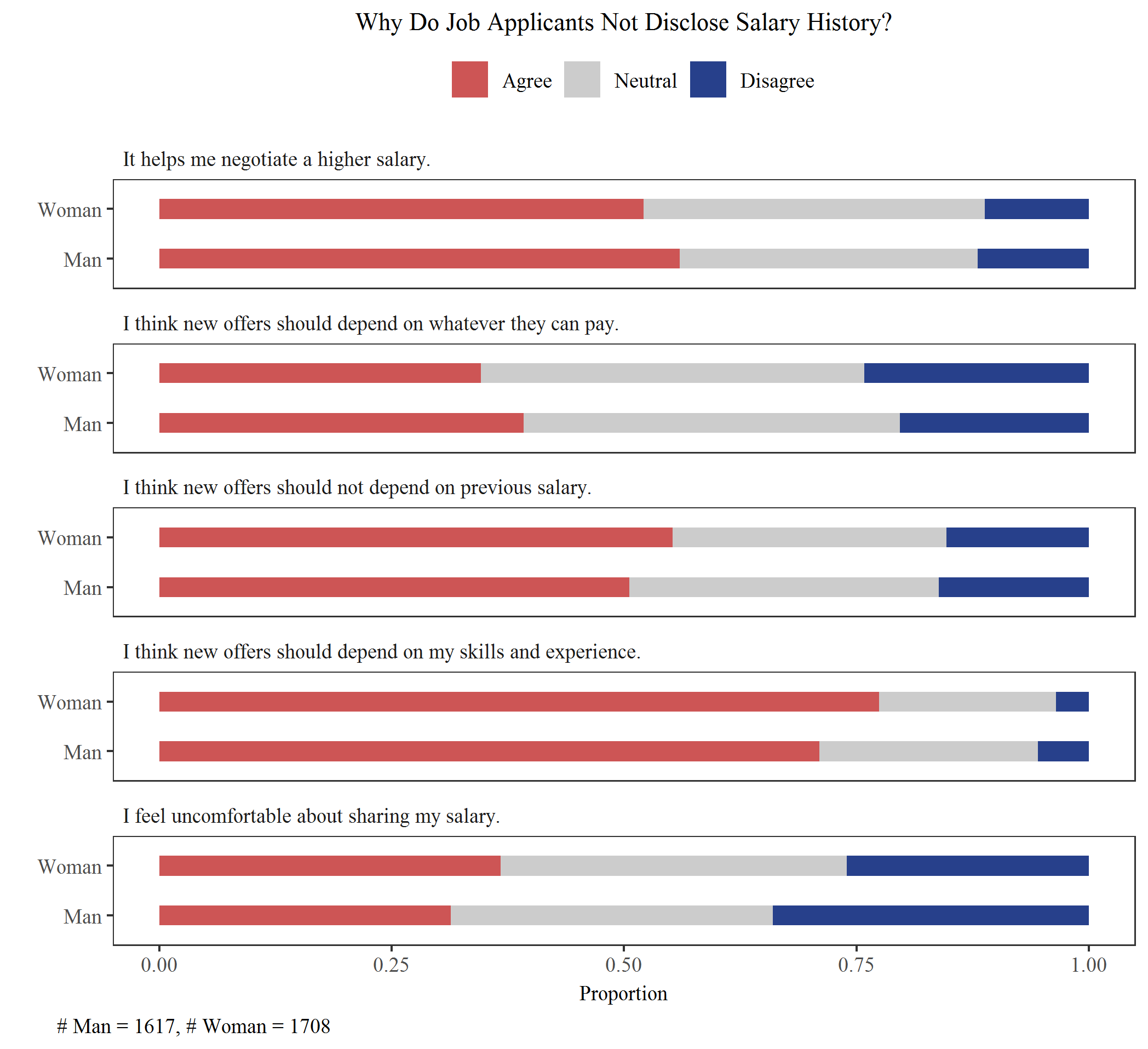}
\end{figure}

\newpage

\begin{figure}[H] 
\caption{\small{Pre-Ban Disclosure Simulation in Simple Framework} \label{fig:fLMfLF_PreBanDisclosure}}
\begin{minipage}{18cm}
\emph{\footnotesize{\newline The figures below show how pre-ban disclosure behavior changes as I vary the fraction of men ($f_{L}^{M}$) and women ($f_{L}^{F}$) who earn the smaller ($w_{L}$) of the two wages in the discrete version of the salary negotiation model. The panel on the left shows the gender gap (female - male) in disclosure rates before the ban, and the panel on the right shows the difference in disclosure rates for those who were asked and those who were not asked before the ban, separately for men and women. The x-axis in the left panel refers to the proportion of low-earning women ($f_{L}^{F}$). The vertical bars in the left panel denote `Low', `Mid', and `High' values of $f_{L}^{F}$ and are the same `Low', `Mid', and `High' values of $f_{L}^{M}$, for which the three lines are drawn in each panel. In the right panel, the vertical bars are drawn for the same `Low', `Mid', and `High' values as in the left panel. } \\}
\end{minipage} 
\begin{subfigure} 
\centering
\includegraphics[width = 0.48\textwidth]{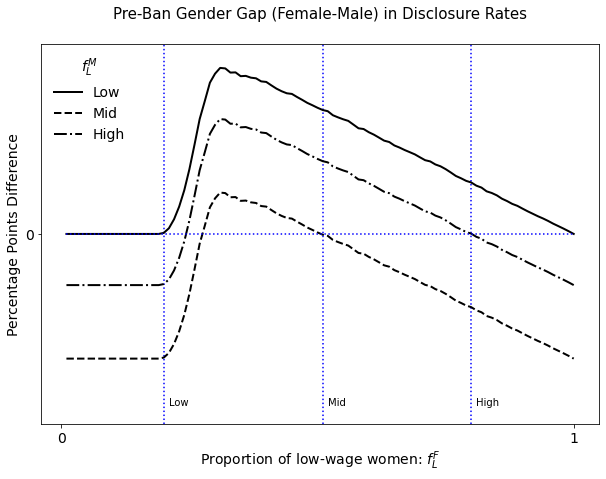} 
\end{subfigure} 
\hfill 
\begin{subfigure}
\centering 
\includegraphics[width = 0.48\textwidth]{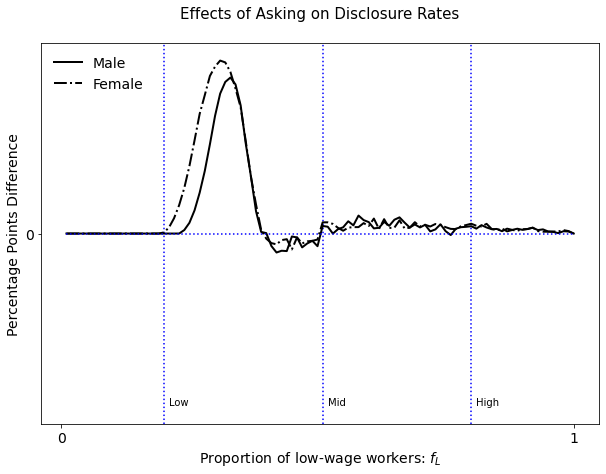} 
\end{subfigure} 
\end{figure}

\begin{figure}[H]
\centering 
\caption{\small{Simulation Effects of SHB on Disclosure Rates in Simple Framework} \label{fig:fLMfLF_SHBEffectOnDisclosure}}
\begin{minipage}{18cm}
\emph{\footnotesize{\newline The figure below shows how the effects of SHB on disclosure rates for men and women vary with the fraction of men ($f_{L}^{M}$) and women ($f_{L}^{F}$) who earn the lower ($w_{L}$) of the two wages in the discrete version of the salary negotiation model. \\}}
\end{minipage} 
\includegraphics[width = 0.48\textwidth]{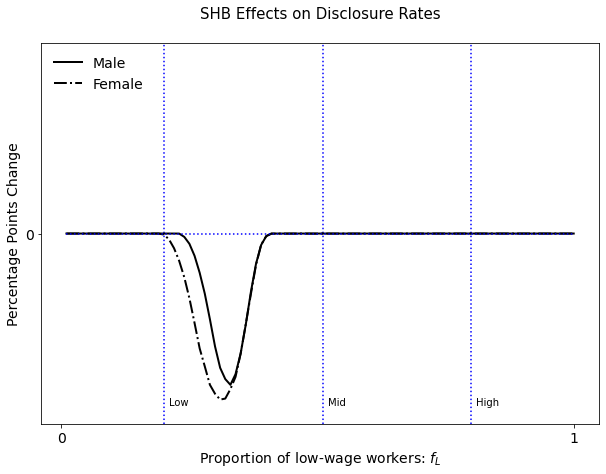}
\end{figure} 

\begin{figure}[H] 
\caption{\small{Gender Pay Gap Simulation in Simple Framework} \label{fig:fLMfLF_GenderPayGap}}
\begin{minipage}{18cm}
\emph{\footnotesize{\newline The figures below show how pre-ban gender pay gap (female - male) and effects of SHB on pay gap change as I vary the fraction of men ($f_{L}^{M}$) and women ($f_{L}^{F}$) who earn the lower ($w_{L}$) of the two wages in the discrete version of the salary negotiation model. The panel on the left shows the gender gap (female - male) in wages before the ban, and the panel on the right shows the SHB effects on gender gap in wages. The x-axis refers to the proportion of low-earning women ($f_{L}^{F}$). The vertical bars denote `Low', `Mid', and `High' values of $f_{L}^{F}$ and are the same `Low', `Mid', and `High' values of $f_{L}^{M}$, for which the three lines are drawn in each panel.} \\}
\end{minipage} 
\begin{subfigure} 
\centering
\includegraphics[width = 0.48\textwidth]{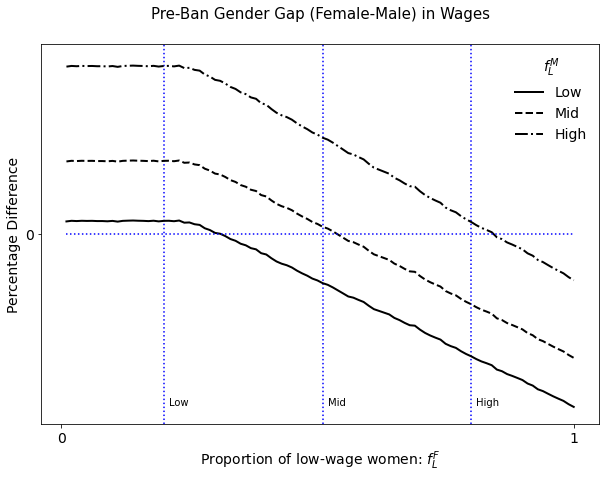} 
\end{subfigure} 
\hfill 
\begin{subfigure}
\centering 
\includegraphics[width = 0.48\textwidth]{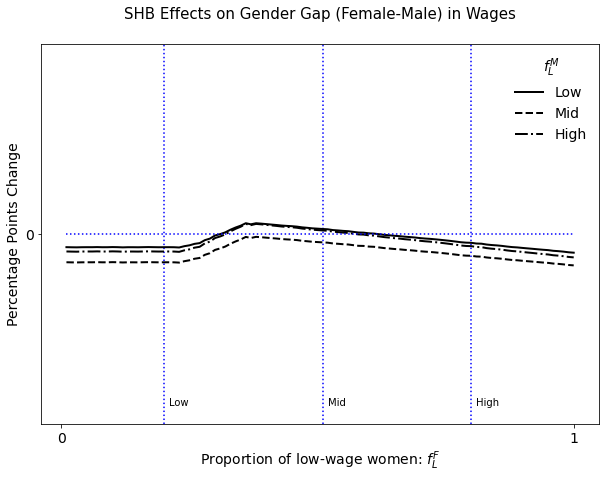} \end{subfigure} 
\end{figure}

\begin{figure}[H]
\centering 
\scriptsize 
\caption{\small{Simulation Effects of SHB on Standard Deviation in Wages in Simple Framework} \label{fig:fLMfLF_EffectOnSDWages}}
\begin{minipage}{18cm}
\emph{\footnotesize{\newline The figure below shows the simulation effects of SHB on standard deviation in wages, separately for men and women and how these effects vary as I change the fraction of men ($f_{L}^{M}$) and women ($f_{L}^{F}$) who earn the lower ($w_{L}$) of the two wages in the discrete version of the salary negotiation model.} \\}
\end{minipage} 
\includegraphics[width = 0.49\textwidth]{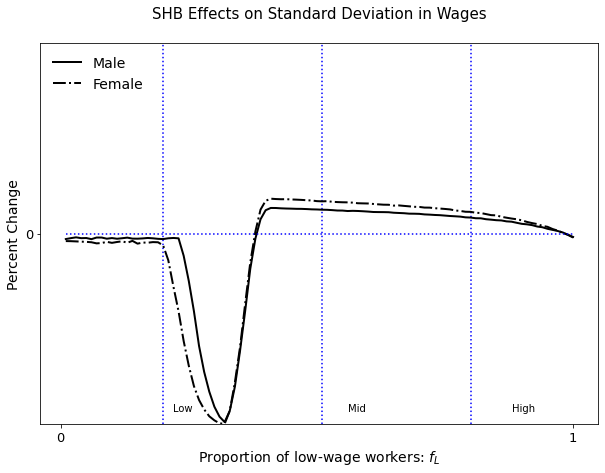}
\end{figure}

\begin{figure}[H]
\centering 
\scriptsize 
\caption{\small{Simulation Effects of SHB on Wages in Simple Framework} \label{fig:fLMfLF_EffectOnWages}}
\begin{minipage}{18cm}
\emph{\footnotesize{\newline The figure below shows the simulation effects of SHB on wages, separately for men and women and how these effects vary as I change the fraction of men ($f_{L}^{M}$) and women ($f_{L}^{F}$) who earn the lower ($w_{L}$) of the two wages in the discrete version of the salary negotiation model. The topmost panel shows the effects on average wages, the middle panel shows the effects on wages for low-earners, and the bottom panel shows the effects on wages for high-earners. } \\}
\end{minipage} 
\includegraphics[width = 0.65\textwidth]{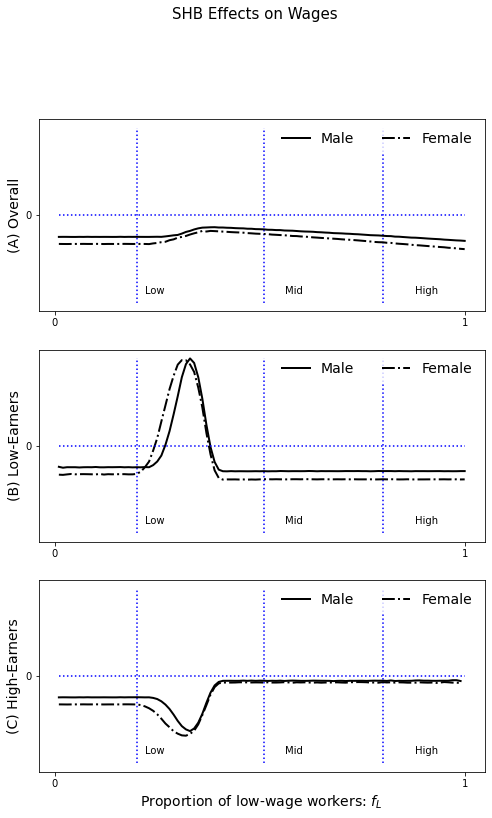}
\end{figure}

%% file: Appendix
\begin{center}\huge{\textbf{Online Appendix}}\end{center}
\addcontentsline{toc}{chapter}{Appendix}


\section{Summary Statistics of Earnings and Disclosure Data \label{AppendixSummaryStat}}
\renewcommand{\thefigure}{A\arabic{figure}}
\renewcommand{\thetable}{A\arabic{table}}
\setcounter{figure}{0}
\setcounter{table}{0}

\begin{figure}[H]
\centering
\caption{\small{Hourly Wage and Weekly Earnings Distribution by Gender (2010 US Dollars)}
 \label{fig:AppendixEarningsDist}}
\begin{minipage}{14cm}
\emph{\footnotesize{\newline \newline The figure below plots the distribution of male and female hourly wages and weekly earnings (in 2010 US\$), pooled across all years from 2010 to 2019. Mean values are represented by $\mu$. Data Source: Current Population Survey.}}
\end{minipage}
\includegraphics[width=15cm, height=8cm]{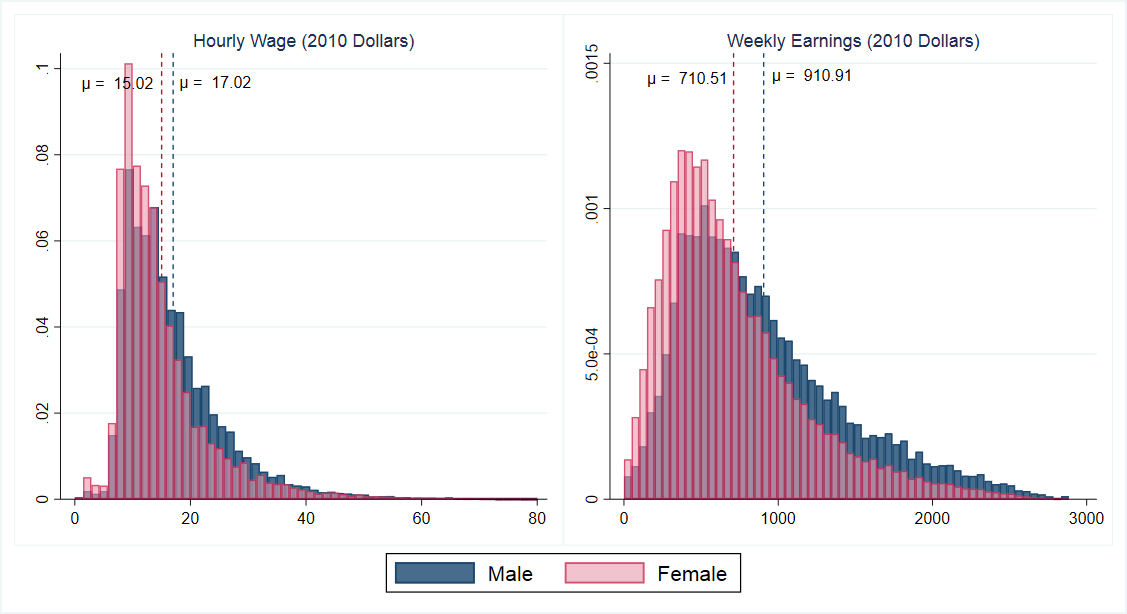}
\end{figure}

\begin{figure}[H]
\centering
\caption{\small{Salary History Disclosure Behavior By Gender}\label{Fig03a}}
\begin{minipage}{12cm}
\emph{\footnotesize{\newline\newline The figure below plots the fraction of men (women) who respond in each category of the salary history question. The data is pooled across all states from April, 2017 to July, 2017, and then from February, 2018 to August, 2019.  Data Source: PayScale.}}
\includegraphics[width=12cm, height = 8cm]{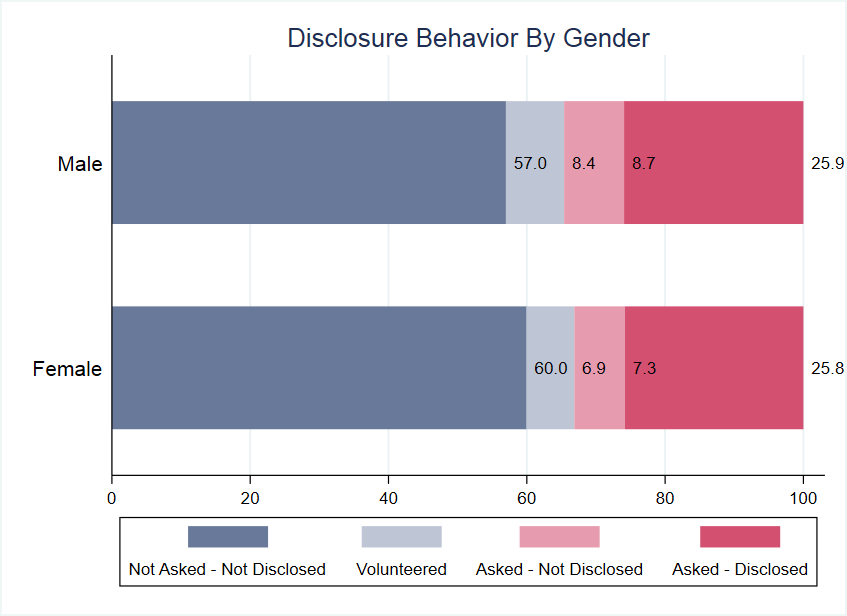}
\end{minipage}
\end{figure}

\begin{figure}[H]
\centering
\caption{\small{Pre-Ban Conditional Gender Differences in Treated and Control Groups of States} \label{Fig04}}
\begin{minipage}{15.5cm}
\emph{\footnotesize{\newline \newline The figure below plots gender differences, conditional on other observables in the pre-ban period (treatment status as of December, 2019), separately for control and treated states for six labor market outcomes. Data Source: Current Population Survey.}}
\end{minipage}
\includegraphics[width=12cm, height=9cm]{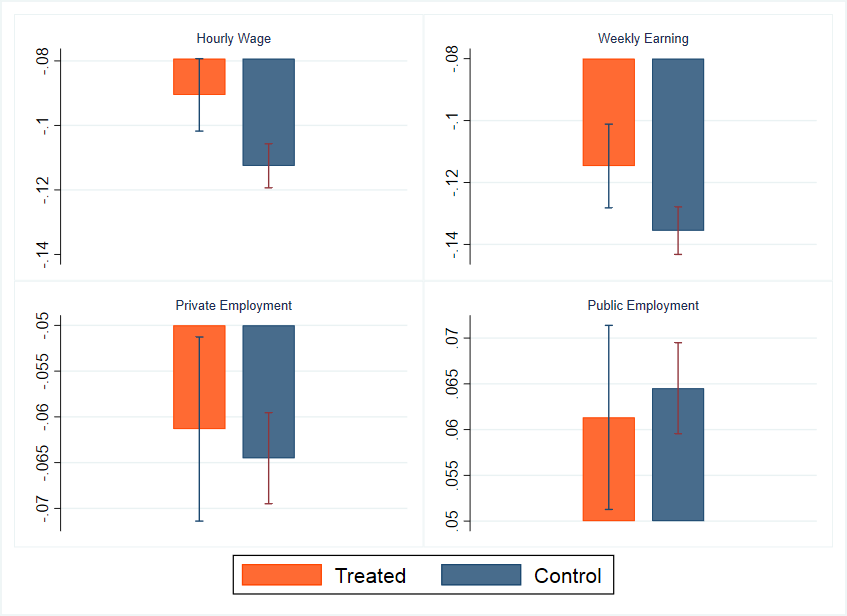}
\end{figure}
 
\begin{figure}[H]
\centering
\scriptsize
\caption{\small{Baseline Gender Pay Gap by Age}
\label{Fig07a}}
\begin{minipage}{15.5cm}
\emph{\footnotesize{\newline \newline The figure below plots the age-specific gender pay gap (female-male) in log(hourly wages) and log(weekly earnings), conditional on other observable.The values combine both within and across-cohort effects. Data Source: Current Population Survey.}}
\end{minipage}
\includegraphics[width=17cm, height=9cm]{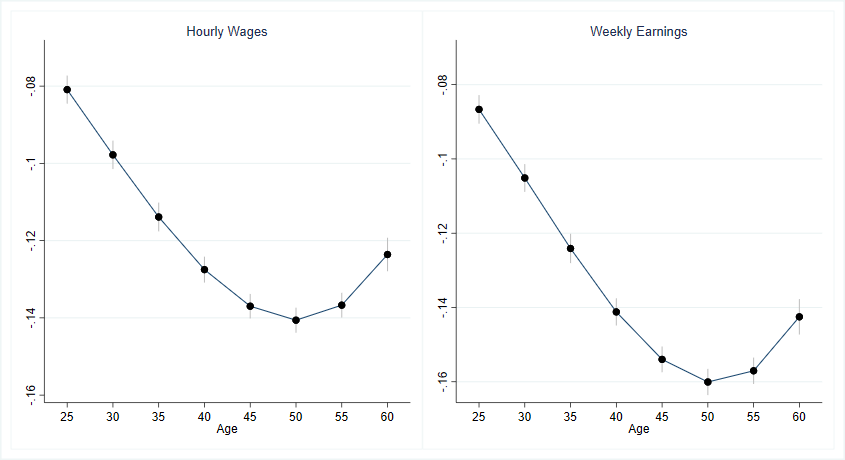}
\end{figure}

\begin{table}[H]
\centering
\scriptsize
\caption{\small{Differences between Hourly Paid and Salaried Workers} \label{Tab01}}
\begin{minipage}{9cm}
\emph{\footnotesize{\newline\newline The table below shows the fraction of salaried and hourly paid workers who belong to each of the groups below. ``\textnormal{No HS}'' implies education level less than high school, ``\textnormal{College}'' implies having at least a four-year college degree. ``\textnormal{Private}'' implies workers employed in the private sector. The data is pooled for January, 2010 to December, 2019, and includes those in our main sample ``AllStateBan''. Data Source: Current Population Survey. \\}}
\end{minipage}
\begin{tabular}{l >{\centering\arraybackslash} m{2cm} >{\centering\arraybackslash} m{2cm} >{\centering\arraybackslash} m{2cm}}
\hline \\
& Salaried & Hourly Paid & p-value of difference \\
\hline \\
Female & 0.463 & 0.517 & 0.000$^{***}$ \\
[1em]
White & 0.831 & 0.803 & 0.000$^{***}$\\
[1em]
Black & 0.080 & 0.115 & 0.000$^{***}$\\
[1em]
Age$<$30 & 0.135 & 0.228 & 0.000$^{***}$\\
[1em]
Age$<$40 & 0.394 & 0.467 & 0.000$^{***}$\\
[1em]
Age$<$50 & 0.663 & 0.695 & 0.000$^{***}$\\
[1em]
Age$<$60 & 0.918 & 0.925 & 0.000$^{***}$\\
[1em]
No HS & 0.029 & 0.094 & 0.000$^{***}$\\
[1em]
College & 0.687 & 0.350 & 0.000$^{***}$\\
[1em]
Full-Time & 0.879 & 0.747 & 0.000$^{***}$\\
[1em]
Private & 0.779 & 0.863 & 0.000$^{***}$\\ \\
\hline 
N & 625,657 &  810,868 & - \\
\hline
\end{tabular}
\end{table}


\section{Effects of SHB on Spillovers Across Employment Sectors\label{AppendixPScoreWeighting}}

\renewcommand{\thefigure}{B\arabic{figure}}
\renewcommand{\thetable}{B\arabic{table}}
\setcounter{figure}{0}
\setcounter{table}{0}

\begin{table}[H]
\centering
\scriptsize
\caption{\small{Effects of Salary History Ban on Earnings By Employment Sector \label{Tab07}}}
\begin{minipage}{13cm}
\emph{\footnotesize{\newline\newline The table below shows the total effects of salary history ban on the two earnings measures for female and male workers, the gender pay gap (GPG), separately for private and public sector workers in our main sample `AllStateBan'. The first two columns are for $log$(Hourly Wage) of hourly-paid workers and the second two are for $log$(Weekly Earnings). Data Source: Current Population Survey.}}
\end{minipage}
\begin{tabular}{l >{\centering\arraybackslash} m{2.5cm} >{\centering\arraybackslash} m{2.5cm} >{\centering\arraybackslash} m{2.5cm} >{\centering\arraybackslash} m{2.5cm}} \\
\hline \hline \\
& \multicolumn{2}{c}{Log(Hourly Wage)} & \multicolumn{2}{c}{Log(Weekly Earning)} \\ \\
& \multicolumn{2}{c}{[N=675,287]} & \multicolumn{2}{c}{[N=1,075,187 ]} \\ \\
& (1) & (2) & (3) & (4) \\
& Public & Private & Public & Private \\
\hline \\ 
Female              &       -0.005   &        0.025$^{***}$  &       0.005  &       0.018$^{**}$ \\
                    &     (0.012)    &     (0.008) &     (0.015)         &     (0.007)        \\ \\
Male                &       0.028$^{***}$&      -0.003 &       0.029$^{**}$         &      -0.009  \\
                    &     (0.010)         &     (0.008)      &     (0.012)         &     (0.007)       \\ \\
GPG                &      -0.033$^{**}$ &       0.027$^{***}$ &      -0.024$^{*}$         &       0.027$^{**}$ \\
                    &     (0.015)         &     (0.012)     &     (0.013)         &     (0.011)      \\
\hline \hline \\
\multicolumn{5}{l}{* p$<$0.10 ** p$<$0.05 *** p$<$0.01} \\
\multicolumn{5}{l}{Standard errors in parentheses}
\end{tabular}
\end{table}

\begin{table}[H]
\centering
\scriptsize
\caption{\small{Spillover Effects of Salary History Bans across Private and Public Sectors} \label{Tab08}}
\begin{minipage}{17cm}
\emph{\footnotesize{\newline\newline The table below shows the total effects of salary history bans on female and male pay, and the gender pay gap (GPG) by sector of employment (public versus private). The first two columns are for the sample `PublicStateBan', while the second two are for the sample `AllStateBan-PP', explained in Section $\ref{ResultsSpillover}$. The estimates are from the specification in ($\ref{Eq02}$) where we interact the treatment indicator with sector of employment. Data Source: Current Population Survey.}}
\end{minipage}
\begin{tabular}{l >{\centering\arraybackslash} m{3cm} >{\centering\arraybackslash} m{3cm} >{\centering\arraybackslash} m{0.5cm}>{\centering\arraybackslash} m{3cm} >{\centering\arraybackslash} m{3cm}} \\
\hline \hline \\
& \multicolumn{2}{c}{\textbf{PublicStateBan}} && \multicolumn{2}{c}{\textbf{AllStateBan-PP}} \\
& (1) & (2) && (3) & (4) \\ \\
& Log(Hourly Wage) & Log(Weekly Earning) && Log(Hourly Wage) & Log(Weekly Earning) \\
\hline \\
FemaleXPublic &  \marktopleft{a1} -0.042 & 0.020 && 0.004 & 0.002 \\ 
& (0.028) & (0.022) && (0.020) & (0.030) \\ \\
MaleXPublic & 0.061$^{**}$ & 0.124$^{***}$ && -0.029 & -0.012 \\ 
& (0.025) & (0.018) && (0.031) & (0.020) \\ \\
GPGXPublic & -0.103$^{***}$ & -0.105$^{***}$ && 0.032 & 0.010 \\
& (0.040) & (0.030) \markbottomright{a1}&& (0.038) & (0.036) \\ \\ \\ 
FemaleXPrivate &  0.016$^{*}$ & 0.016$^{**}$ && \marktopleft{b1} 0.023$^{**}$ & 0.021$^{**}$ \\ 
& (0.009) & (0.007) && (0.009) & (0.009) \\ \\
MaleXPrivate & -0.002 & -0.005 && -0.007 & -0.004 \\ 
& (0.008) & (0.008) && (0.007) & (0.007) \\ \\
GPGXPrivate & 0.018 & 0.021$^{**}$ && 0.030$^{**}$ & 0.025$^{**}$ \\
& (0.014) & (0.010) && (0.013) & (0.012) \markbottomright{b1}\\ \\
\hline \\ 
N & 513,755 & 806,014 && 261,575 & 420,952 \\
\hline \hline \\
\multicolumn{6}{l}{* p$<$0.10 ** p$<$0.05 *** p$<$0.01} \\
\multicolumn{6}{l}{Standard errors in parentheses} 
\end{tabular}
\end{table}

\subsection*{Methodology for Constructing Propensity Score-Based Weights}

We use the following method to conduct the propensity score-based weighting of the three samples used to analyze spillover effects across sectors. 
\begin{itemize}
\item[1.] We first generate three samples as described in the main text: (1) \textbf{AllStateBan} (A) which is used to compare states with bans for all employers with states that have bans for no employers, (2) \textbf{PublicStateBan} (B) which is used to compare states with bans for only public employers with states with no bans for any employer, and (3) \textbf{AllStateBan-PP} (C) which is used to compare states with bans for all employers with states that have bans for only public employers. 
It should be kept in mind that these three samples are not mutually exclusive. 
\item[2.] Then we assign to each observation an indicator ($S_{i}\in\left\{A,B,C\right\}$) that specifies the sample to which it belongs and append all observations into a single dataset. 
\item[3.] Next, we split this combined sample into four separate datasets, each for a specific gender (male, female) and sector of employment (public, private). Since the sectoral spillover analyses compares gender pay gaps in public versus private employment, this sample splitting is essential to ensure that the background covariates of workers are balanced within each of these groups. 
\item[4.] Then for each of the four datasets, we use a multinomial logit (or, probit) model where we predict the probabilities that each observations belongs to each of the three samples (A, B, C). 
This is done usng the following specification:
\begin{equation}\label{EQCovBal}S_{i} = X_{i}\beta+\epsilon_{i}\nonumber\end{equation}
Here $X_{i}$ is a vector of individual characteristics and job characteristics and excludes the state of residence and the calendar year. 
These two variables are excluded because they define the final treatment and can therefore strongly predict the sample identifier without any contribution from other demographic characteristics. 
This gives us, for each observation, a set of three predicted probabilities: $(p_{i}^{A},p_{i}^{B},p_{i}^{C})$, where $p_{i}^{Z}$ denotes the predicted probability that the observation belong to Sample $Z$, and $p_{i}^{A}+p_{i}^{B}+p_{i}^{C}=1$. 
It should be kept in mind, that in a binary classification problem, the inverse of the predicted probability of belonging to a group can be interpreted as an inverse probability weight that can be used to ensure covariate balance between the two groups. 
We extend this method to a three group analyses as in our case.  
\item[5.] To do this, we consider the \textbf{AllStateBan} (or, A) as our reference sample, and assign to each observation in this sample a weight of 1. 
For each observation in the other two samples, we modify their sample weight as $w_{i} = \frac{p_{i}^{A}}{p_{i}^{Z}},\  \forall\  i\in Z,\  \forall\  Z\in\left\{B,C\right\}$.
\item[6.] We further modify the weights within each sample as follows: $\hat{w}_{i} = w_{i}*\frac{\sum_{i\in Z} \textbf{1}}{\sum_{i\in Z} w_{i}},\  \forall\  Z\in\left\{A,B,C\right\}$. 
This is done to ensure that when we collate observations belonging to the same sample across gender and sector types, one or more groups do not bias the statistics in our analyses. 
\item[7.] Finally, we collate observations belonging to the same sample across gender and sector of employment types and create three different weighted samples. 
\item[8.] We then use these weighted samples to conduct our sectoral spillover analyses.  
\item[9.] For inference we bootstrap the process. 
To do this, we first draw randomly from each of the base samples separately and then repeat the steps (1)-(8) above.  
\end{itemize}

\begin{table}[H]
\centering 
\scriptsize 
\caption{\small{Covariate Balance Check across Weighted Samples in Spillover Analyses [All Workers]} \label{Tab-CovBalWeek}}
\begin{minipage}{17cm}
\emph{\footnotesize{\newline The table below shows the covariate balance checks across the three samples used for the analyses of spillovers between public and private sectors. The samples are weighted using the propensity score based method outlined in Appendix Section \ref{AppendixPScoreWeighting}. Balance tests are conducted separately for men and women, in the public and private sectors. The t-tests in columns (4), (5), (6) show p-values for pairwise comparison between samples. Column (7) shows the p-value for the F-test for joint orthogonality. Data Source: Current Population Survey.}}
\end{minipage}
\begin{tabular}{l >{\centering\arraybackslash} m{2.5cm} >{\centering\arraybackslash} m{2.5cm} >{\centering\arraybackslash} m{2.5cm} >{\centering\arraybackslash} m{1cm} >{\centering\arraybackslash} m{1cm} >{\centering\arraybackslash} m{1cm} >{\centering\arraybackslash} m{1cm}} \\
\hline\hline \\
& (1) & (2) & (3) & (4) & (5) & (6) & (7) \\ \\
& \textbf{AllStateBan} & \textbf{PublicStateBan} & \textbf{AllStateBan-PP} & p-val & p-val & p-val & p-val  \\
Variable & \textbf{A} & \textbf{B} & \textbf{C} & A$=$B & A$=$C & B$=$C & A$=$B$=$C \\  
\hline \hline  \\
\multicolumn{8}{c}{\textbf{Male - Public}} \\
\hline 
Age & 44.43 & 44.46 & 44.47 & 0.75 & 0.77 & 0.96 & 0.92 \\
Frac-HighSchool & 0.22 & 0.22 & 0.22 & 0.30 & 0.88 & 0.60 & 0.59 \\
Frac-College & 0.61 & 0.61 & 0.61 & 0.74 & 0.45 & 0.67 & 0.73 \\
Frac-Black & 0.10 & 0.10 & 0.10 & 0.30 & 0.99 & 0.50 & 0.58 \\
Frac-White & 0.81 & 0.82 & 0.81 & 0.17 & 0.55 & 0.64 & 0.35 \\
Frac-PaidHourly & 0.47 & 0.47 & 0.47 & 0.70 & 0.33 & 0.56 & 0.60 \\
Frac-FullTime & 0.86 & 0.87 & 0.87 & 0.37 & 0.52 & 0.97 & 0.58 \\
Frac-PhysicalDifficulty & 0.04 & 0.04 & 0.04 & 0.75 & 0.64 & 0.54 & 0.83 \\
Frac-Citizen & 0.97 & 0.97 & 0.97 & 0.81 & 0.28 & 0.26 & 0.51 \\
Frac-Married & 0.70 & 0.70 & 0.70 & 0.94 & 0.62 & 0.64 & 0.87 \\ 
N & 80791 & 21289 & 10346 & & & &  \\
\hline \\
\multicolumn{8}{c}{\textbf{Male - Private}}\\
\hline  
Age & 41.60 & 41.58 & 41.62 & 0.56 & 0.31 & 0.15 & 0.36 \\ 
Frac-HighSchool & 0.40 & 0.40 & 0.40 & 0.64 & 0.88 & 0.84 & 0.90 \\ Frac-College & 0.42 & 0.42 & 0.42 & 0.43 & 0.92 & 0.60 & 0.71 \\ 
Frac-Black & 0.08 & 0.08 & 0.08 & 0.69 & 0.19 & 0.32 & 0.41 \\ 
Frac-White & 0.84 & 0.84 & 0.84 & 0.82 & 0.04$^{**}$ & 0.07$^{*}$ & 0.09$^{*}$ \\
Frac-PaidHourly & 0.56 & 0.56 & 0.57 & 0.45 & 0.13 & 0.04$^{**}$ & 0.12 \\
Frac-FullTime & 0.88 & 0.88 & 0.88 & 0.98 & 0.74 & 0.73 & 0.94 \\
Frac-PhysicalDifficulty & 0.03 & 0.03 & 0.03 & 0.95 & 0.64 & 0.67 & 0.89 \\
Frac-Citizen & 0.90 & 0.90 & 0.90 & 0.28 & 0.13 & 0.61 & 0.28 \\
Frac-Married & 0.61 & 0.61 & 0.61 & 0.94 & 0.69 & 0.74 & 0.92 \\
N & 455617 & 387262 & 201760 & & & & \\ 
\hline \\
\multicolumn{8}{c}{\textbf{Female - Public}} \\ 
\hline 
Age & 44.67 & 44.69 & 44.59 & 0.83 & 0.43 & 0.41 & 0.69 \\
Frac-HighSchool & 0.18 & 0.18 & 0.18 & 0.41 & 0.66 & 0.36 & 0.61 \\ Frac-College & 0.68 & 0.68 & 0.68 & 0.90 & 0.67 & 0.66 & 0.90 \\ 
Frac-Black & 0.12 & 0.12 & 0.12 & 0.08$^{*}$ & 0.21 & 0.95 & 0.13 \\
Frac-White & 0.80 & 0.80 & 0.80 & 0.41 & 0.16 & 0.56 & 0.30 \\ 
Frac-PaidHourly & 0.45 & 0.44 & 0.45 & 0.76 & 0.76 & 0.64 & 0.90 \\
Frac-FullTime & 0.78 & 0.78 & 0.78 & 0.37 & 0.88 & 0.65 & 0.67 \\ 
Frac-PhysicalDifficulty & 0.03 & 0.03 & 0.03 & 0.63 & 0.59 & 0.86 & 0.80 \\ 
Frac-Citizen & 0.98 & 0.97 & 0.97 & 0.28 & 0.10$^{*}$ & 0.43 & 0.17 \\
Frac-Married & 0.64 & 0.64 & 0.63 & 0.87 & 0.23 & 0.36 & 0.49 \\
N & 113880 & 33218 & 16197 & & & & \\
\hline \\
\multicolumn{8}{c}{\textbf{Female - Private}} \\
\hline 
Age & 41.83 & 41.82 & 41.88 & 0.63 & 0.15 & 0.08$^{*}$ & 0.20 \\
Frac-HighSchool & 0.32 & 0.32 & 0.32 & 0.93 & 0.74 & 0.80 & 0.95 \\
Frac-College & 0.49 & 0.49 & 0.49 & 0.88 & 0.44 & 0.39 & 0.67 \\
Frac-Black & 0.10 & 0.10 & 0.10 & 0.78 & 0.35 & 0.49 & 0.65 \\
Frac-White & 0.81 & 0.81 & 0.81 & 0.96 & 0.05$^{**}$ & 0.06$^{*}$ & 0.10 \\
Frac-PaidHourly & 0.64 & 0.64 & 0.65 & 0.72 & 0.72 & 0.53 & 0.81 \\ Frac-FullTime & 0.75 & 0.74 & 0.74 & 0.89 & 0.59 & 0.67 & 0.86 \\
Frac-PhysicalDifficulty & 0.03 & 0.03 & 0.03 & 0.89 & 0.81 & 0.90 & 0.97 \\
Frac-Citizen & 0.93 & 0.93 & 0.93 & 0.13 & 0.013$^{**}$ & 0.29 & 0.04$^{**}$ \\
Frac-Married & 0.55 & 0.55 & 0.55 & 0.99 & 0.78 & 0.78 & 0.95 \\
N & 424899 & 364245 & 192649 & & & & \\
\hline 
\end{tabular}
\end{table}

\begin{table}[H]
\centering\scriptsize 
\caption{\small{Covariate Balance Check across Weighted Samples in Spillover Analyses [Hourly-Paid Workers]} \label{Tab-CovBalHour}}
\begin{minipage}{17cm}
\emph{\scriptsize{\newline The table below shows the covariate balance checks across the three samples (only hourly paid workers) used for the analyses of spillovers between public and private sectors. The samples are weighted using the propensity score based method outlined in Appendix Section \ref{AppendixPScoreWeighting}. Balance tests are conducted separately for men and women, in the public and private sectors. The t-tests in columns (4), (5), (6) show p-values for pairwise comparison between samples. Column (7) shows the p-value for the F-test for joint orthogonality. Data Source: Current Population Survey.}}
\end{minipage} 
\begin{tabular}{l >{\centering\arraybackslash} m{2.5cm} >{\centering\arraybackslash} m{2.5cm} >{\centering\arraybackslash} m{2.5cm} >{\centering\arraybackslash} m{1cm} >{\centering\arraybackslash} m{1cm} >{\centering\arraybackslash} m{1cm} >{\centering\arraybackslash} m{1cm}} \\
\hline\hline \\
& (1) & (2) & (3) & (4) & (5) & (6) & (7) \\ \\
& \textbf{AllStateBan} & \textbf{PublicStateBan} & \textbf{AllStateBan-PP} & p-val & p-val & p-val & p-val  \\
Variable & \textbf{A} & \textbf{B} & \textbf{C} & A$=$B & A$=$C & B$=$C & A$=$B$=$C \\  
\hline \hline  \\
\multicolumn{8}{c}{\textbf{Male - Public}} \\
\hline 
Age & 43.91 & 43.87 & 43.90 & 0.79 & 0.98 & 0.89 & 0.97 \\
Frac-HighSchool & 0.33 & 0.34 & 0.33 & 0.14 & 0.77 & 0.50 & 0.34 \\
Frac-College & 0.44 & 0.44 & 0.43 & 0.57 & 0.23 & 0.51 & 0.44 \\
Frac-Black & 0.11 & 0.11 & 0.11 & 0.45 & 0.63 & 0.94 & 0.70 \\
Frac-White & 0.80 & 0.80 & 0.80 & 0.84 & 0.98 & 0.91 & 0.98 \\
Frac-FullTime & 0.83 & 0.83 & 0.83 & 0.31 & 0.75 & 0.72 & 0.58 \\ 
Frac-PhysicalDifficulty & 0.05 & 0.05 & 0.05 & 0.53 & 0.53 & 0.35 & 0.65 \\ 
Frac-Citizen & 0.98 & 0.98 & 0.98 & 0.52 & 0.86 & 0.79 & 0.81 \\ 
Frac-Married & 0.64 & 0.65 & 0.64 & 0.78 & 0.85 & 0.74 & 0.94 \\ 
N & 41046 & 9208 & 4469 & & & & \\
\hline \\
\multicolumn{8}{c}{\textbf{Male - Private}} \\
\hline 
Age & 40.21 & 40.19 & 40.24 & 0.55 & 0.51 & 0.27 & 0.54 \\
Frac-HighSchool & 0.53 & 0.53 & 0.53 & 0.93 & 0.97 & 0.98 & 0.99 \\ 
Frac-College & 0.27 & 0.27 & 0.27 & 0.73 & 0.95 & 0.74 & 0.92 \\ 
Frac-Black & 0.09 & 0.09 & 0.09 & 0.61 & 0.21 & 0.40 & 0.46 \\ 
Frac-White & 0.83 & 0.83 & 0.82 & 0.69 & 0.13 & 0.27 & 0.32 \\ 
Frac-FullTime & 0.83 & 0.83 & 0.83 & 0.99 & 0.88 & 0.89 & 0.99 \\ 
Frac-PhysicalDifficulty & 0.04 & 0.04 & 0.04 & 0.89 & 0.53 & 0.61 & 0.82 \\ 
Frac-Citizen & 0.88 & 0.88 & 0.88 & 0.47 & 0.65 & 0.86 & 0.75 \\
Frac-Married & 0.53 & 0.53 & 0.53 & 0.86 & 0.75 & 0.86 & 0.95 \\
N & 285952 & 240393 & 122729 & & & & \\
\hline \\ 
\multicolumn{8}{c}{\textbf{Female - Public}} \\
\hline 
Age & 44.87 & 44.91 & 44.73 & 0.73 & 0.41 & 0.35 & 0.64 \\ 
Frac-HighSchool & 0.29 & 0.29 & 0.28 & 0.46 & 0.49 & 0.29 & 0.56 \\ 
Frac-College & 0.50 & 0.50 & 0.50 & 0.78 & 0.70 & 0.62 & 0.88 \\ 
Frac-Black & 0.13 & 0.14 & 0.14 & 0.12 & 0.09$^{*}$ & 0.64 & 0.09$^{*}$ \\ 
Frac-White & 0.78 & 0.77 & 0.77 & 0.28 & 0.11 & 0.57 & 0.19 \\ 
Frac-FullTime & 0.68 & 0.69 & 0.69 & 0.37 & 0.47 & 0.98 & 0.55 \\ 
Frac-PhysicalDifficulty & 0.04 & 0.04 & 0.04 & 0.87 & 0.93 & 0.86 & 0.98 \\ 
Frac-Citizen & 0.97 & 0.97 & 0.97 & 0.38 & 0.98 & 0.55 & 0.68 \\
Frac-Married & 0.60 & 0.60 & 0.60 & 0.64 & 0.27 & 0.52 & 0.51 \\ 
N & 53468 & 14097 & 7057 & & & & \\ 
\hline \\
\multicolumn{8}{c}{\textbf{Female - Private}} \\ 
\hline 
Age & 41.14 & 41.13 & 41.17 & 0.80 & 0.49 & 0.39 & 0.68 \\ 
Frac-HighSchool & 0.40 & 0.40 & 0.40 & 0.91 & 0.95 & 0.98 & 0.99 \\ 
Frac-College & 0.39 & 0.39 & 0.39 & 0.93 & 0.71 & 0.67 & 0.91 \\ 
Frac-Black & 0.11 & 0.11 & 0.11 & 0.71 & 0.28 & 0.44 & 0.56 \\
Frac-White & 0.80 & 0.80 & 0.80 & 0.74 & 0.09$^{*}$ & 0.17 & 0.22 \\ 
Frac-FullTime & 0.67 & 0.67 & 0.67 & 0.89 & 0.79 & 0.88 & 0.97 \\ 
Frac-PhysicalDifficulty & 0.04 & 0.04 & 0.04 & 0.83 & 0.89 & 0.98 & 0.98 \\
Frac-Citizen & 0.92 & 0.92 & 0.92 & 0.30 & 0.08$^{*}$ & 0.43 & 0.19 \\ 
Frac-Married & 0.52 & 0.52 & 0.52 & 0.97 & 0.76 & 0.79 & 0.95 \\ 
N & 294821 & 250057 & 127320 & & & & \\
\hline 
\end{tabular}
\end{table}


\section{Heterogeneous Effects of SHB by Race, Education, and Age \label{AppendixResultsRace}}
\renewcommand{\thefigure}{C\arabic{figure}}
\renewcommand{\thetable}{C\arabic{table}}
\setcounter{figure}{0}
\setcounter{table}{0}

\begin{table}[H]
\centering
\scriptsize
\caption{Effect of SHB on Pay and Gender Pay Gap [By Race] \label{tab:GPGRace}}
\begin{minipage}{16cm}
\emph{\small{\newline\newline The table below shows the effect of SHB on the hourly wages and weekly earnings and the gender pay gap in these measures separatey for white and black workers. `WhiteGap' and `BlackGap' refers to the gender pay gap (Female-Male) for white and black workers respectively. Data Source: Current Population Survey. }}
\end{minipage}
\begin{tabular}{l >{\centering\arraybackslash} m{3cm} >{\centering\arraybackslash} m{3cm} >{\centering\arraybackslash} m{3cm} >{\centering\arraybackslash} m{3cm}} \\
\hline \hline \\
& (1) & (2) & (3) & (4) \\
& Hourly Wage & Hourly Wage & Weekly Earn & Weekly Earn \\ 
\hline \\
TreatXWhite & 0.010$^{*}$ & & 0.003 & \\
& (0.005) & & (0.005) & \\ \\
TreatXBlack & -0.005 & & -0.003 & \\
& (0.008) & & (0.011) & \\ \\
TreatXWhiteGap & & 0.021$^{*}$ & & 0.035$^{*}$ \\
& & (0.012) & & (0.020) \\ \\
TreatXBlackGap & & -0.009 & & -0.031 \\
& & (0.038) & & (0.040) \\ \\
BaseWhiteGap & & -0.144$^{***}$ & & -0.323$^{***}$ \\
& & (0.005) & & (0.007) \\ \\
BaseBlackGap & & -0.072$^{***}$ & & -0.165$^{***}$ \\
& & (0.007) & & (0.009) \\ \\
\hline
N & 675287 & 675287 & 1075187 & 1075187 \\
\hline\hline \\   
\end{tabular}
\end{table}

\begin{table}[H]
\centering
\scriptsize
\caption{Effect of SHB on Race Gap [White-Black] in Earnings \label{tab:RaceGap}}
\begin{minipage}{16cm}
\emph{\small{\newline\newline The table below shows the effect of SHB on the race gap (White-Black) in earnings for hourly wages and weekly earnings. `RaceGap' refers to the premium in earnings for white workers over black workers. `TreatXRaceGap' therefore denotes the effect of ban on the racial gap. The effect on RaceGap is further decomposed into effect on white and black workers. These estimates are denoted as `TreatXWhite' and `TreatXBlack' respectively. Data Source: Current Population Survey. }}
\end{minipage}
\begin{tabular}{l >{\centering\arraybackslash} m{2cm} >{\centering\arraybackslash} m{2cm} >{\centering\arraybackslash} m{2cm} >{\centering\arraybackslash} m{2cm} >{\centering\arraybackslash} m{2cm} >{\centering\arraybackslash} m{2cm}} \\
\hline \hline \\
& \multicolumn{2}{c}{Overall} & \multicolumn{2}{c}{Male} & \multicolumn{2}{c}{Female} \\ 
& (1) & (2) & (3) & (4) & (5) & (6) \\ 
& Hourly Wage & Weekly Earn & Hourly Wage & Weekly Earn & Hourly Wage & Weekly Earn \\
\hline \\
TreatXRaceGap & 0.015$^{*}$ & 0.006 & -0.001 & -0.015 & 0.030$^{**}$ & 0.025 \\
& (0.009) & (0.011) & (0.021) & (0.014) & (0.014) & (0.017) \\ \\
TreatXWhite & 0.010$^{*}$ & 0.003 & -0.001 & -0.007 & 0.020$^{***}$ & 0.013 \\ 
& (0.005) & (0.005) & (0.008) & (0.008) & (0.008) & (0.008) \\ \\
TreatXBlack & -0.005 & -0.003 & 0.000 & 0.008 & -0.010 & -0.012 \\
& (0.008) & (0.011) & (0.009) & (0.013) & (0.014) & (0.016) \\ \\
Baseline RaceGap & 0.042$^{***}$ & 0.082$^{***}$ & 0.076$^{***}$ & 0.115$^{***}$ & 0.015$^{**}$ & 0.056$^{***}$ \\
& (0.006) & (0.010) & (0.006) & (0.011) & (0.006) & (0.010) \\ \\
\hline
N & 675287 & 1075187 & 675287 & 1075187 & 675287 & 1075187 \\ 
\hline \hline \\
\end{tabular}
\end{table}

\begin{figure}[H]
\centering 
\caption{\small{Effects of Salary History Ban on $log$(Hourly Wage) [By Age and Education]}\label{fig:SHBEffectOnHWByAgeXEduc}}
\begin{minipage}{17cm}
\emph{\footnotesize{\newline\newline The figure below shows the effects of SHB on log(hourly wage) of male and female workers and on the gender gap in log(hourly wage). Each row denotes one of the five education groups. The panels on the left show effects on male and female wages, while those on the right show effects on gender wage gap. Standard error spikes represent 95\% confidence intervals. Data Source: Current Population Survey. \newline }}
\end{minipage} 
\includegraphics[width=0.7\textwidth]{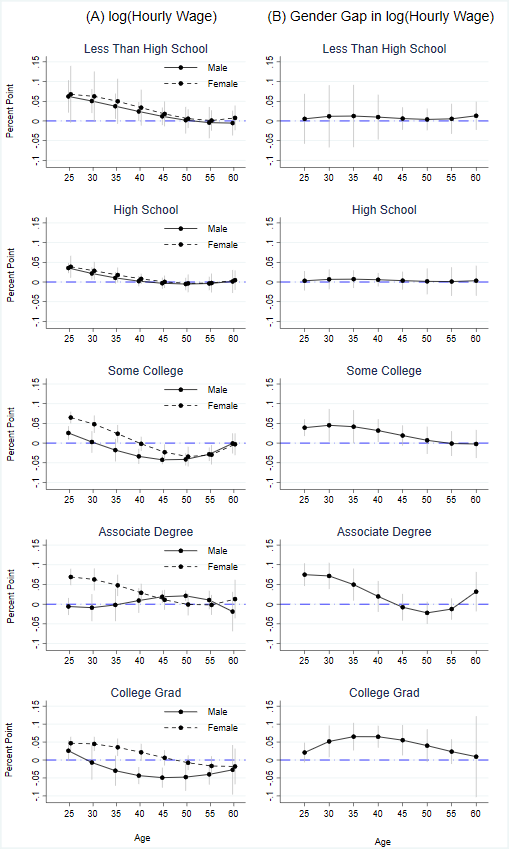}
\end{figure} 

\begin{figure}[H]
\centering 
\caption{\small{Effects of Salary History Ban on $log$(Weekly Earning) [By Age and Education]}\label{fig:SHBEffectOnWEByAgeXEduc}}
\begin{minipage}{17cm}
\emph{\footnotesize{\newline\newline The figure below shows the effects of SHB on log(weekly earning) of male and female workers and on the gender gap in log(weekly earning). Each row denotes one of the five education groups. The panels on the left show effects on male and female earnings, while those on the right show effects on gender earnings gap. Standard error spikes represent 95\% confidence intervals. Data Source: Current Population Survey. \newline }}
\end{minipage} 
\includegraphics[width=0.7\textwidth]{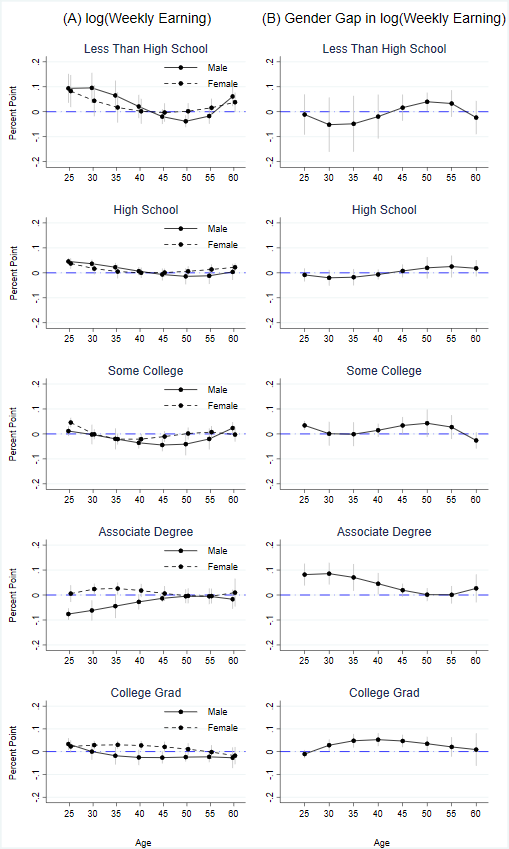}
\end{figure} 

\section{Effects of SHB on Other Labor Market Outcomes \label{AppendixResultsPT}}
\renewcommand{\thefigure}{D\arabic{figure}}
\renewcommand{\thetable}{D\arabic{table}}
\setcounter{figure}{0}
\setcounter{table}{0}

\begin{figure}[H]
\centering
\caption{\small{Pre-Trends in Labor Market Participation Outcomes by Gender} \label{Fig07}}
\begin{minipage}{18cm}
\emph{\scriptsize{\newline\newline The figure below shows the difference in gender gap (female-male) between treated and control states for nine labor market outcomes before and after the salary history ban. These outcomes are labor force participation rate, unemployment rate, private sector employment rate, public sector employment rate, U2E - unemployment to employment rate, E2U - employment to unemployment rate, J2J - job to job transition rate, Pr2Pu - private sector to public sector transition rate, Pu2Pr - public sector to private sector transition rate. All transition rates are monthly transitions. Data Source: Current Population Survey. \\}}
\end{minipage}
\includegraphics[width=18cm, height=14cm]{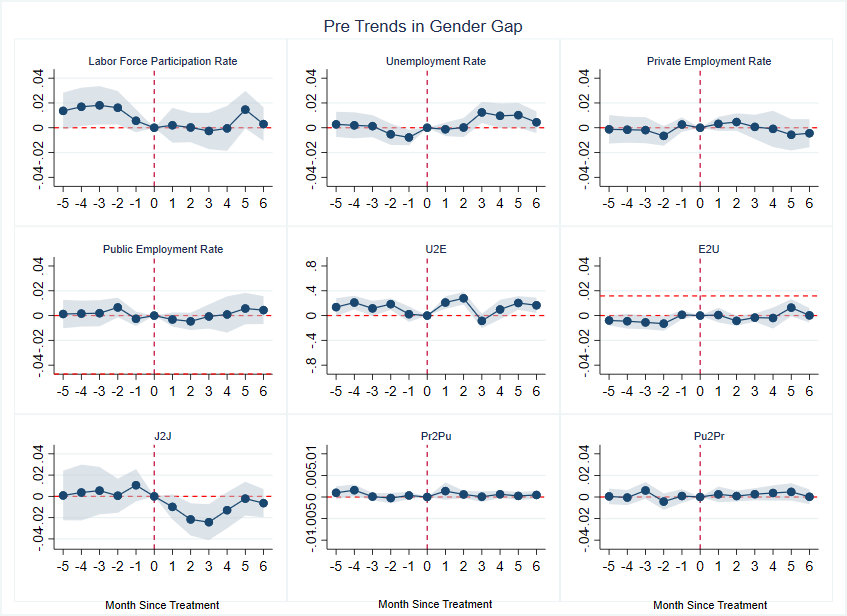}
\end{figure}

\begin{sidewaystable}
\begin{table}[H]
\centering
\scriptsize
\caption{\small{Effects of Salary History Ban on Labor Market Outcomes and Turnover} \label{Tab10}}
\begin{minipage}{18cm}
\emph{\footnotesize{\newline\newline The table below shows the effects of salary history bans on nine outcome variables related to labor market participation, and turnover using a baseline specification similar to ($\ref{Eq01}$) and our main sample `AllStateBan'. The interpretation of coefficients is analogous to Table $\ref{Tab03}$. The seven outcome variables are: LFP - labor 
force participation rate, Unemp - unemployment rate, Private Emp - private sector employment rate, Public Emp - public sector employment rate, U2E - unemployment to employment 
transition rate, E2U - employment to unemployment transition rate, J2J - Job-to-Job transition rate, Pr2Pu - Private sector to public sector transition rate, and Pu2Pr - Public sector to private sector 
transition rate. All transition rates are monthly. Data Source: Current Population Survey. }}
\end{minipage}
\begin{tabular}{l >{\centering\arraybackslash} m{1.5cm} >{\centering\arraybackslash} m{1.5cm} >{\centering\arraybackslash} m{1.5cm} >{\centering\arraybackslash} m{1.5cm} >{\centering\arraybackslash} m{1.5cm} >{\centering\arraybackslash} m{1.5cm} >{\centering\arraybackslash} m{1.5cm} >{\centering\arraybackslash} m{1.5cm} >{\centering\arraybackslash} m{1.5cm}} \\
\hline \hline \\
& (1) & (2) & (3) & (4) & (5) & (6) & (7) & (8) & (9) \\
& LFP & Unemp & Private Emp & Public Emp & U2E & E2U & J2J & Pr2Pu & Pu2Pr \\ 
\hline \\
TreatXFemale & 0.004      &    -0.001      &    -0.005       &    0.005 &0.005 & -0.000 & 0.003 & 0.000 & 0.000 \\ 
& (0.004) & (0.002) & (0.003) & (0.003) & (0.013) & (0.001) & (0.002) & (0.000) & (0.001) \\ \\
Treat & -0.001 &          0.004$^{**}$   &      0.003    &      -0.003 & -0.005 & 0.000 & 0.0003 & -0.000 & -0.000 \\
& (0.003)    &     (0.002)    &     (0.002)     &    (0.002)   &(0.012) & (0.000) & (0.004) & (0.000) & (0.001) \\ \\
Female & -0.118$^{***}$ &       -0.010$^{***}$    &   -0.002     &      0.002 &-0.019$^{***}$ & -0.002$^{***}$ & 0.001 & -0.000 & 0.000 \\ 
& (0.004)     &    (0.001)   &      (0.001)    &     (0.001) & (0.005) & (0.000) & (0.001) & (0.000) & (0.000) \\ \\
FemaleXEverTreat & -0.002    &      -0.000    &       0.002    &      -0.002&-0.004      &    -0.000      &     0.001        &   0.000        &   0.000\\ 
&(0.003)    &     (0.001)  &       (0.002)    &     (0.002) & (0.005)     &    (0.000)       &  (0.001)        & (0.000)     &    (0.000) \\ \\
Constant & 0.954$^{***}$    &    0.400$^{***}$  &      0.980$^{***}$ &       0.020$^{*}$ & 0.431$^{***}$   &     0.082$^{***}$      &  0.749$^{***}$    &    0.001$^{*}$      &    0.003\\
& (0.012)      &   (0.011)     &    (0.012)    &     (0.012) & (0.046)    &     (0.004)     &    (0.014)      &   (0.000)     &    (0.006)\\ \\
\hline \\
N & 6486454    &     4915131    &     4915131   &      4915131 & 184206 & 3244701 & 3148966 & 2584465 & 564501 \\
\hline \hline \\
\multicolumn{10}{l}{* p$<$0.10 ** p$<$0.05 *** p$<$0.01} \\
\multicolumn{10}{l}{Standard errors in parentheses} \\ \\
\end{tabular}
\end{table}
\end{sidewaystable}

\newpage

\section{PayScale Sample Weighting \label{AppendixPayScaleWeighting}}
\renewcommand{\thefigure}{E\arabic{figure}}
\renewcommand{\thetable}{E\arabic{table}}
\setcounter{figure}{0}
\setcounter{table}{0}

To make the PayScale salary history data representative, we use the data from the NY Fed Survey of Consumer Expectations (SCE) Job Search Module (2013-2017). Data from this module can be linked to the monthly core data from the Survey of Consumer Expectations, where respondents are followed for 12 consecutive months. However, since the Job Search module is administered only once a year, this module is essentially a repeated cross-section of individuals. The size of this sample in the publicly available data from 2013-2017 is 5917. Within this module we look at the subsample of workers who have been contacted by any employer in the previous 4 weeks, which reduces our sample size to 1139. This subsample is chosen because the PayScale salary history survey is administered to only those users who wish to evaluate a new job offer. To construct weights we first divide the SCE data into cells defined by a combination of worker demographics. For this subsample we choose a combination of gender, an indicator variable for residence in a state which has any type of salary history ban, and age quartiles within the range 22-64 years. We choose gender and treatment status to ensure that the data is representative along our two main treatment arms. We choose age quartiles, because the unweighted PayScale data is considerably left skewed in age distribution compared to the NY Fed subsample. It is difficult to consider additional dimensions when defining these cells because the small size of the SCE subsample will result in many empty cells. Then we construct the distribution of these cells within the SCE subsample, accounting for given sample weights. These constitute our `population' distributions. We construct the same cells in the PayScale salary history data, and compute their distributions within this data. Then by comparing the `population' distributions with PayScale's sample distribution, we construct inverse probability weights for the PayScale salary history data. \\

To checkhow our weighting strategy performs for worker characteristics which we use to construct cells (gender, treatment status, age) and other characteristics (race, education, total annual earnings), we plot the distribution of these variables for the SCE subsample, unweighted PayScale data, and the weighted PayScale data. These are shown in Appendix Figure $\ref{FigB01}$ below. 
It is evident that the PayScale survey oversamples young workers, who reside predominantly in states which have salary history bans, are more likely to have at least a four-year college degree, and the income distribution is left shifted, in comarison to the SCE subsample. Upon correction the weighted PayScale data matches the SCE subsample along gender, treatment status and age, the three variables which were used to construct weights. The weighted data is also marginally closer to the SCE subsample than the unweighted data along other dimensions like total annual earnings, race and education.  
\begin{figure}[H]
\centering
\caption{\small{Comparison of NY Fed SCE Job Search Module with (Un)Weighted PayScale Salary History Data} \label{FigB01}}
\begin{minipage}{15cm}
\emph{\scriptsize{\newline\newline The figure below compares the distributions of six variables: gender, treatment status, age, total annual earnings, race, education, for three samples: SCE subsample of workers who were contacted by any employer in the last 4 weeks, unweighted PayScale salary history data, and weighted PayScale salary history data where weightes are constructed using the method outlined above. Only three of these variables (gender, treatment status, and age) were used to construct inverse probability weights for the PayScale data.}}
\end{minipage}
\includegraphics[width=16cm, height=12cm]{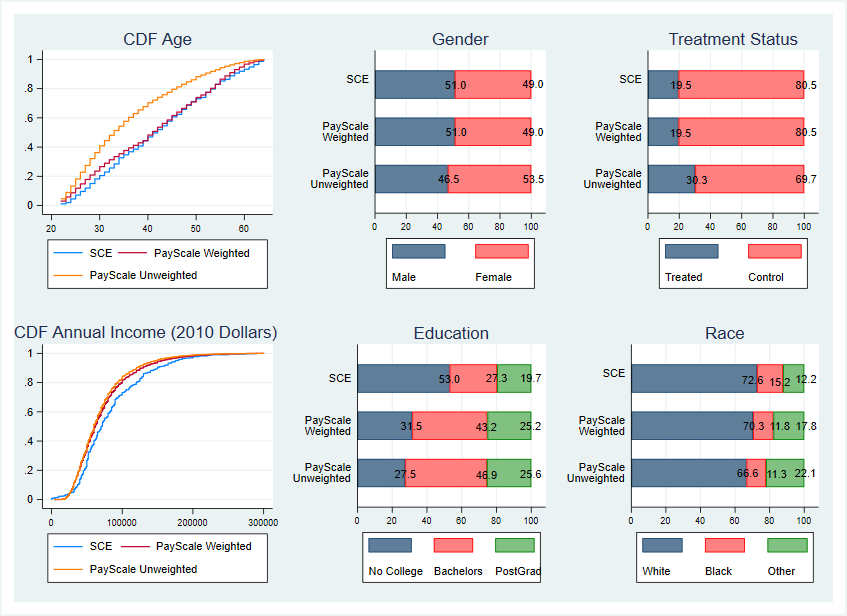}
\end{figure} 

\section{Enquiry and Disclosure Among Job Applicants \label{AppendixPayScaleDisclosure}}
\renewcommand{\thefigure}{F\arabic{figure}}
\renewcommand{\thetable}{F\arabic{table}}
\setcounter{figure}{0}
\setcounter{table}{0}

\begin{table}[H]
\centering 
\scriptsize 
\caption{\small{Disclosure Behavior by Income and Gender} \label{tab:AppendixDiscByIncomeGender}}
\begin{minipage}{18cm}
\emph{\footnotesize{\newline\newline The table below shows the gender gap in disclosure rates by income and income gap in disclosure rates by gender, both before and after the ban. In Panel A, we do not control for enquiry, while in Panels B and C we run regressions separately for those who were asked and not asked respectively. \textit{(Female - Male)} refers to gender gap and \textit{(AboveMedian - BelowMedian)} refers to the gap between those earnings higher and lower than the nationwide occupation-year specific median income. Results are shown for the Linear Probability Model and using both weighted and unweighted PayScale data. Data Source: PayScale. \\}}
\end{minipage} 
\begin{tabular}{l >{\centering\arraybackslash} m{2.5cm} >{\centering\arraybackslash} m{2.5cm}  >{\centering\arraybackslash} m{2.5cm}  >{\centering\arraybackslash} m{2.5cm}} \\
\hline\hline \\
& \multicolumn{2}{c}{Pre Ban} & \multicolumn{2}{c}{\textbf{Post Ban}} \\ \\
& (1) & (2) & (3) & (4) \\
& Weighted & UnWeighted & Weighted & UnWeighted \\ 
\hline \\ 
\textbf{Panel A: OVERALL} [N=31410] &&&& \\ \\ 
(Female - Male)XBelowMedian & 0.026$^{**}$ & 0.029$^{***}$ & -0.014 & -0.023$^{***}$ \\ 
& (0.011) & (0.009) & (0.016) & (0.006) \\ \\
(Female - Male)XAboveMedian & 0.003 & -0.000 & 0.012 & 0.001 \\ 
& (0.013) & (0.011) & (0.013) & (0.010) \\ \\ 
FemaleX(AboveMedian - BelowMedian) & 0.055$^{***}$ & 0.055$^{***}$ & 0.022$^{***}$ & 0.020$^{***}$ \\ 
& (0.012) & (0.011) & (0.006) & (0.005) \\ \\
MaleX(AboveMedian - BelowMedian) & 0.078$^{***}$ & 0.084$^{***}$ & -0.004 & -0.005 \\ 
& (0.013) & (0.011) & (0.008) & (0.012) \\ \\
\hline \\ 
\textbf{Panel B: ASKED [N=9061]} &&&& \\ \\ 
(Female - Male)XBelowMedian & 0.065$^{***}$ & 0.066$^{***}$ & &  \\ 
& (0.022) & (0.014) & &  \\ \\
(Female - Male)XAboveMedian & 0.000 & 0.002 & &  \\ 
& (0.019) & (0.016) & & \\ \\
FemaleX(AboveMedian - BelowMedian) & -0.008 & -0.011 & & \\ 
& (0.013) & (0.010) & & \\ \\
MaleX(AboveMedian - BelowMedian) & 0.056$^{**}$ & 0.052$^{**}$ & \\
& (0.025) & (0.020) & & \\ \\
\hline \\
\textbf{Panel C: NOT ASKED [N=17293]} &&&& \\ \\ 
(Female - Male)XBelowMedian & -0.016$^{*}$ & -0.011 & & \\ 
& (0.008) & (0.008) & &  \\ \\
(Female - Male)XAboveMedian & -0.022$^{**}$ & -0.023$^{***}$ & & \\ 
& (0.009) & (0.008) & & \\ \\
FemaleX(AboveMedian - BelowMedian) & 0.020$^{***}$ & 0.022$^{***}$ & & \\ 
& (0.008) & (0.008) & & \\ \\
MaleX(AboveMedian - BelowMedian) & 0.026$^{**}$ & 0.034$^{***}$ & & \\
& (0.011) & (0.011) & & \\ \\
\hline 
Controls & $\checkmark$ & $\checkmark$ & $\checkmark$ & $\checkmark$ \\
\hline\hline \\
\multicolumn{5}{l}{* p$<$0.10 ** p$<$0.05 *** p$<$0.01} \\
\multicolumn{5}{l}{Standard errors in parentheses}
\end{tabular}
\end{table} 

\newpage
\section{Job Interview Behavior Survey \label{JobInterviewBehaviorSurvey}}
\renewcommand{\thefigure}{G\arabic{figure}}
\renewcommand{\thetable}{G\arabic{table}}
\renewcommand{\theequation}{G\arabic{equation}} 
\setcounter{figure}{0}
\setcounter{table}{0}
\setcounter{equation}{0}

\begin{flushleft}\textit{[Responses were required for questions marked with $^{\ast}$.]}\end{flushleft}
\begin{itemize}
    \item[$^{\ast}$1.] Would you disclose your salary to a recruiter during the hiring process?
    \begin{itemize}[leftmargin=10pt]
        \item[(a)] Yes, if they ask. No, if they don't. \hfill [Go to 2a]
        \item[(b)] Yes, even if they don't ask. \hfill [Go to 2a]
        \item[(c)] No, never. \hfill [Go to 2b]
    \end{itemize}
    \item[$^{\ast}$2a.] To what extent do you agree or disagree with the following reasons for disclosing your salary? 
    \newline \textit{(Measured on a 3-point scale: Disagree, Neutral, Agree.)}
    \begin{itemize}[leftmargin=10pt]
        \item[(a)] I can ask for more since my current salary is high. 
        \item[(b)] I can avoid lengthy negotiations. 
        \item[(c)] I can avoid any confrontation. 
        \item[(d)] If I refuse, the recruiter might think my current salary is low. 
        \item[]  \hfill [Go to 2b if (a) in 1, else go to Q3]
    \end{itemize}
    \item[$^{\ast}$2b.] To what extent do you agree or disagree with the following reasons for not disclosing your salary? \newline \textit{(Measured on a 3-point scale: Disagree, Neutral, Agree.)}
    \begin{itemize}[leftmargin=10pt]
        \item[(a)] Not disclosing helps me negotiate better. 
        \item[(b)] I think new offers should depend on whatever the firm can pay. 
        \item[(c)] I think new offers should depend on my skills and experience. 
        \item[(d)] I think new offers should not depend on my previous salary. 
        \item[(e)] I don't want to share confidential salary information. 
        \item[] \hfill [Go to 3]
    \end{itemize}
    \item[3.] If a recruiter asks for your previous salary, how likely would you think of the following? \newline \textit{(Measured on a 3-point scale: Disagree, Neutral, Agree.)}
    \begin{itemize}[leftmargin=10pt]
        \item[(a)] They are trying to lowball me. 
        \item[(b)] They care only about how little they can pay me. 
        \item[(c)] They don't care about my skills and experience. 
        \item[(d)] Wouldn't think too much of it - it's a hiring strategy. 
        \item[] \hfill [Go to 4]
    \end{itemize}
    \item[4.] How do you think your salary compares with the market average for your position? 
    \begin{itemize}[leftmargin=10pt]
        \item[(a)] Above Average. 
        \item[(b)] Just Average. 
        \item[(c)] Below Average. 
        \item[(d)] Don't Know. 
    \end{itemize}
\end{itemize}

\newpage
\section{Proofs of Propositions \label{ProofsOfPropositions}}
\renewcommand{\thefigure}{H\arabic{figure}}
\renewcommand{\thetable}{H\arabic{table}}
\renewcommand{\theequation}{H\arabic{equation}} 
\setcounter{figure}{0}
\setcounter{table}{0}
\setcounter{equation}{0}

\begin{proposition}{1}{}\label{propalt:existnondisc}
Given $(z,c,z',e,c_{e1},c_{e0},w_{L},w_{H})$ $\exists\  \tilde{f}_{e}\in\left(0,1\right)$ such that:
\begin{itemize}
    \item[1.] $\forall\  f_{L}>\tilde{f}_{e}$ there is a \textbf{Separating Equilibrium} where the firm commits to offering $\hat{w}_{e1}$ to high-earners if they disclose, and $\tilde{w}_{e0}$ to everyone else regardless of their disclosure. 
    Low-earners (i.e., $w=w_{L}$) optimally choose to not disclose ($d=0$) and high-earners (i.e., $w=w_{H}$) choose to disclose ($d=1$).  
    \item[2.] $\forall\  f_{L}\leq\tilde{f}_{e}$ there is a \textbf{Pooling Equilibrium} where the firm offers a flat wage $\hat{w}_{e0}$ regardless of disclosure $d$ and initial wage $w$ (if disclosed).
    Both high and low-earners optimally choose to not disclose ($d=0$).  
    \item[3.] $\tilde{f}_{1}<\tilde{f}_{0}$. 
\end{itemize}
\end{proposition} 

\begin{flushleft}\textbf{Proof:}\end{flushleft} 
We prove this proposition by first characterizing the optimal wage offers under which workers find it optimal to not deviate from equilibrium strategies and then show why these wage offers are optimal for the firm. 
Given $(z,c,z',c_{e1}^{'},c_{e0}^{'},w_{L},w_{H},u(.))$ we first define four threshold wages $(\hat{w}_{e1},\hat{w}_{e0},\tilde{w}_{e1},\tilde{w}_{e0})$ that satisfy the following equations\footnote{We show later what happens when one or more of these threshold wage values do not exist.}:
\begin{align}
u(\hat{w}_{e1}, z')-c_{e1} &= u(w_{H},z)- c \label{eq:highdisc} \\
u(\hat{w}_{e0}, z')-c_{e0} &= u(w_{H},z)- c \label{eq:highnodisc} \\
u(\tilde{w}_{e1}, z')-c_{e1} &= u(w_{L},z)- c \label{eq:lowdisc} \\
u(\tilde{w}_{e0}, z')-c_{e0} &= u(w_{L},z)- c \label{eq:lownodisc} 
\end{align}
$\because\  c_{e1}>c_{e0}$, (\ref{eq:highdisc}) and (\ref{eq:highnodisc}) $\Rightarrow\  \hat{w}_{e1}>\hat{w}_{e0}$. 
Similarly, we can show that $\tilde{w}_{e1}>\tilde{w}_{e0}$ using (\ref{eq:lowdisc}) and (\ref{eq:lownodisc}). 
Again, $\because\  w_{H}>w_{L}$, using (\ref{eq:highdisc}), (\ref{eq:lowdisc}), and Assumption \ref{Assumputil} we can argue that $\hat{w}_{e1}>\tilde{w}_{e1}$.
Following similar logic, we can show $\hat{w}_{e0}>\tilde{w}_{e0}$. 

Therefore, in a separating equilibrium, for any worker with $w=w_{H}$:
\begin{align}
& u(\hat{w}_{e1}, z')-c_{e1} > u(w_{L},z)-c & \  (\text{using }\ref{eq:highdisc}\text{ and }\text{Assumption}\  \ref{Assumputil}) &\nonumber \\
\Rightarrow\  & u(\hat{w}_{e1}, z')-c_{e1} > u(\tilde{w}_{e0}, z')-c_{e0} &\  (\text{using }\ref{eq:lownodisc}) \nonumber \\
\Rightarrow\  & \mathbf{d}(w=w_{H}, \text{Separating}) = 1\  \qedsymbol \nonumber 
\end{align}
Again, in a separating equilibrium for a worker with $w=w_{L}$:
\begin{align}
& u(\tilde{w}_{e0}, z')-c_{e0} < u(\tilde{w}_{e0}, z')-c_{e1} & (\text{using Assumption }\ref{Assumpprivacycost}) \nonumber \\
\Rightarrow\  & \mathbf{d}(w=w_{L}, \text{Separating}) = 0\  \qedsymbol \nonumber 
\end{align}
In a pooling equilibrium for any worker:
\begin{align}
& u(\hat{w}_{e0}, z')-c_{e0} > u(\hat{w}_{e0}, z')-c_{e1} & (\text{using Assumption }\ref{Assumpprivacycost}) \nonumber \\
\Rightarrow\  & \mathbf{d}(w, \text{Pooling}) = 0\  \forall\  w\in\left\{w_{L},w_{H}\right\}\  \qedsymbol \nonumber 
\end{align}
Therefore, in both separating and pooling equilibria, given the new firm's optimal wage offers, both types of workers have no incentive to deviate from the equilibria strategies above. 
Next, we verify that the the wage offers that the new firm commits to result in the maximum expected profit for the firm.

First, we show why there can be only the two types of equilibria above: a separating equilibria where high-earners disclose but low-earners do not disclose, and a pooling equilibria where no types of workers disclose. 
To see this, let's consider the case where both types of workers disclose. In order to incentivize them to disclose while still accepting the new job, the firm has to offer $\hat{w}_{e1}$ to the high-earner and $\tilde{w}_{e1}$ to the low-earner. 
However, once high-earners disclose the firm can identify low-earners and therefore instead of offering them $\tilde{w}_{e1}$ they can be offered a lower wage $\tilde{w}_{e0}$ while incentivizing them to not disclose. 
Therefore, incentivizing both types of workers to disclose is not optimal for the new firm. 
Similarly, let's consider the case where the high-earners do not disclose and the low-earners disclose. 
In order to incentivize them to follow these disclosure decisions in equilibrium and still accept the new job, the new firm has to offer the high-earner $\hat{w}_{e0}$ and the low-earner $\tilde{w}_{e1}$. 
Both high and low-earners would be receiving just their outside option surplus at these offered wages. 
Then the low-earner has an incentive to not disclose and earn $\hat{w}_{e0}$ which will give them a utility value of ($u(w_{H},z)-c$) which is greater than their outside option. 
Again, there cannot be an equilibrium where the high-earner does not disclose and the low-earner discloses.

Second, we show why it's optimal for the firm to offer the separating and pooling wages as above. 
Consider the separating equilibria where high-earners disclose and low-earners do not disclose. 
In order to incentivize high-earners to disclose while still accepting the new offer, they have to be paid at least $\hat{w}_{e1}$ which gives them just enough utility to match their outside option. 
Similarly, in order to incentivize low-earners to accept the offer while not disclosing, they need to be paid a wage slightly higher than $\tilde{w}_{e0}$. 
Any wages higher than these offers would imply less expected profits for the firm, while any wages lower than these values would incentivize workers to not accept the offer and leave the firm with 0 output. 
Now consider the pooling equilibria where the firm offers both workers $\hat{w}_{e0}$, by which the firm extracts full surplus from the high-earner and provides some rent to the low-earner. 
For any wages higher than this offer, the firm would lose out on profits. 
For any wage between $\tilde{w}_{e0}$ and $\hat{w}_{e0}$, the high-earners would not accept the offer, the low-earners would still accept the offer and the firm would lose out on expected profits. 
For any wage less than $\tilde{w}_{e0}$, even low-earners would no longer accept the offer and the firm would generate 0 output. 

Finally, we show when it's optimal for the firm to offer separating versus pooling wages as $f_{L}$ varies. 
Define the threshold value $\tilde{f}_{e}$ as follows:
\begin{equation}\label{eq:cutofffrac}\tilde{f}_{e} = \frac{\hat{w}_{e1}-\hat{w}_{e0}}{\hat{w}_{e1}-\tilde{w}_{e0}}\in(0,1),\  (\text{using }\ref{eq:highnodisc},\ref{eq:lownodisc}\text{ and Assumption }\ref{Assumputil})\end{equation}
Therefore, $\forall\  f_{L}>\tilde{f}_{e}$: 
\begin{align}
& f_{L} > \frac{\hat{w}_{e1}-\hat{w}_{e0}}{\hat{w}_{e1}-\tilde{w}_{e0}} \nonumber \\
\Rightarrow\  & f_{L}(\hat{w}_{e1}-\tilde{w}_{e0}) > \hat{w}_{e1}-\hat{w}_{e0} \nonumber \\
\Rightarrow\  & (1-f_{L})\hat{w}_{e1}+f_{L}\tilde{w}_{e0} < \hat{w}_{e0} \nonumber \\
\Rightarrow\  & z'-\Big[(1-f_{L})\hat{w}_{e1}+f_{L}\tilde{w}_{e0}\Big] > z' - \hat{w}_{e0} \nonumber \\
\Rightarrow\  & \text{Offer Separating Wages} \nonumber 
\end{align} 
Similarly, it can be proved that $\forall\  f_{L}\leq\tilde{f}_{e}$ it is optimal for the new firm to offer pooling wages. 

The third part of this proposition can be proved as follows.
\begingroup
\allowdisplaybreaks
\begin{align}
& u(x,z')-c' = u(w,z)-c \nonumber \\
\Rightarrow\  & \frac{\partial u}{\partial x}\frac{\partial x}{\partial c'} - 1 = 0 \nonumber \\
\Rightarrow\  & \frac{\partial x}{\partial c'} = -(\frac{\partial u}{\partial x})^{-1} \nonumber \\
\Rightarrow\  & \frac{{\partial}^{2}x}{\partial c'\partial w} = -\underbrace{(\frac{\partial u}{\partial x})^{-2}}_{>0}\underbrace{\frac{\partial^{2}u}{\partial x^{2}}}_{<0}\underbrace{\frac{\partial x}{\partial w}}_{>0}\  (\text{using Assumption }\ref{Assumputil}) \nonumber \\
\Rightarrow\  & \frac{{\partial}^{2}x}{\partial c'\partial w} > 0 \nonumber \\ 
\Rightarrow\  & \hat{w}_{10}-\tilde{w}_{10} > \hat{w}_{00}-\tilde{w}_{00} \nonumber \\
\Rightarrow\ & \frac{\hat{w}_{10}-\tilde{w}_{10}}{\hat{w}_{11}-\tilde{w}_{10}} > \frac{\hat{w}_{00}-\tilde{w}_{00}}{\hat{w}_{11}-\tilde{w}_{00}},\  (\text{using Assumptions }\ref{Assumpequalgender}\text{ and }\ref{Assumpnonconformity},\text{ and }\ref{eq:lownodisc}) \nonumber \\ 
\Rightarrow\  & \frac{\hat{w}_{10}-\tilde{w}_{10}}{\hat{w}_{11}-\tilde{w}_{10}} > \frac{\hat{w}_{00}-\tilde{w}_{00}}{\hat{w}_{01}-\tilde{w}_{00}} \nonumber \\ 
\Rightarrow\  & \frac{\hat{w}_{11}-\tilde{w}_{10}}{\hat{w}_{11}-\tilde{w}_{10}} < \frac{\hat{w}_{01}-\tilde{w}_{00}}{\hat{w}_{01}-\tilde{w}_{00}} \nonumber \\
\Rightarrow\  & \tilde{f}_{1} < \tilde{f}_{0}\  \qedsymbol\nonumber
\end{align}
\endgroup

\medskip

\begin{proposition}{2}{}\label{propalt:SHBchangedisc}
Given $(z,c,z',e,c_{11}=c_{01}>c_{00},w_{L},w_{H})$ $\exists\  \bar{c},\tilde{f}_{1},\tilde{f}_{0}$ such that\\
\textbf{Case A: }$\forall\  c_{10}\in(c_{00},\bar{c}),\  \forall\  f_{L}\in(\tilde{f}_{1},\tilde{f}_{0})$:
\begin{itemize}
    \item[1.] The new firm offers separating wages when $e=1$ and pooling wages when $e=0$. 
    \item[2.] $\text{Pr}(d=1|e=1)>\text{Pr}(d=1|e=0)$. 
    \item[3.] $\mathbb{E}(w^{'}|e=1, w=w_{H})>\mathbb{E}(w^{'}|e=0, w=w_{H})$. 
    \item[4.] $\mathbb{E}(w^{'}|e=1, w=w_{L})<\mathbb{E}(w^{'}|e=0, w=w_{L})$
\end{itemize}
\textbf{Case B: }$\forall\  c_{10}\in(c_{00},\bar{c}),\  \forall\  f_{L}\leq\tilde{f}_{1}$:
\begin{itemize}
    \item[1.] The new firm offers pooling wages both when $e=1$ and $e=0$. 
    \item[2.] $\text{Pr}(d=1|e=1)=\text{Pr}(d=1|e=0)=0$. 
    \item[3.] $\mathbb{E}(w^{'}|e=1, w=w_{H})>\mathbb{E}(w^{'}|e=0, w=w_{H})$. 
    \item[4.] $\mathbb{E}(w^{'}|e=1, w=w_{L})>\mathbb{E}(w^{'}|e=0, w=w_{L})$
\end{itemize}
\textbf{Case C: }$\forall\  c_{10}\in(c_{00},\bar{c}),\  \forall\  f_{L}\geq\tilde{f}_{0}$:
\begin{itemize}
    \item[1.] The new firm offers separating wages both when $e=1$ and $e=0$. 
    \item[2.] $\text{Pr}(d=1|e=1)=\text{Pr}(d=1|e=0)=(1-f_{L})$. 
    \item[3.] $\mathbb{E}(w^{'}|e=1, w=w_{H})>\mathbb{E}(w^{'}|e=0, w=w_{H})$. 
    \item[4.] $\mathbb{E}(w^{'}|e=1, w=w_{L})>\mathbb{E}(w^{'}|e=0, w=w_{L})$
\end{itemize}
\textbf{Case D: }$\forall\  c_{10}\geq\bar{c},\  \forall\  f_{L}\in(\tilde{f}_{1},\tilde{f}_{0})$:
\begin{itemize}
    \item[1.] The new firm offers separating wages when $e=1$ and pooling wages when $e=0$. 
    \item[2.] $\text{Pr}(d=1|e=1)>\text{Pr}(d=1|e=0)$. 
    \item[3.] $\mathbb{E}(w^{'}|e=1, w=w_{H})>\mathbb{E}(w^{'}|e=0, w=w_{H})$. 
    \item[4.] $\mathbb{E}(w^{'}|e=1, w=w_{L})\geq\mathbb{E}(w^{'}|e=0, w=w_{L})$
\end{itemize}
\textbf{Case E: }$\forall\  c_{10}\geq\bar{c},\  \forall\  f_{L}\leq\tilde{f}_{1}$:
\begin{itemize}
    \item[1.] The new firm offers pooling wages both when $e=1$ and $e=0$. 
    \item[2.] $\text{Pr}(d=1|e=1)=\text{Pr}(d=1|e=0)=0$. 
    \item[3.] $\mathbb{E}(w^{'}|e=1, w=w_{H})>\mathbb{E}(w^{'}|e=0, w=w_{H})$. 
    \item[4.] $\mathbb{E}(w^{'}|e=1, w=w_{L})>\mathbb{E}(w^{'}|e=0, w=w_{L})$
\end{itemize}
\textbf{Case F: }$\forall\  c_{10}\geq\bar{c},\  \forall\  f_{L}\geq\tilde{f}_{0}$:
\begin{itemize}
    \item[1.] The new firm offers separating wages both when $e=1$ and $e=0$. 
    \item[2.] $\text{Pr}(d=1|e=1)=\text{Pr}(d=1|e=0)=(1-f_{L})$. 
    \item[3.] $\mathbb{E}(w^{'}|e=1, w=w_{H})>\mathbb{E}(w^{'}|e=0, w=w_{H})$. 
    \item[4.] $\mathbb{E}(w^{'}|e=1, w=w_{L})>\mathbb{E}(w^{'}|e=0, w=w_{L})$
\end{itemize}
\end{proposition} 

\begin{flushleft}\textbf{Proof:}\end{flushleft}
First let's prove the results on disclosure rates. \\
Following Proposition \ref{propalt:existnondisc} we can see that $\forall\  f_{L}>\tilde{f}_{1}$ the new firm offers separating wages $(\hat{w}_{11},\tilde{w}_{10})$ when workers are asked $(e=1)$ and the firm offers pooling wages $(\hat{w}_{00})$ when workers are not asked $(e=0)$. 
Since high-earners disclose when separating wages are offered we have $\text{Pr}(d=1|e=1)=(1-f_{L})>0=\text{Pr}(d=0|e=0)\  \  \qedsymbol$. \\
Similarly, when $f_{L}<\tilde{f}_{1}\leq\tilde{f}_{0}$, the new firm offers pooling wages both when workers are asked and when not asked. 
Since both types of workers do not disclose in pooling equilibria, disclosure rates are zero, regardless of enquiry. 
Finally, when $f_{L}\geq\tilde{f}_{0}>\tilde{f}_{1}$, the firm offers separating wages both when workers are asked and when not asked. 
Since only high-earners disclose and the proportion of high-earners is fixed at $(1-f_{L})$, there are no differences in disclosure rates between the two cases.

Next, we show the effects on wage offers for high-earners.
For Cases A and D, high-earners are offered $\hat{w}_{11}$ when asked and $\hat{w}_{00}$ when not asked. 
For Cases B and E, high-earners are offered $\hat{w}_{10}$ when asked and $\hat{w}_{00}$ when not asked. 
Finally, for Cases C and F, high-earners are offered $\hat{w}_{11}$ when asked and $\hat{w}_{01}$ when not asked. 
Using (\ref{eq:highdisc}), ($\ref{eq:highnodisc}$), and Assumptions \ref{Assumpprivacycost} and \ref{Assumpnonconformity} we can show $\hat{w}_{11}=\hat{w}_{01}>\hat{w}_{10}>\tilde{w}_{00}$. 
Therefore, $\mathbb{E}(w^{'}|e=1, w=w_{H})\geq\mathbb{E}(w^{'}|e=0, w=w_{H})\  \  \qedsymbol$.

Finally, we want to impose an upper bound on the value of $c_{10}$ which will ensure $\tilde{w}_{10}<\hat{w}_{00}$. 
This can be done as follows:
\begin{align}
& u(\tilde{w}_{00},z')-c_{00}=u(w_{L},z)-c < u(w_{H},z)-c = u(\hat{w}_{00},z')-c_{00} \nonumber \\
\Rightarrow\  & \tilde{w}_{00} < \hat{w}_{00} \label{Prop2eq01}
\end{align}
Furthermore,
\begin{align}
& u(x,z')-c' = u(w,z)-c \nonumber \\
\Rightarrow\  & \frac{\partial x}{\partial c'} = (\frac{\partial u}{\partial x})^{-1} > 0 \label{Prop2eq02}
\end{align}
The inequalities in (\ref{Prop2eq01}) and (\ref{Prop2eq02}) imply that $\exists\  \bar{c}$ such that $\forall\  c_{10}\in(c_{00},\bar{c})$ we have $\tilde{w}_{00}<\tilde{w}_{10}<\hat{w}_{00}$. 
And $\forall\  c_{10}\geq\bar{c}$ we have $\tilde{w}_{00}<\hat{w}_{00}\leq\tilde{w}_{10}$. \\
For Case A, low-earners are offered $\tilde{w}_{10}$ when asked and $\hat{w}_{00}$ when not asked. 
Therefore $\mathbb{E}(w^{'}|e=1, w=w_{L})<\mathbb{E}(w^{'}|e=0, w=w_{L})\  \  \qedsymbol$. \\
For Cases B and E, low-earners are offered $\hat{w}_{10}$ when asked and $\hat{w}_{00}\  (<\hat{w}_{10})$ when not asked. 
Therefore, $\mathbb{E}(w^{'}|e=1, w=w_{L})>\mathbb{E}(w^{'}|e=0, w=w_{L})\  \  \qedsymbol$. \\
For Cases C and F, low-earners are offered $\tilde{w}_{10}$ when asked and $\tilde{w}_{00}$ when not asked. 
Therefore, $\mathbb{E}(w^{'}|e=1, w=w_{L})>\mathbb{E}(w^{'}|e=0, w=w_{L})\  \  \qedsymbol$. \\
For Case D, low-earners are offered $\tilde{w}_{10}$ when asked and $\hat{w}_{00}$ when not asked. 
Therefore, $\mathbb{E}(w^{'}|e=1, w=w_{L})\geq\mathbb{E}(w^{'}|e=0, w=w_{L})\  \  \qedsymbol$.

\medskip
\begin{proposition}{3}{}\label{propalt:diffgenderdist}
Given ($z,c,z',c_{11}=c_{01}>c_{10}^{M},c_{00},w_{L},w_{H}$) $\exists\  \underline{c}_{F}$ ($>c_{10}^{M}$) such that \\ 
$\forall\  c_{10}^{F}\in\left(c_{10}^{M},\underline{c}_{F}\right)$, $\exists\  \tilde{f}_{1M},\bar{f},\tilde{f}_{0F}$ such that $\forall\  f_{L}^{M}<\tilde{f}_{1M}<\bar{f}<f_{L}^{F}<\tilde{f}_{0F}$:
\begin{itemize}
    \item[1.] Firm offers separating wages to female workers and pooling wages to male workers when asked, and pooling wages to all workers when not asked. 
    \item[2.] $\text{Pr}(d=1|e=1,k=F)>\text{Pr}(d=1|e=1,k=M)$.
    \item[3.] $\text{Pr}(d=1|e=1)>\text{Pr}(d=1|e=0)=0$.
    \item[4.] $\mathbb{E}(w'|e=1,w=w_{L},k=F)<\mathbb{E}(w'|e=0,w=w_{L},k=F)$, \\ $\mathbb{E}(w'|e=1,w=w_{L},k=M)>\mathbb{E}(w'|e=0,w=w_{L},k=M)$. 
    \item[5.] $\mathbb{E}(w'|e=1,w=w_{H},k)>\mathbb{E}(w'|e=0,w=w_{H},k)\  \forall\  k\in\left\{F,M\right\}$.
    \item[6.] $\mathbb{E}(w'|e=0,k=F)-\mathbb{E}(w'|e=0,k=M)>\mathbb{E}(w'|e=1,k=F)-\mathbb{E}(w'|e=1,k=M)$.
\end{itemize}
\end{proposition}

\begin{flushleft}\textbf{Proof:}\end{flushleft}
First we prove the results on disclosure rates and wages for men. 
Given $(z,c,z',c_{11}=c_{01}>c_{10}^{M}>c_{00},w_{L},w_{H})$ define $\tilde{f}_{1M}$ following (\ref{eq:cutofffrac}) as:
\begin{equation}\tilde{f}_{1M} = \frac{\hat{w}_{11}-\hat{w}_{10}^{M}}{\hat{w}_{11}-\tilde{w}_{10}^{M}}\in(0,1)\nonumber\end{equation}
If $f_{L}^{M}<\tilde{f}_{1M}$ then by Proposition \ref{propalt:SHBchangedisc}, the firm would offer pooling wages to men both when they are asked and when they are not asked. 
Therefore, both when they are asked and when they're not, disclosure rates among men are zero. 
Regardless of whether $c_{10}^{M}$ is high, Cases B and E in Proposition \ref{propalt:SHBchangedisc} show that for both high and low-earning men, wage offers are lower when workers are not asked in comparison to when they're asked. 

Second, we prove the results on disclosure rates and wages for women. 
To do this, we argue that $\tilde{f}_{0F}\text{ and }\tilde{f}_{0M}$, which are the $\tilde{f}_{0}$ equivalent (from Proposition \ref{propalt:SHBchangedisc}) for women and men respectively, are equal. 
To see this, we use use (\ref{eq:cutofffrac}) as follows:
\begin{equation}\tilde{f}_{0F} = \frac{\hat{w}_{01}-\hat{w}_{00}}{\hat{w}_{01}-\tilde{w}_{00}}=\tilde{f}_{0M}\nonumber\end{equation}
Intuitively, since there are no differences in disclosure and non-disclosure costs between men and women when they are not asked, and so the proportion of low-earners below which the firm offers pooling wages ($\tilde{f}_{0F},\tilde{f}_{0M}$) are the same. 

Next, we argue that $\tilde{f}_{1M}\in(\tilde{f}_{1F},\tilde{f}_{0F}$ where $\tilde{f}_{1F}\text{ and }\tilde{f}_{1M}$ are the $\tilde{f}_{1}$ (from Proposition \ref{propalt:SHBchangedisc}) for women and men respectively. 
The first inequality $\tilde{f}_{1F}<\tilde{f}_{1M}$ can be proved using the single crossing property of $u(w,z)-c$ in the same way as the last result of Proposition \ref{propalt:existnondisc}. The second inequality (i.e., $\tilde{f}_{1M}<\tilde{f}_{0F}$ can also be proved in the same way by replacing $\tilde{f}_{0F}$ with $\tilde{f}_{0M}$. 
Therefore $f_{L}^{F}\in(\tilde{f}_{1M},\tilde{f}_{0F})\Rightarrow f_{L}^{F}\in(\tilde{f}_{1F},\tilde{f}_{0F})$ and so the new firm offers separating wages to female workers when they are asked and pooling wages when not asked (by Proposition \ref{propalt:SHBchangedisc}).
Since only high-earners disclose when separating wages are offered, the disclosure rate is $(1-f_{L}^{F})$ when women are asked and 0 when not asked. 
This shows that disclosure rates are higher among women than men on average, when workers are asked. 
It also shows that disclosure rates are higher on average when workers are asked than when not asked. 
Cases A and D in Proposition \ref{propalt:SHBchangedisc} show that for high earning women, wages decrease when workers are not asked. 
Finally, we prove the results on wages for low-earning women and use this for the result on gender wage gap.
Define $\bar{f}$ as follows:
\begin{equation}\label{frac:cutoffgwg}\bar{f} = \frac{\hat{w}_{11}-\hat{w}_{10}^{M}}{\hat{w}_{11}-\tilde{w}_{10}^{F}}\end{equation}
First, $c_{10}^{M}<c_{10}^{F}\Rightarrow\  \tilde{w}_{10}^{M}<\tilde{w}_{10}^{F}\Rightarrow\  \frac{\hat{w}_{11}-\hat{w}_{10}^{M}}{\hat{w}_{11}-\tilde{w}_{10}^{M}}<\frac{\hat{w}_{11}-\hat{w}_{10}^{M}}{\hat{w}_{11}-\tilde{w}_{10}^{F}}\Rightarrow\  \tilde{f}_{1M}<\bar{f}$. \\
Next, define $\bar{f}(x)$ as:
\begin{equation}\bar{f}(x) = \frac{\hat{w}_{11}-\hat{w}_{10}^{M}}{\hat{w}_{11}-x}\nonumber\end{equation}
Note that $\bar{f}(\tilde{w}_{10}^{M}) = \tilde{f}_{1M}$, $\bar{f}(\tilde{w}_{10}^{F})=\bar{f}$ and $\frac{\partial}{\partial x}\bar{f}(x)>0$. We also know that $\tilde{f}_{1M}<\tilde{f}_{0F}$ and $\frac{\partial}{\partial c_{10}^{F}}\tilde{w}_{10}^{F}>0$. Therefore $\exists\  \hat{c}_{F}>c_{10}^{M}$ such that $\bar{f}(\tilde{w}_{10}^{F})\in(\tilde{f}_{1M},\tilde{f}_{0F})$. 
Following Proposition \ref{propalt:SHBchangedisc} define $\bar{c}_{F}>c_{10}^{M}$ as the upper-bound on $c_{10}^{F}$ such that $\tilde{w}_{10}^{F}<\hat{w}_{00}$. 
Finally, define $\underline{c}_{F}:=\min(\bar{c}_{F},\hat{c}_{F})>c_{10}^{M}$. 

Then, by Case D of Proposition \ref{propalt:SHBchangedisc}, wage offers of low-earning women are higher when workers are not asked than when they're asked. Finally, 
\begin{align}
& f_{L}^{F} > \bar{f} \nonumber \\ 
\Rightarrow\  & f_{L}^{F} > \frac{\hat{w}_{11}-\hat{w}_{10}^{M}}{\hat{w}_{11}-\tilde{w}_{10}^{F}} \nonumber \\ 
\Rightarrow\  & f_{L}^{F}\tilde{w}_{10}^{F} + (1-f_{L}^{F})\hat{w}_{11} < \hat{w}_{10}^{M} \nonumber \\
\Rightarrow\  & \hat{w}_{00} - \Big[f_{L}^{F}\tilde{w}_{10}^{F} + (1-f_{L}^{F})\hat{w}_{11}\Big] > \hat{w}_{00} - \hat{w}_{10}^{M} \nonumber \\
\Rightarrow\  & \mathbb{E}(w'|e=0,k=F)-\mathbb{E}(w'|e=0,k=M)>\mathbb{E}(w'|e=1,k=F)-\mathbb{E}(w'|e=1,k=M)\  \qedsymbol \nonumber 
\end{align}

\subsection*{Non-Existence of Threshold Values \label{proofnonexistence}}

The existence of the threshold values $\tilde{f}_{ek},\bar{c}_{k},\underline{c}_{k}$ in the propositions above depend on the existence of the threshold values of wages $\tilde{w}_{ed}^{k},\hat{w}_{ed}^{k}$. 
In this section we show what happens when these threshold wage values do not exist. To recap, the four threshold wages for each enquiry status are as follows:
\begin{align}
u(\hat{w}_{e1}, z')-c_{e1} &= u(w_{H},z)- c \nonumber \\
u(\hat{w}_{e0}, z')-c_{e0} &= u(w_{H},z)- c \nonumber \\
u(\tilde{w}_{e1}, z')-c_{e1} &= u(w_{L},z)- c \nonumber \\
u(\tilde{w}_{e0}, z')-c_{e0} &= u(w_{L},z)- c \nonumber 
\end{align}

Since the firm is risk neutral it has no incentive to offer wages higher than $z'$. 
The following can be proved:
\begin{align}
\exists\  \hat{w}_{e1}\Rightarrow\  \exists\  \hat{w}_{e0}\Rightarrow\  \exists\  \tilde{w}_{e0} \label{threshexist01}
\end{align}
Intuitively, the cost of disclosing information $c_{e1}$ is higher than the cost of not disclosing information $c_{e0}$ and high-earners have to be compensated more than low-earners. 
Therefore, if the firm can offer a wage to incentivize high-earners to accept the offer while disclosing their wages, they can also incentivize both high and low-earners when they do not disclose information. 
This is true both when workers are asked and when they are not asked. 

Therefore, if $\nexists\  \tilde{w}_{e0}$ implies the new firm would not be able to offer wages that would incentivize even low-earners to accept the offer and therefore there will be no turnover. \\
If $\nexists\  \hat{w}_{10}$ implies that the new firm would not be able to compensate high earners and only low-earners would accept the offer. 
Since it costs the firm to incentivize workers to disclose the firm would offer low-earners $\tilde{w}_{e0}$. \\
Finally, if $\nexists\  \hat{w}_{11}$ then again the firm would offer the pooling wage $\hat{w}_{10}$ to all workers since it cannot compensate high-earners for disclosing information. 

\newpage
\section{Numerical Simulations of Model}
\renewcommand{\thefigure}{I\arabic{figure}}
\renewcommand{\thetable}{I\arabic{table}}
\renewcommand{\theequation}{I\arabic{equation}} 
\setcounter{figure}{0}
\setcounter{table}{0}
\setcounter{equation}{0}

\subsection{Discrete Initial Wage Distribution with Zero Correlation between Variables}
\label{AppendixSimResultsDiscreteFrameworkfLMfLF}

\begin{figure}[H]
\centering 
\caption{\small{Existence of Initial Male and Female Wage Distributions} \label{fig:AppendixfLMfLF_exist}}
\begin{minipage}{18cm}
\emph{\footnotesize{The figure below shows (in dark green) the combinations of ($f_{L}^{F},f_{L}^{M}$), i.e., fraction of low-earning women and men, for which the following simulation results go through: (a) female wages were lower than male wages in pre-SHB period, (b) gender wage gap improved with SHB, (c) workers disclosed more when asked than when not asked, (d) women disclosed more in pre-SHB period than men, (e) high-earners disclosed more than low-earners, (f) SHB reduced disclosure rates on average. The figure shows that these valid regions include combinations where $f_{L}^{M}<f_{L}^{F}$. }\\}
\end{minipage}
\includegraphics[width=12cm, height = 10cm]{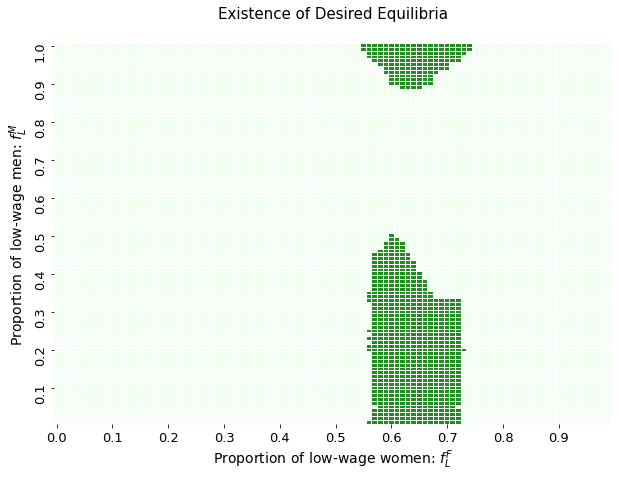}
\end{figure}

\newpage
\subsection{Discrete Initial Wage Distribution Correlated with Initial Match-Output Distribution
\label{AppendixSimResultsDiscreteFrameworkCorrwoldzold}}

In the main simulation results shown in Figures (\ref{fig:fLMfLF_PreBanDisclosure})-(\ref{fig:fLMfLF_EffectOnWages}), we do not account for any correlation between the distributions of the underlying variables $(w,z,c,w',z',c_{e1})$. 
In this section, we consider a setting where the initial wage distribution ($w$) can be correlated with initial match-output distribution ($z$). 
This does not fundamentally affect the results of Propositions (\ref{prop:existnondisc})-(\ref{prop:diffgenderdist}).
The only difference is that now agents evaluate the proportion of low-earners ($f_{L}$) and all thresholds conditional on the observed $z$. 
With this adjustment, all other results go through. 
In the figures below we show the same graphs as in Figures (\ref{fig:fLMfLF_PreBanDisclosure})-(\ref{fig:fLMfLF_EffectOnWages}), but now vary the correlation between initial $w$ and $z$ distributions. 
In particular, we choose parameter values in a way to ensure that at zero-correlation between these two distributions, pre-SHB gender gap in disclosure rates is positive, pre-SHB gender pay gap is negative, and SHB reduces both disclosure rates and gender pay gap. 
Then we allow this correlation to vary between negative and positive ranges. 

\newpage
\begin{figure}[H]
\centering 
\caption{\small{Pre-Ban Disclosure Simulation in Discrete Framework} \label{fig:corrwoldzold_preshbgdgaskeffect}}
\begin{minipage}{18cm}
\emph{\footnotesize{The figure below shows the pre-SHB gender difference (female - male) in disclosure rates in the left panel and the difference in disclosure rates between those who are asked and not asked before SHB in the right panel. Both figures show how these results change as we vary the correlation between initial $w$ and initial $z$. In particular, the figure on the left has the correlation between initial $w$ and $z$ on the x-axis, and we show how the gender gap in disclosure rate varies as we vary the correlation for men from negative to zero and positive.}\\}
\end{minipage} 
\begin{subfigure}
\centering 
\includegraphics[width=0.48\textwidth]{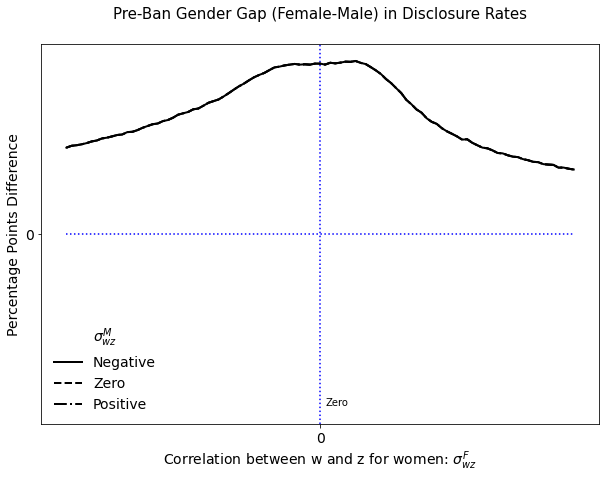}
\end{subfigure}
\hfill 
\begin{subfigure}
\centering 
\includegraphics[width=0.48\textwidth]{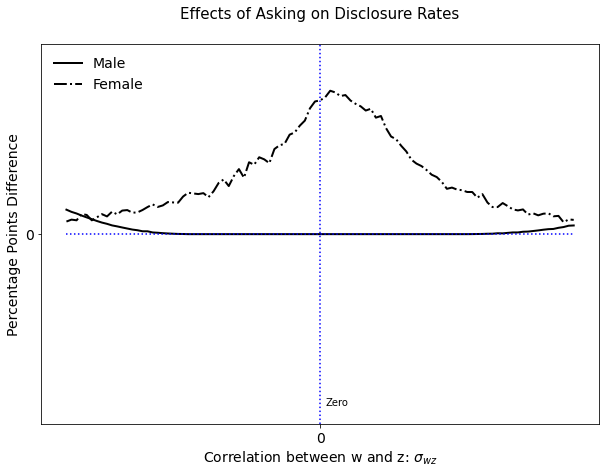}
\end{subfigure}
\end{figure} 

\begin{figure}[H]
\centering 
\caption{\small{Simulation Effects of SHB on Disclosure Rates in Simple Framework} \label{fig:corrwoldzold_shbeffectdisc}}
\begin{minipage}{18cm}
\emph{\footnotesize{The figure below shows how the effects of SHB on disclosure rates vary as we change the correlation between initial wage ($w$) and initial match-output ($z$) from negative, to zero, to positive.} \\}
\end{minipage}
\includegraphics[width=0.5\textwidth]{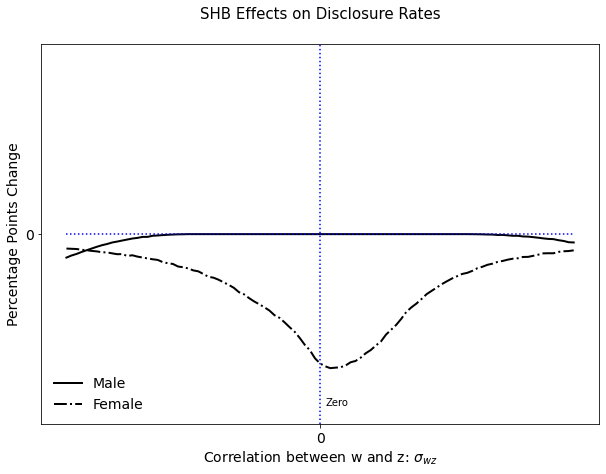}
\end{figure} 

\begin{figure}[H]
\centering 
\caption{\small{Gender Pay Gap Simulations in Simple Framework} \label{fig:corrwoldzold_gpg}} 
\begin{minipage}{18cm}
\emph{\footnotesize{The figures below show the pre-SHB gender pay gap (or female premium) in wages in the left panel and the effects of SHB on gender pay gap in the right panel. We show how these results vary as we change both the correlation of initial $w$ and initial $z$ for women (on the x-axis) and the correlation for men alongside. } \\}
\end{minipage} 
\begin{subfigure}
\centering 
\includegraphics[width=0.48\textwidth]{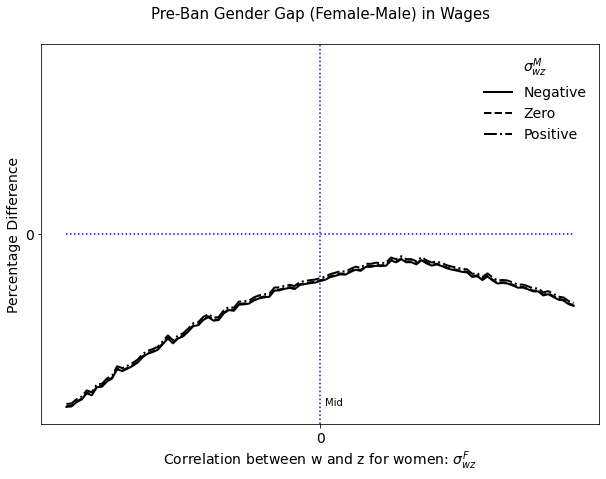}
\end{subfigure}
\hfill
\begin{subfigure}
\centering 
\includegraphics[width=0.48\textwidth]{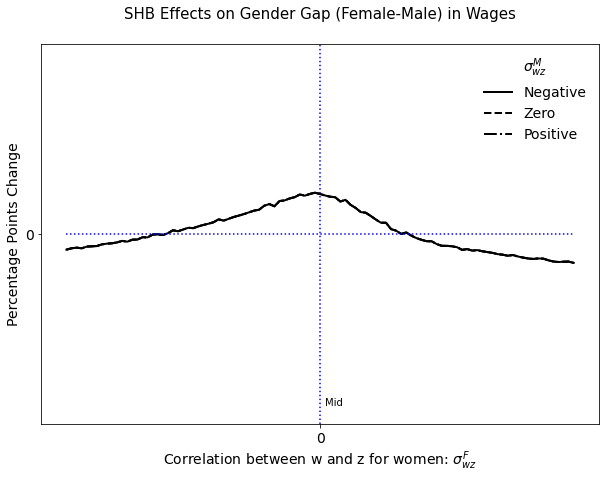}
\end{subfigure}
\end{figure} 

\begin{figure}[H]
\centering 
\caption{\small{Simulation Effects of SHB on Standard Deviation in Wages in Simple Framework}\label{fig:corrwoldzold_effectsdwages}}
\begin{minipage}{18cm}
\emph{\footnotesize{The figure below shows how the standard deviation in wages change as we vary the correlation between initial wage ($w$) and initial match-output ($z$) distributions for men and women independently.}\\}
\end{minipage} 
\includegraphics[width=0.5\textwidth]{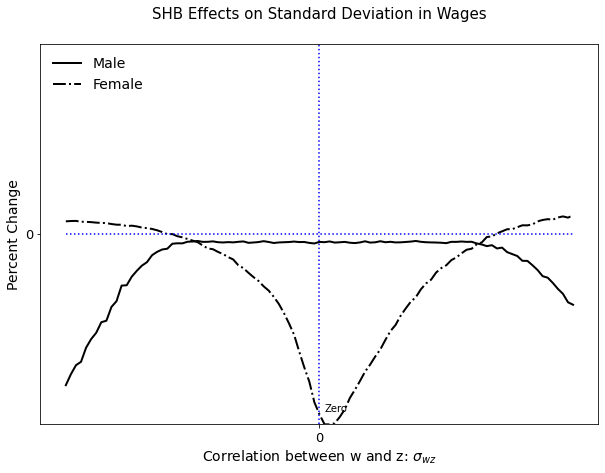}
\end{figure} 

\begin{figure}[H]
\centering 
\caption{\small{Simulation Effects of SHB on Wages in Simple Framework}\label{fig:corrwoldzold_effectwages}}
\begin{minipage}{18cm}
\emph{\footnotesize{The figure below shows how the effects of SHB on wages change as we vary the correlation between initial wage ($w$) and initial match-output ($z$) distributions for men and women independently. The topmost panel shows the overall effects for all workers, the middle panel shows the effects for low-earners, and the third panel shows the effects for high-earners.}\\}
\end{minipage} 
\includegraphics[width=0.65\textwidth]{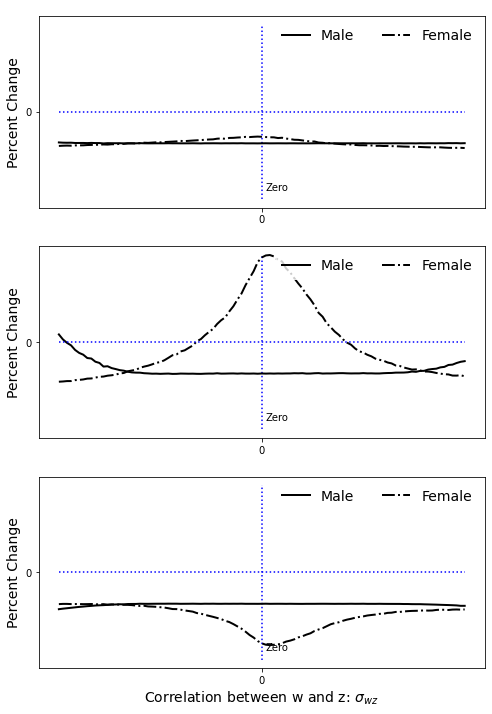}
\end{figure}

\newpage
\subsection{Simulation Results with Continuous Initial Wage Distributions \label{SimResultsContinuousFramework}}

\begin{figure}[H]
\centering 
\scriptsize 
\caption{\small{Pre-SHB Gender Gap in Disclosure Rates in Continuous Framework} \label{fig:OldWageDist_PreSHBGDG}}
\begin{minipage}{18cm}
\emph{\footnotesize{\newline The figure below shows how the pre-ban gender gap (female - male) in disclosure rate changes, as we vary the skewness of the initial male ($\mu^{M}$) and female ($\mu^{F}$) wage distributions. The first three panels (A,B,C) show simulation results for all workers, the next three panels (D,E,F) show results for workers who were asked, and the last three panels (G,H,I) show results for workers who were not asked. For each set of panels, the first (A,D,G) shows results for all workers, the second (B,E,H) shows results for workers who earn below the median wage, and the third (C,F,I) shows results for workers who earn above the median wage. All 9 panels are plotted on the same scale. The vertical dotted blue lines marks the $\mu^{F}$ values equal to the three $\mu^{M}$ values for the three lines in each panel. } \\}
\end{minipage} 
\begin{subfigure}
\centering
\includegraphics[width=0.99\textwidth]{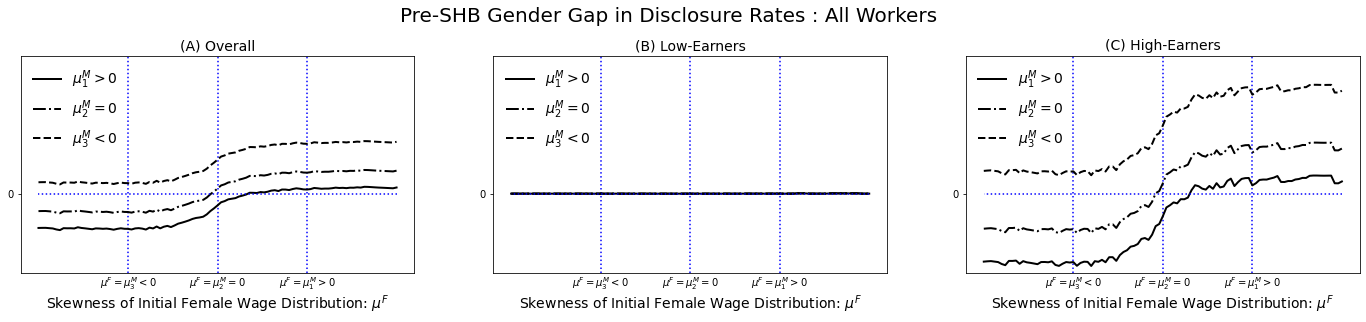}
\end{subfigure}
\begin{subfigure}
\centering
\includegraphics[width=0.99\textwidth]{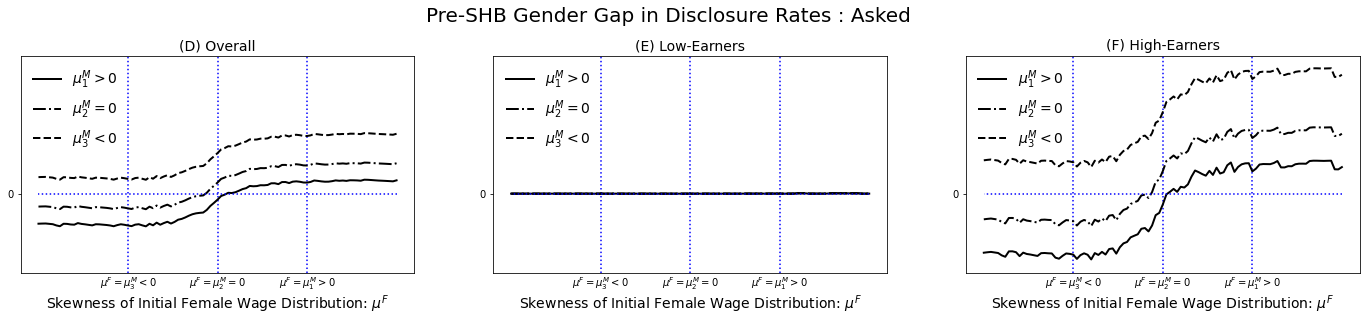}
\end{subfigure}
\begin{subfigure}
\centering
\includegraphics[width=0.99\textwidth]{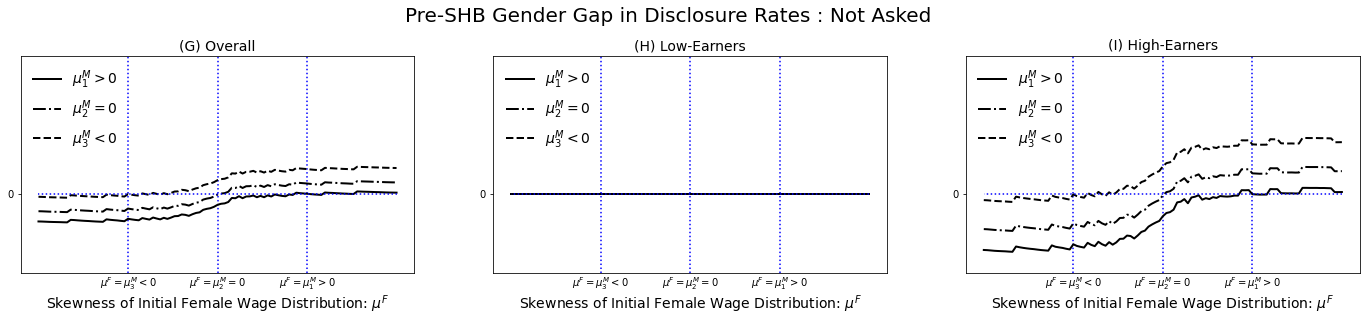}
\end{subfigure}
\end{figure} 

\begin{figure}[H]
\centering
\scriptsize 
\caption{\small{Pre-SHB Effects of Enquiry on Disclosure Rates in Continuous Framework} \label{fig:OldWageDist_PreSHBAskEffect}}
\begin{minipage}{18cm}
\emph{\footnotesize{The figure below shows how the difference between disclosure rates among those who are asked and those who are not asked, changes as we change the skewness ($\mu$) of the initial wage distribution. Panel (A) shows the simulation results for all workers, while panels (B) and (C) show results for workers who earn below and above the median wage respectively. All three panels are plotted on the same scale.} \\}
\end{minipage} 
\includegraphics[width=\textwidth]{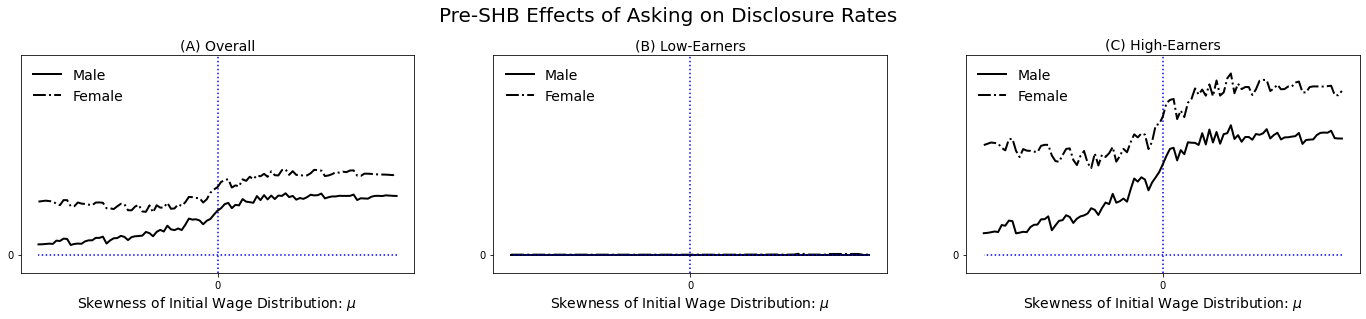}
\end{figure} 

\begin{figure}[H]
\centering
\scriptsize 
\caption{\small{Pre-SHB Income Difference in Disclosure Rates in Continuous Framework} \label{fig:OldWageDist_PreSHBIncEffect}}
\begin{minipage}{18cm}
\emph{\footnotesize{The figure below shows how the difference between disclosure rates among those who earn above and below the median wage, changes as we change the skewness ($\mu$) of the initial wage distribution. Panel (A) shows the simulation results for all workers, while panels (B) and (C) show results for workers who were asked and not asked respectively. All three panels are plotted on the same scale.}\\}
\end{minipage} 
\includegraphics[width=\textwidth]{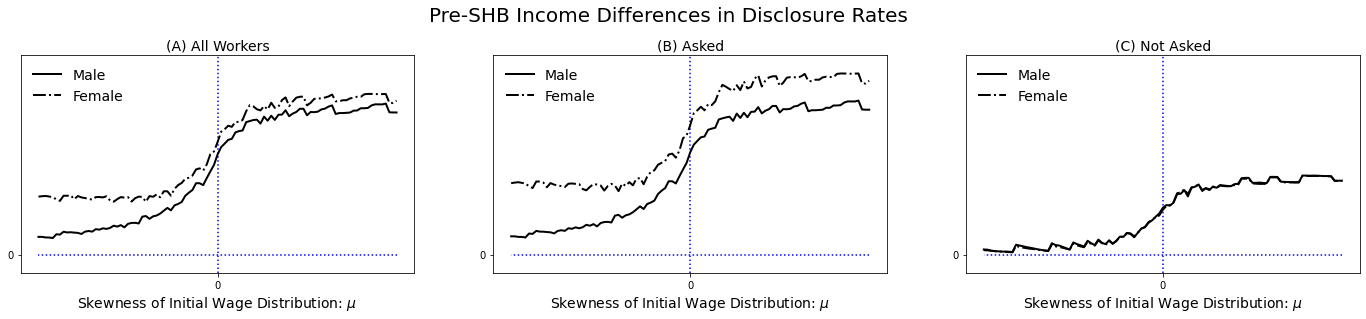}
\end{figure} 

\begin{figure}[H]
\centering 
\scriptsize 
\caption{\small{SHB Effects on Disclosure Rates in Continuous Framework}\label{fig:OldWageDist_SHBEffectonDiscRates}}
\begin{minipage}{18cm}
\emph{\footnotesize{The figure below shows how the effects of SHB on disclosure rates changes as we vary the skewness ($\mu$) of initial wage distribution. Panel (A) shows the simulation results for all workers, while panels (B) and (C) show the simulation results for workers who earn below and above the median wage respectively. All three panels are plotted on the same scale.}\\}
\end{minipage} 
\includegraphics[width=0.99\textwidth]{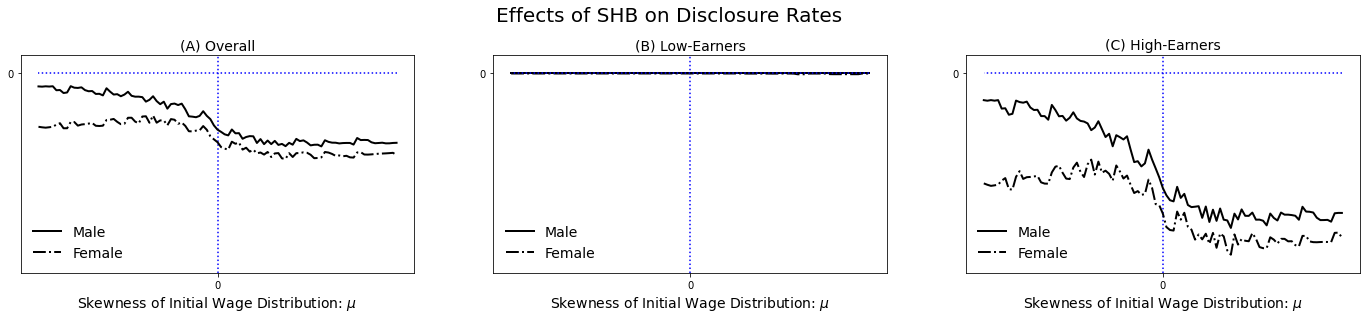}
\end{figure} 

\begin{figure}[H]
\centering 
\scriptsize 
\caption{\small{Post-SHB Income Difference in Disclosure Rates in Continuous Framework}\label{fig:OldWageDist_PostSHBIncEffect}}
\begin{minipage}{18cm}
\emph{\footnotesize{The figure below shows how the post-SHB difference in disclosure rates between those who earn above and below the median wage, changes as we vary the skewness ($\mu$) of the initial wage distribution.}\\}
\end{minipage} 
\includegraphics[width=0.99\textwidth]{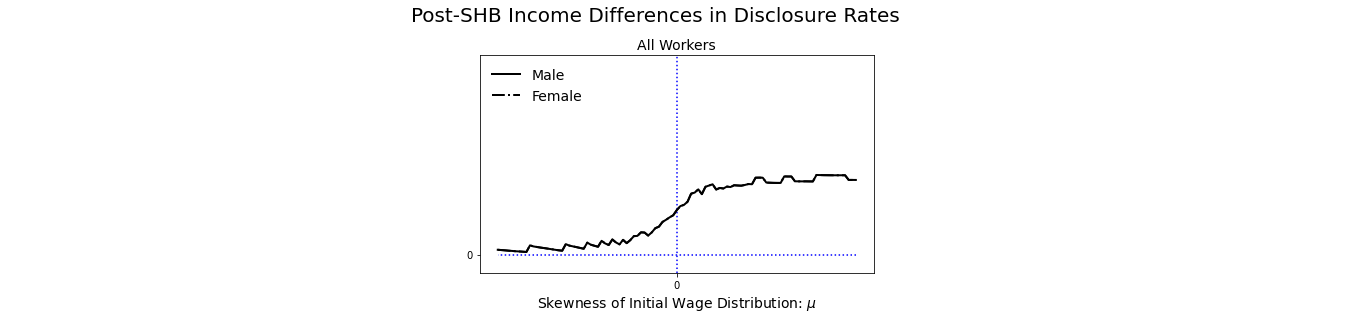}
\end{figure} 

\begin{figure}[H]
\centering 
\scriptsize 
\caption{\small{Post-SHB Gender Gap in Disclosure Rates in Continuous Framework}\label{fig:OldWageDist_PostSHBGDG}}
\begin{minipage}{18cm}
\emph{\footnotesize{The figure below shows how the post-SHB gender gap (female-male) in disclosure rates changes as we vary the skewness of the initial wage distributions for men ($\mu^{M}$) and women ($\mu^{F}$). Panel (A) shows the results for all workers, while panels (B) and (C) show the effects for those who earn below and above the median wage respectively. For each panel, the three vertical lines mark three values where $\mu^{F}$ equals the three values of $\mu^{M}$ at which the three lines are drawn. All three panels are plotted on the same scale. }\\}
\end{minipage} 
\includegraphics[width=0.99\textwidth]{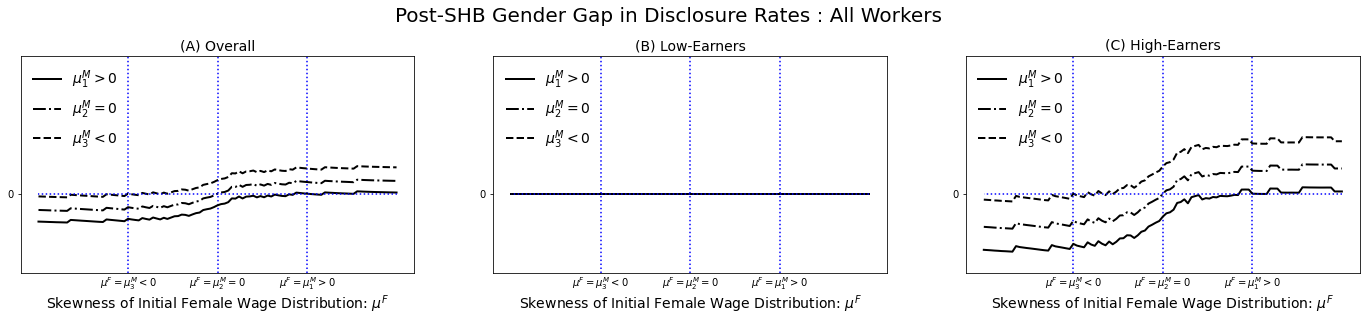}
\end{figure} 

\begin{figure}[H]
\centering 
\scriptsize 
\caption{\small{Gender Gap in Wages in Continuous Framework}\label{fig:OldWageDist_PreEffectSHBGPG}}
\begin{minipage}{18cm}
\emph{\footnotesize{The figure below shows how the pre-SHB gender wage gap (left panel) and SHB Effects on gender wage gap (right panel), change as we vary the skewness of initial wage distributions of men ($\mu^{M}$) and women ($\mu^{F}$). The three vertical lines mark three values at which $\mu^{F}$ equal to the three values of $\mu^{M}$ for which we plot the lines.}\\}
\end{minipage}
\begin{subfigure} 
\centering
\includegraphics[width=0.49\textwidth]{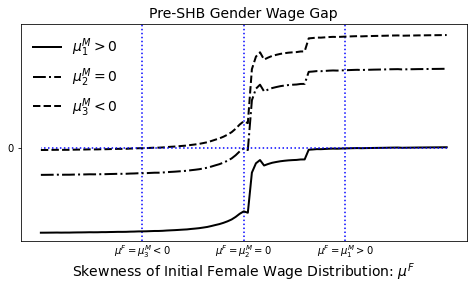} 
\end{subfigure}
\hfill
\begin{subfigure} 
\centering
\includegraphics[width=0.49\textwidth]{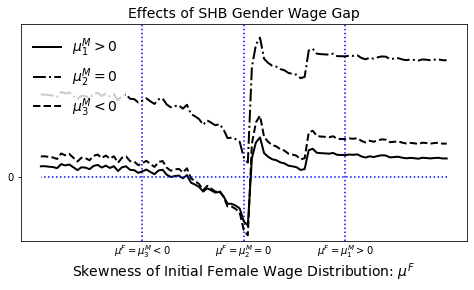} 
\end{subfigure}
\end{figure} 

\begin{figure}[H]
\centering 
\scriptsize 
\caption{\small{SHB Effects on Mean Wages in Continuous Framework}\label{fig:OldWageDist_SHBEffectonMeanWage}}
\begin{minipage}{18cm}
\emph{\footnotesize{The figure below shows how the simulation effects of SHB on mean wages change as we vary the skewness ($\mu$) of initial wage distributions. Panel (A) shows results for all workers, while panels (B) and (C) show results for those who earn below and above the median wage respectively. All three panels are plotted on the same scale.}\\}
\end{minipage} 
\includegraphics[width=0.99\textwidth]{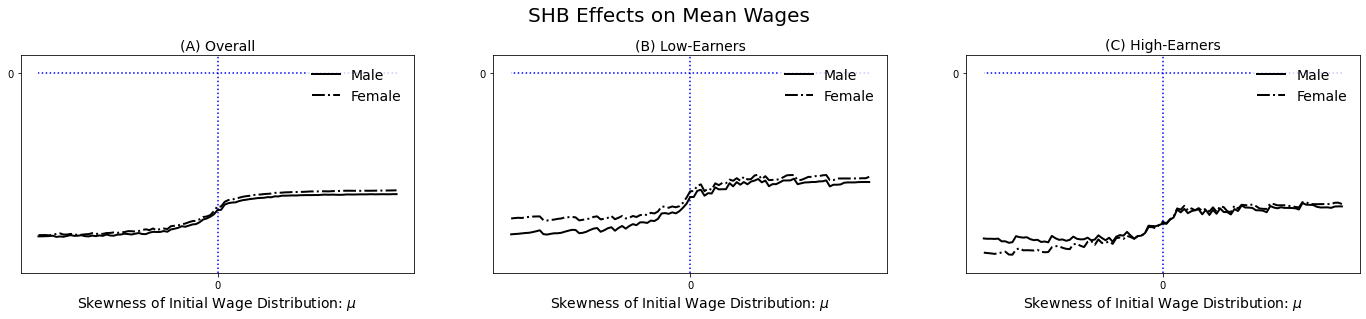}
\end{figure} 

\begin{figure}[H]
\centering 
\scriptsize 
\caption{\small{SHB Effects on Auto-Correlation in Wages in Continuous Framework}\label{fig:OldWageDist_SHBEffectonAutoCorrWage}}
\begin{minipage}{18cm}
\emph{\footnotesize{The figure below shows how the simulation effects of SHB on how the auto-correlation in wages changes as we vary the skewness ($\mu$) of initial wage distributions. Panel (A) shows results for all female workers while panels (B) shows results for male workers. In each panel, I show the pre-SHB and post-SHB auto-correlation coefficient, along with the post/pre ratio. Both panels are plotted on the same scale.}\\}
\end{minipage} 
\includegraphics[width=0.60\textwidth]{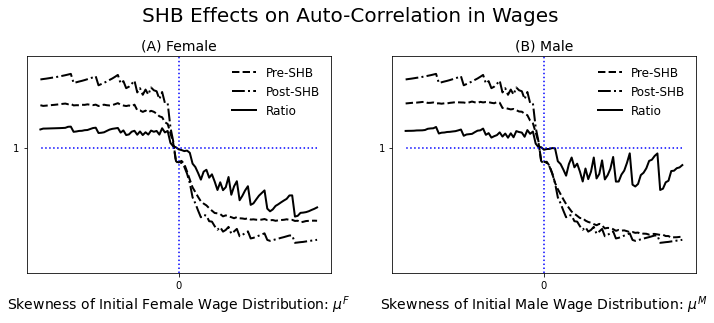}
\end{figure} 

\begin{figure}[H]
\centering 
\scriptsize 
\caption{\small{SHB Effects on Standard Deviation in Wages in Continuous Framework}\label{fig:OldWageDist_SHBEffectonMeanWage}}
\begin{minipage}{18cm}
\emph{\footnotesize{The figure below shows how the simulation effects of SHB on standard deviation in wages change as we vary the skewness ($\mu$) of initial wage distributions. Panel (A) shows results for all workers, while panels (B) and (C) show results for those who earn below and above the median wage respectively. All three panels are plotted on the same scale.}\\}
\end{minipage} 
\includegraphics[width=0.99\textwidth]{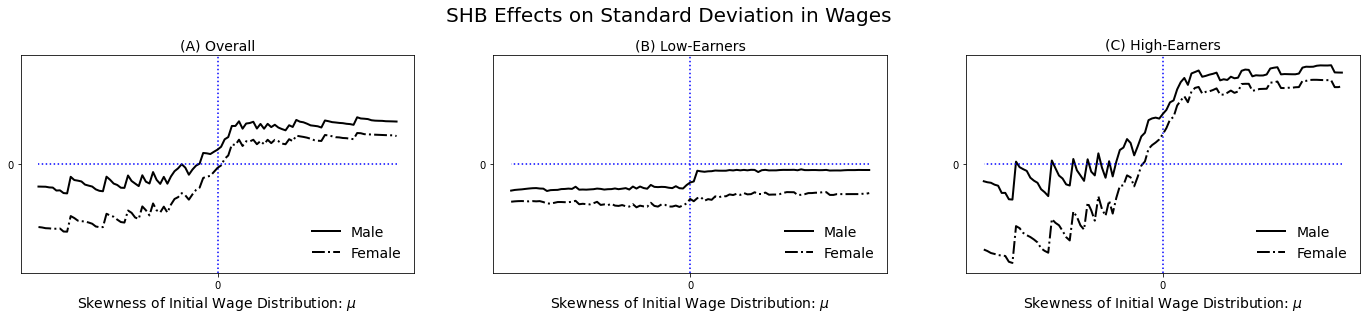}
\end{figure}